\documentclass{jkas}

\def\year{2014} % publication year
\def\month{May} % publication month
\def\volume{00} % journal volume
\def\issue{0} % journal issue
\def\beginpage{1} % first page of article
\def\endpage{99} % last page of article
\def\received{February 30, 2014} % date paper was received by JKAS
\def\accepted{February 31, 2014} % date of acceptance
\date{Received \received; accepted \accepted}

\newcommand{\degr}{^{\circ}}
\usepackage[dvips]{color}

\title{
Interferometric Monitoring of Gamma--Ray Bright Active Galactic Nuclei II: Frequency Phase Transfer\thanks{Part of a special issue on the Korean VLBI Network (KVN)}%\thanks{This document deals with the technical aspects of writing a JKAS paper, \emph{not} with editorial policy or scientific questions.}
}

\author[1]{Juan-Carlos~Algaba}%\thanks{The actual author of this document, additional dummy authors have been added for illustration purposes only.}}
\author[1]{Guang-Yao~Zhao}
\author[1,2]{Sang-Sung~Lee}
\author[1]{Do-Young Byun}
\author[1,2]{Sin-Cheol Kang}
\author[3]{Dae-Won Kim}
\author[3]{Jae-Young Kim}
\author[4]{Jeong-Sook Kim}
\author[1,2]{Soon-Wook Kim}
\author[1]{Motoki Kino}
\author[1,5]{Atsushi Miyazaki}
\author[3]{Jong-Ho Park}
\author[3]{Sascha Trippe}
\author[1]{Kiyoaki Wajima}

\affil[1]{Korea Astronomy and Space Science Institute, 776, Daedeokdae-ro, Yuseong-gu, Daejeon, Republic of Korea 305-348; \email{algaba@kasi.re.kr}}
\affil[2]{Korea University of Science and Technology, 176 Gajeong-dong, Yuseong-gu, Daejeon, 305, 350, Korea}
\affil[3]{Department of Physics and Astronomy, Seoul National University, 599 Gwanak-ro, Gwanak-gu, Seoul 151, 742, Korea}
\affil[4]{National Astronomical Observatory of Japan, 2-21-1 Osawa, Mitaka, Tokyo 181-8588, Japan}
\affil[5]{Faculty of Science and Engineering, Hosei University, 3-7-2 Kajino-cho, Koganei, Tokyo 184-8584, Japan}
\def\corrauthor{
J. C. Algaba
}

\def\runningauthor{
J. C. Algaba et al
}

\def\runningtitle{
iMOGABA II: Frequency Phase Transfer
}

\def\keywords{
galaxies: active -- methods: data analysis -- techniques: high angular resolution -- techniques: interferometric
}

\def\abstracttext{
The Interferometric Monitoring of Gamma--ray Bright Active galactic nuclei (iMOGABA) program provides not only simultaneous multifrequency observations of bright gamma--ray detected active galactic nuclei (AGN), but also covers the highest Very Large Baseline Interferometry (VLBI) frequencies ever being systematically monitored, up to 129~GHz. However, observation and imaging of weak sources at the highest observed frequencies is very challenging. In the second paper in this series, we evaluate the viability of the frequency phase transfer technique to iMOGABA in order to obtain larger coherence time at the higher frequencies of this program (86 and 129 GHz) and image additional sources that were not detected using standard techniques. We find that this method is applicable to the iMOGABA program even under non--optimal weather conditions.
}

\begin{document}
\jkashead %% set title, authors, abstract, etc.

%%%%%%%%%%%%%%%%%%%%%%%%%%%%%%%%%%%%%%%%%%%%%%%%%%%%%%%%%%%%%%%%%%%%%
%%% BEGIN MAIN TEXT HERE %%%%%%%%%%%%%%%%%%%%%%%%%%%%%%%%%%%%%%%%%%%%
%%%%%%%%%%%%%%%%%%%%%%%%%%%%%%%%%%%%%%%%%%%%%%%%%%%%%%%%%%%%%%%%%%%%%

\section{Introduction\label{sec:intro}}

Active galactic nuclei (AGN) emit over a wide range of wavelengths from radio to gamma-rays. Most of the $\gamma-$ray sources detected by EGRET telescope were identified as blazars \citep{Mattox01,Sowards-Emmerd03,Sowards-Emmerd04}. This was later confirmed with the Fermi LAT \citep[see e.g.,][]{Abdo09,Kovalev09,Lister09,Abdo10,Ackermann11}. Variability, which seems to be common in these objects, was also seen to be correlated \citep{Kovalev+09}. In some cases a time delay was found, with the flares leading at higher frequencies \citep{Stevens94,Pushkarev10}. Furthermore, monitoring of a sample of EGRET blazars with VLBI at 43~GHz suggested the association with $\gamma-$ray events and the emergence of parsec--scale radio knots moving downstream the jet \citep{Jorstad01}. These findings were corroborated by single--dish and VLBA observations at 22, 37 and 43~GHz \citep{Savolainen02,Lahteenmaki03}.

Based on these findings, it appears that gamma-ray emission is associated with emission arising from the upstream millimeter radio--emiting regions of the jet. The exact location of the gamma--ray flares is nonetheless yet unclear. On one hand, rapid timescales seem to suggest an innermost origin, within the broad line region. On the other hand, connection with parsec scale events suggests larger scales. Additionally, the cause behind the gamma-ray flares  is also not always clear: they might be produced due to a local compression and heating of the plasma, in--situ generation of relativistic particles or variability of flux and magnetic fields. 

The shock--in--jet model proposed by \cite{MarscherGear85} aimed to explain these observations. In this model, a shock wave travels downstream a steady jet and relativistic particles crossing the shock front gain energy. The flaring flux is produced by accelerated particles behind the shock front. Modifications of this model including jet geometry or multiple Compton scattering were later performed and used to successfully model flaring observations \citep[see e.g.,][and references therein]{Fromm11}.  Alternatively, some turbulent multi-zone models can also explain the multi-frequency variability in blazars \citep{Marscher14}. 

The observed properties of synchrotron flares may depend on the peak frequency and its relation with the observed frequency \cite[e.g.,][]{Valtaoja92}. Given that turnover frequencies at the beginning of flares can be $\nu_m>30$~GHz \citep{Fromm13}, observations at millimeter and sub-millimeter wavelengths are crucial in probing the most compact regions, where the radiation is optically thin, and investigate flares in their growth state. Multi-frequency multi-epoch observations of AGN have been shown to be very useful to study the physical properties of these objects. In this sense, monitoring programs can help us understand the emission properties and follow-up the behavior of the innermost regions of AGN, both for individual objects and in a statistical sense. Traditionally, however, such programs have been limited to single dish (hence, resolution--limited) (e.g., UMRAO database\footnote{https://dept.astro.lsa.umich.edu/datasets/umrao.php}, Mets\"ahovi Quasar Research\footnote{http://www.metsahovi.fi/quasar/}) or to relatively low frequency  observations $\lesssim43$~GHz (e.g., MOJAVE\footnote{http://www.physics.purdue.edu/astro/MOJAVE/index.html}, TANAMI\footnote{http://pulsar.sternwarte.uni-erlangen.de/tanami/} or VLBA-BU-BLAZAR\footnote{http://www.bu.edu/blazars/VLBAproject.html}).

The interferometric MOnitoring of GAmma-ray Bright AGN (iMOGABA) does not suffer from these drawbacks, as it provides the highest radio frequencies ever being systematically monitored (22, 43, 86 and 129 GHz) and uses the unique simultaneous multi--frequency capabilities of the Korean VLBI Network (KVN). This program is novel with regards to previous VLBI observations at 86 and 129~GHz and has thus a clear advantage over other similar programs to investigate the validity of the various models to explain $\gamma-$ray flares in AGN. The iMOGABA specifications and science goals are discussed in \cite{Lee13,Lee15} and \cite{Wajima15}.

Among a series of early results in the iMOGABA program, in this paper we particularly evaluate the viability of the frequency phase transfer (FPT) technique. We prove its usefulness in order to detect and obtain images of faint sources at hight frequencies that could not be obtained otherwise with standard techniques. The organization of this paper is as follows. In Section 2 we briefly introduce the methodology used for the calibration technique; in Section 3 we summarize the observations and data reduction; in Section 4 we present the results of the FPT and in Section 5 we discuss them. We summarize and provide our conclusions in Section 6.

\section{Frequency Phase Transfer}

High frequency VLBI observations are challenging. First, the antenna, electronics and optics are much more demanding for high frequencies, which results in much higher receiver temperatures. Likewise, atmospheric effects, and particularly tropospheric effects, become much more important both in terms of opacity and coherence time. For a review of the system performance, in particular in K and Q bands, see \cite{Lee14}. Moreover, due to the optically thin spectra of most of the AGN at such frequencies, their fluxes become relatively weak and the number of effectively observable sources becomes very limited \citep{Lee08}.

Standard VLBI data handling includes data reduction in \textsc{Aips} consisting of amplitude and phase calibration, splitting the various sources at each frequency and independently imaging them in \textsc{Difmap} using hybrid mapping with phase--only self-calibration. In practice, this is not always possible for all iMOGABA target sources. Given a system equivalent flux density (SEFD) larger than a few thousand at frequencies $\gtrsim43$GHz for KVN antennas, a longer integration time (few minutes) is needed for the faintest sources (few tens of mJy) for secure detection. However, the coherence times are normally limited to shorter scales by the sky conditions.

In order to overcome tropospheric limitations in VLBI observations, phase reference techniques have been typically used \citep[see e.g.,][]{Alef89,BeasleyConway95}. These methods are however very challenging over frequencies above 43~GHz and only a single observation has been proved successful at 86~GHz for a pair of sources separated by only 14 arcseconds \citep{Porcas02}. The frequency phase transfer (FPT) technique aims to circumvent this limitation. Under the assumption that the tropospheric path delay is independent of the frequency, we can correct for it at high frequencies by extrapolating from lower frequencies.

Early applications of this technique were performed on connected arrays such as the Nobeyama millimeter array \citep{Asaki98} or the VLA \citep{Carilli99}. 
Using fast frequency switching with the VLBA, mm-VLBI observations were performed by \cite{Middelberg05} to increase the coherence time; furthermore, successful astrometric calibration using source frequency phase reference were performed by \cite{Dodson09} and \cite{Rioja11}. 
The multi--channel receivers at the KVN allow for simultaneous observations at 22, 43, 86 and 129 GHz \citep{Oh11,Han13} and are therefore ideal to apply this technique, as they avoid the need of frequency switching and interpolation. Performance of the Korean VLBI Network was first tested by \cite{Rioja14a,Rioja14b,Rioja15}, \cite{Jung14} or \cite{Zhao15} on specifically designed observations.

\subsection{Method}
Frequency phase transfer (FPT) uses the observations of a source at a low frequency to calibrate the tropospheric errors at a higher frequency. This method is a simplification of the source-frequency phase reference described in detail in \cite{Rioja11} and \cite{Rioja14a}, and is particularly useful when we do not require accurate astrometry of the sources. We briefly discuss the application of this technique to our case here. The residual phase errors for the reference (low) frequency can be described as:
\begin{equation}
\label{phaselow}
\phi_{low}=\phi^{str}_{low}+\phi^{geo}_{low}+\phi^{trop}_{low}+\phi^{ion}_{low}+\phi^{inst}_{low}+\phi^{th}_{low}+2\pi n_{low},
\end{equation}
where the indices indicate contributions from structure, geometric, tropospheric, ionospheric, instrumental, and thermal effects.  The last term originates from the $2\pi$ ambiguity for the phase, with $n_{low}$ an integer. A point source has $\phi^{str}_{low}=0$, and the $\phi^{th}_{low}$ term has a negligible contribution compared with the rest. A similar expression can be written for the residual phase errors for the target (high) frequency $\phi_{high}$.

Our low-frequency self-calibrated map contains a set of antenna-based corrections $\phi^{selfcal}_{low}$, which account for the structure. If we scale by the frequency ratio $R=\nu_{high}/\nu_{low}$ and transfer for the tropospheric calibration of the high-frequency data, the residual target phases are
\begin{eqnarray}
\phi^{FPT}=\phi_{high}-R\cdot\phi^{self cal}_{low}=\nonumber\\
=\phi^{str}_{high}
+(\phi^{geo}_{high}-R\cdot\phi^{geo}_{low})
+(\phi^{tro}_{high}-R\cdot\phi^{tro}_{low})+\nonumber\\
+(\phi^{ion}_{high}-R\cdot\phi^{ion}_{low})
+(\phi^{ins}_{high}-R\cdot\phi^{ins}_{low})+\nonumber\\
+2\pi (n_{high}-R \cdot n_{low}),
\end{eqnarray}
where we have already excluded the thermal term for simplicity.

The tropospheric residual errors scale linearly with frequency, so we can approximate this term to become negligible. The ionospheric errors are inversely proportional to the frequency, the geometric residual errors cancelling out except for a frequency--dependent source position shift $\theta$, and depending on the baseline $D$. Thus, the residual target phases become
\begin{eqnarray}
\phi^{FPT}=\phi^{str}_{high}+ 2\pi D\cdot\theta
+\left(\frac{1}{R}-R\right)\phi^{ion}_{low}+\nonumber\\
+(\phi^{ins}_{high}-R\cdot\phi^{ins}_{low})+
+2\pi (n_{high}-R \cdot n_{low}).
\end{eqnarray}

Thus, high frequency visibilities are free of random and tropospheric errors but contain un--modelled ionospheric and instrumental residual phase errors that are blended with the core shift or the positional phases. In general, this limits the coherence time to timescales of several tens of minutes, which is in any case much longer than the initial coherence time. The removal of long term drifts  and other offsets requires self-calibration and hybrid mapping and in any case removes any positional information included in $\theta$. Alternatively, including the observations of a second source, it is possible to recover the astrometric signature; this is the basis for source frequency phase reference (SFPR).

The ratio $R$ is highly recommended to be an integer, to avoid unreasonable extrapolations when phase wraps occur in the data. For example, if the frequency ratio is a non integer (e.g., R=2.5), and the reference frequency wraps from 359$\degr$ to 0$\degr$, then the scaled frequency wraps from 897.5$\degr$ (=177.5$\degr$) to 0$\degr$, introducing an undesired phase wrap in the target frequency of 177.5$\degr$ \citep{Middelberg05}.

As discussed in depth in \cite{Rioja11}, the errors induced due to the FPT technique are negligible: first, given the identical line of sight at the two frequencies, other than micro--arcsecond core--shifts, both geometric and tropospheric (static component) errors can be proven to be almost identical to zero. The errors of the dynamic component of the troposphere also cancels out for our synchronous multi--frequency observations. Thus, based on this error analysis, the errors associated with the FPT high frequency images are dominated by the usual thermal, ionospheric, and instrumental errors and no further error analysis is needed.

\section{Observations and Data Reduction}

In order to properly evaluate the conditions for FPT, we selected two different epochs from the iMOGABA monitoring: epoch 9 (hereafter iMOGABA9), observed on 2013/11/19, which is one of the epochs with the best overall weather on KVN stations, and epoch 15 (hereafter (iMOGABA15), observed on 2014/06/13, which accounts for one of the most overall poor weather. For both epochs, we observed up to 31 gamma-ray bright AGN\footnote{For a complete list of sources, see \url{sslee.kasi.re.kr}} during a 22-hours session at 21.7, 43.4, 86.8 and 129.3~GHz simultaneously using the snapshot mode. The total 16~IFs (spectral windows) with a full bandwidth of 256~MHz were equally distributed among the four frequencies, with 64~MHz each. The typical scan length was about 5 minutes, and sources were observed over 2--10 scans (about 5 scans on average), depending on their flux and declination. For example, in iMOGABA15, 0827+243 and 3C84 were observed for 2 and 10 scans respectively.

Data reduction was separated in two steps. For the first step, typical VLBI data reduction was employed: after loading the data in the NRAO \textsc{Aips} package, data was flagged according to the observing logs. In the case of iMOGABA15, unusual pointing offsets as large as 50 arc seconds in azimuth direction and 400 arc seconds in elevation on the KVN Tamna antenna during the beginning (UT 05:42--16:02) of the epoch were detected, causing large amplitude losses for the baselines with Tamna. Furthermore, we found an additional amplitude loss at 86 and 129~GHz of about 30\% related with Tamna that was not associated with the antenna offset. We checked for other possible causes and ruled out its origin arising from source structure, source coordinates or different polarization feeds in this antenna. We suggest the cause may be related to inaccurate system temperature measurements.

Correction for amplitudes in cross-correlation spectra due to errors in sampler thresholds and amplitude calibration were performed with \textsc{Aips} tasks ACCOR and APCAL respectively. Bandpass correction was applied using the sources 3C454.3 and 3C84. Manual phase calibration was done to estimate the residual antenna-based residuals phases using a 10 seconds time interval on a scan of 3C454.3 as calibrator. A final fringe fitting was performed with a solution interval of 30 seconds, signal--to--noise ratio cutoff of 5, and combining all the IFs for a given frequency. The reference antennas used were Ulsan  for iMOGABA9 and Tamna for iMOGABA15, which performed better than the other two antennas despite the amplitude problems due to comparatively better weather conditions. The data were then split and imaged with \textsc{Difmap}.

The second step included the implementation of the FPT technique. We removed the source structure contribution by using the low (22 or 43~GHz) frequency hybrid maps from the previous step as input models for the fringe-fitting. We then scaled the phases using the parameter `XFER' of the task SNCOR and applied the solutions to the high frequency (86 or 129~GHz) visibilities of the same source. Given that the various frequencies are simultaneously observed in the KVN system, no interpolation was needed. The results of this calibration are free of tropospheric errors, with increased coherence time. We finally re--fringe--fitted the FPT--calibrated data with a larger solution interval of 300 seconds and signal--to--noise ratio threshold equal to 3. 

For iMOGABA9, when solutions were found, the high frequency data were then split and imaged with \textsc{Difmap}. This was not possible for iMOGABA15 because of the amplitude offset problems which do not affect significantly the FPT technique other than in terms of SNR, but may have a dramatic impact on imaging. This is specially true for the case of KVN for which poor UV coverage and no closure amplitudes can be obtained, due to the low number of baselines. See \cite{Rioja15} for alternative ways to estimate the global amplitude gain corrections with KVN observations.

\section{Results}
\subsection{iMOGABA9}
\subsubsection{Frequency Phase Transfer to 86~GHz}\label{sec:FPT86}
The system equivalent flux density (SEFD) for iMOGABA9 at 86~GHz was around $3-4\times10^3$ Jy. The baseline--based sensitivity limit was then $\sigma^{theo}_{min}\sim$60 mJy and $\sigma^{theo}_{min}\sim$20 mJy for 30 and 300 seconds integration time, respectively, considering a 2-bit sampling, two antennas, and 64~MHz bandwidth. 
Three sources could not be imaged in a robust way at 86~GHz with the standard methods presented in \cite{Wajima15}: 0235+164, 0827+243 and 1343+451.  In Table \ref{theoreticalflux86} we show these sources, their expected peak flux based on an extrapolation from lower frequencies and/or posteriori knowledge, and the detection SNR for an integration time of 30 and 300 seconds.

\begin{table}[t!]
\caption{iMOGABA9 Expected Detection Limits at 86~GHz\label{theoreticalflux86}}
\centering
\begin{tabular}{ccccc}
\toprule
Source   & SEFD (Jy) & $S^{86}$ (mJy)  & SNR$^{86}_{30s}$ & SNR$^{86}_{300s}$ \\
\midrule
0235+164 & 3300 & 400 & 6 & 20 \\
0827+243 & 3900 & 220 & 3 & 10 \\
1343+451 & 3100 & 210 & 4 & 11 \\
\bottomrule
\end{tabular}
\end{table}

Based on these data, it seems clear that the typical reduction analysis used in \cite{Lee15} is not capable to detect these sources with a high SNR$^{86}\gtrsim3$ level, typically used as a cutoff in the fringe fitting. However, if we use instead the FPT technique, we can increase the coherence time to about 5 minutes and integrate the signal within such solution interval, leading to clear detection and imaging of these sources.  

\begin{figure*}[h]
\centering
\includegraphics[angle=-90,width=57mm]{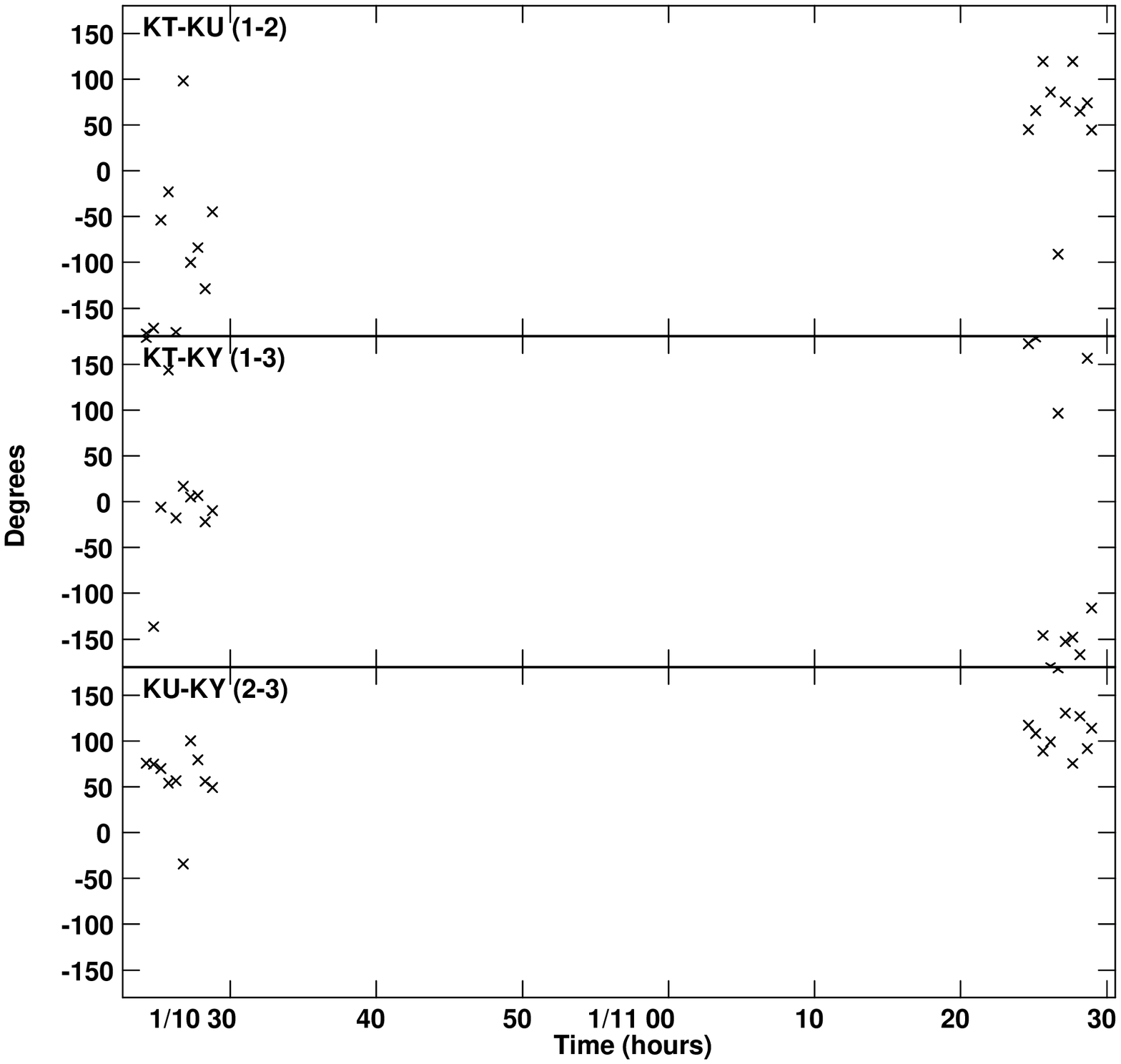}
\includegraphics[angle=-90,width=57mm]{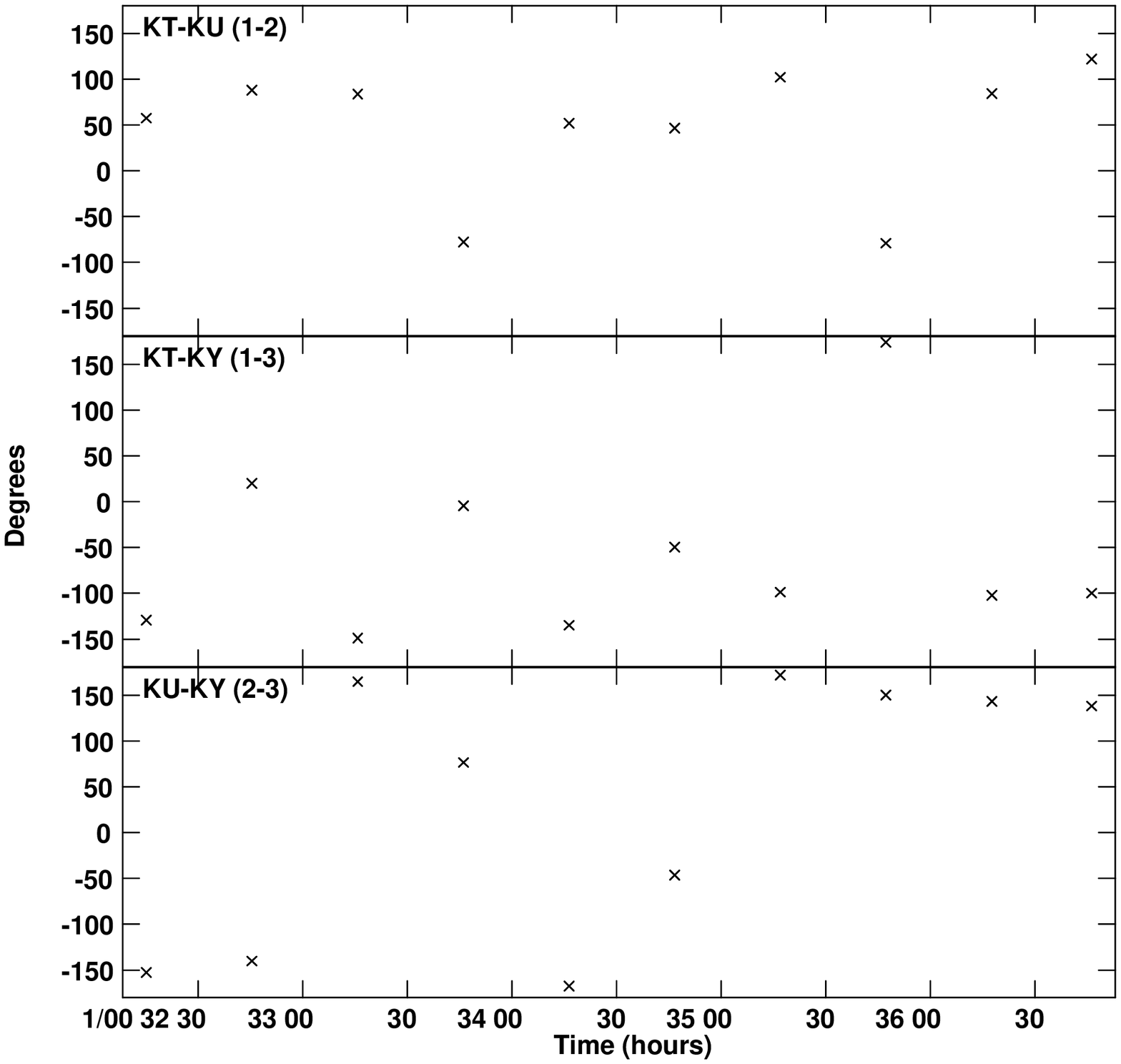}
\includegraphics[angle=-90,width=57mm]{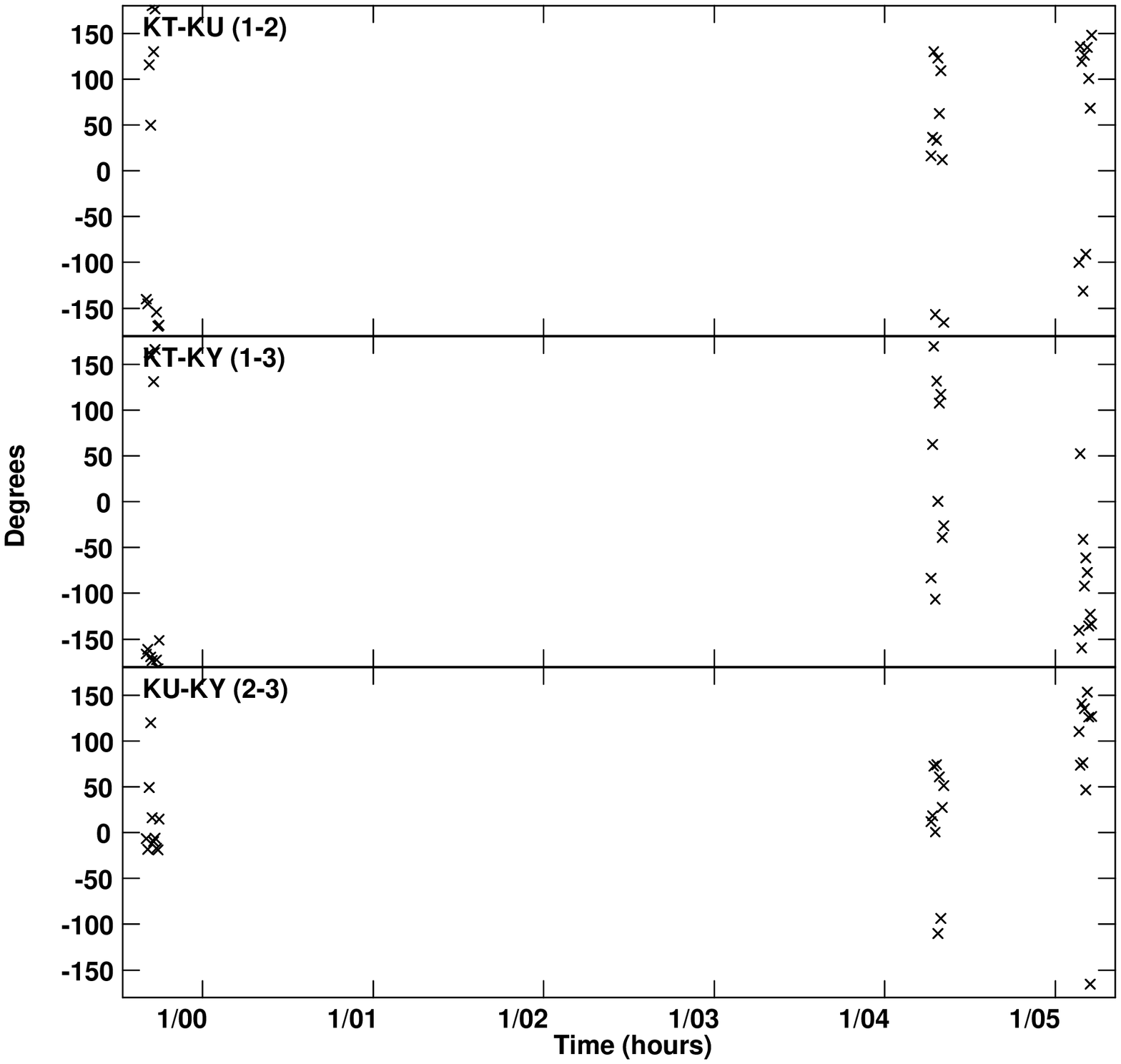}
\begin{tabular}{lll}
\hspace{0.8cm}(a) 0235+164 &\hspace{3.3cm} (b) 0827+243 &\hspace{3.3cm} (c) 1343+451\\
\end{tabular}
\caption{FPT visibility phases for iMOGABA9 observations at 86 GHz. Each data set was calibrated using the scaled solutions from the analysis of the same source at 22 GHz. Plots show temporal average of 30 seconds.\label{FPTphase22-86}}
\end{figure*}

We calibrated the phases at 22~GHz, and transferred the scaled solutions of each source to 86~GHz ($R=\nu_{high}/\nu_{low}=4$). We then applied these solutions to the 86~GHz visibilities. Plots of the visibility phases can be seen in Figure \ref{FPTphase22-86}. Although the limited number of scans prevents us from properly following the phase trends, it is clear that there is a reasonable alignment of the phases around a certain value. We remind here that such phase solutions were either very poor or non-existent for typical self--calibration analysis, while the FPT analysis allows us to find solutions with high SNR for longer integration times (up to the duration of the scans,  $\sim$5 minutes) and therefore permits us to find fringe solutions and perform posterior mapping. 

\begin{figure*}[h]
\centering
\includegraphics[angle=-90,width=57mm]{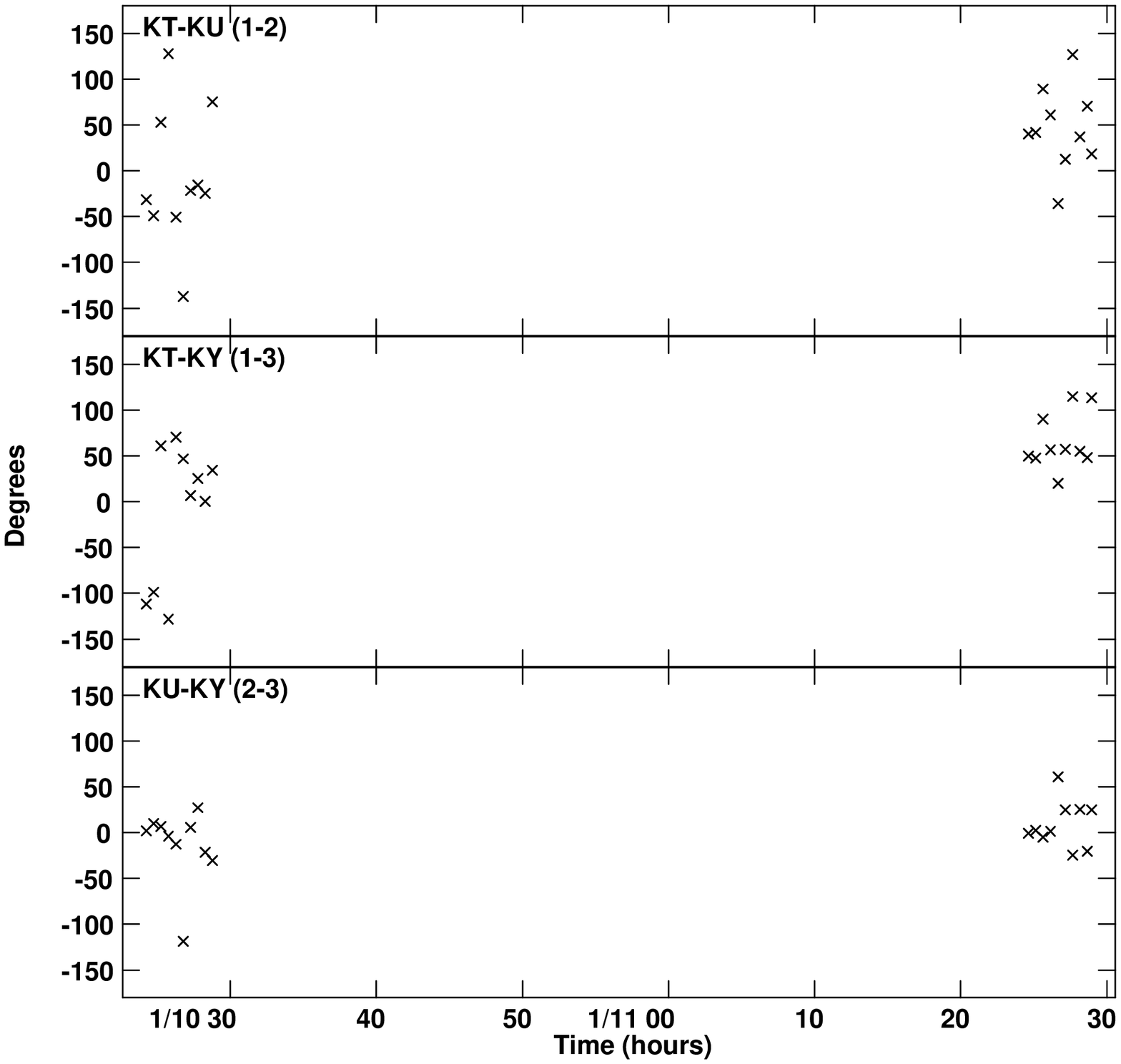}
\includegraphics[angle=-90,width=57mm]{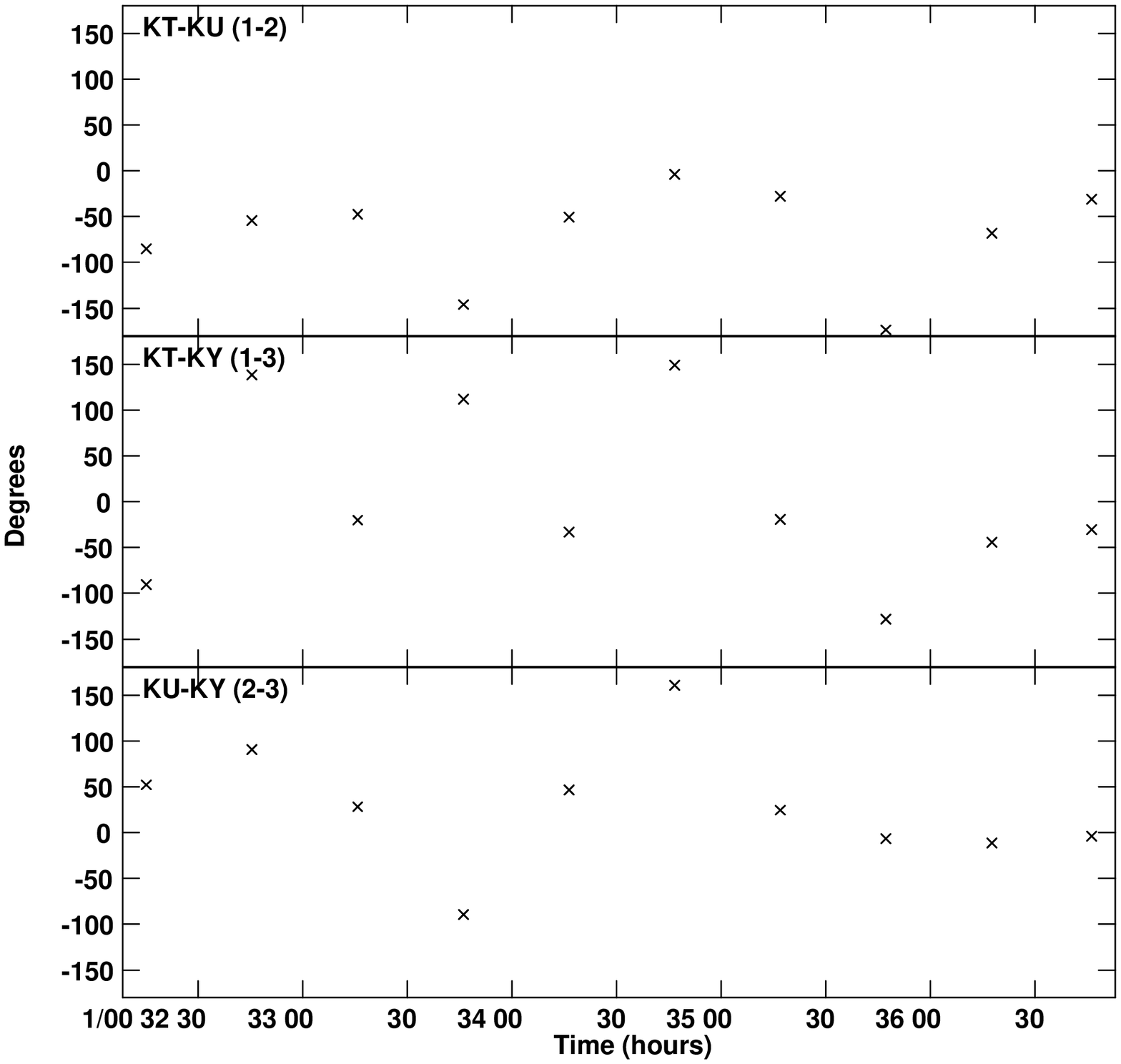}
\includegraphics[angle=-90,width=57mm]{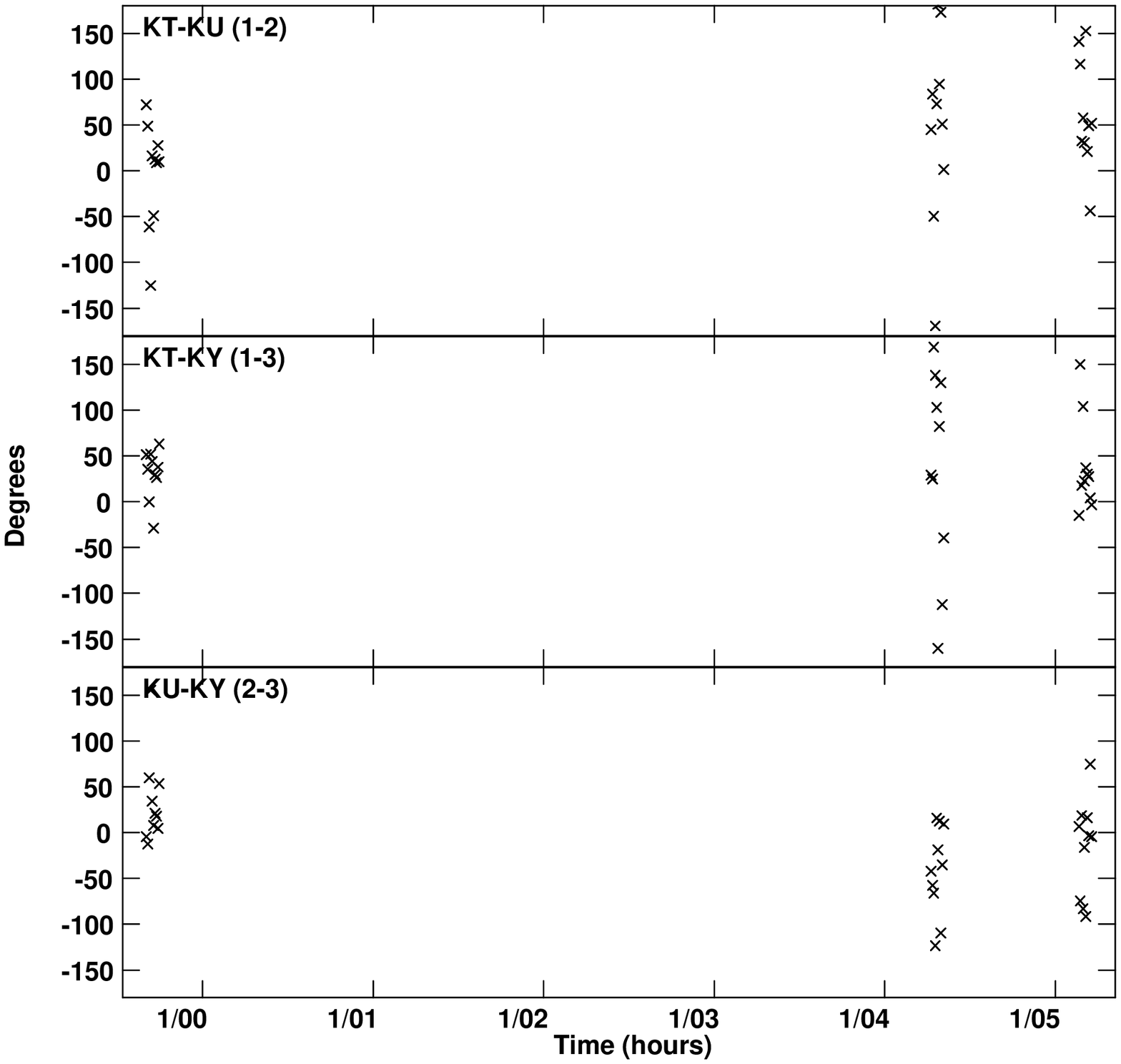}
\begin{tabular}{lll}
\hspace{0.8cm}(a) 0235+164 &\hspace{3.3cm} (b) 0827+243 &\hspace{3.3cm} (c) 1343+451\\
\end{tabular}
\caption{Same as Figure \ref{FPTphase22-86}, but using the scaled solutions from the analysis of the same source at 43 GHz.\label{FPTphase43-86}}
%\vspace{5mm} %% add extra space ONLY when figures/tables are "colliding"!
\end{figure*}

In order to check for consistency and compare our results, we also transferred scaled solutions from phases calibrated at 43~GHz (R=2). We plot such calibrated visibility phases in Figure \ref{FPTphase43-86}, which can be directly compared with Figure \ref{FPTphase22-86}. Although the scaling factor is 2 times smaller for this transfer, naively suggesting a factor of 2 lower scatter in the scaled solutions, it is clear that the fading of these sources at 43~GHz compared to 22~GHz due to its spectral index is degrading the SNR of the phase solutions and hence increasing the residual errors. Indeed, we found that fringe SNR$^22\sim$45 and SNR$^43\sim$28 for 0235+164, 0528+134 and 0735+178. Combined with the overall performance of the KVN system (higher SEFD, larger pointing errors) at higher frequencies, this explains the similarities between the phase solutions transferred from 22 and 43~GHz.

After the FPT, we re-fringed the data to align the phases with a solution interval that effectively covers the whole scan ($\sim$5 minutes). Due to the still significant scatter in some of the calibrated visibilities, some solutions were not found. This was the case, for example, for Tamna baselines on the first scan for 0235+164; or all baselines on the second scan for 1343+451. This effect was particularly severe here, as we obtained only a single solution for each scan due to the long solution interval. This means that a failed solution automatically discarded all the data for a scan. Additionally, this implied that, during imaging, self--calibration also failed for the scans without solutions for a given baseline. In some cases, visibilities of only a single scan were left.

\begin{figure*}[h]
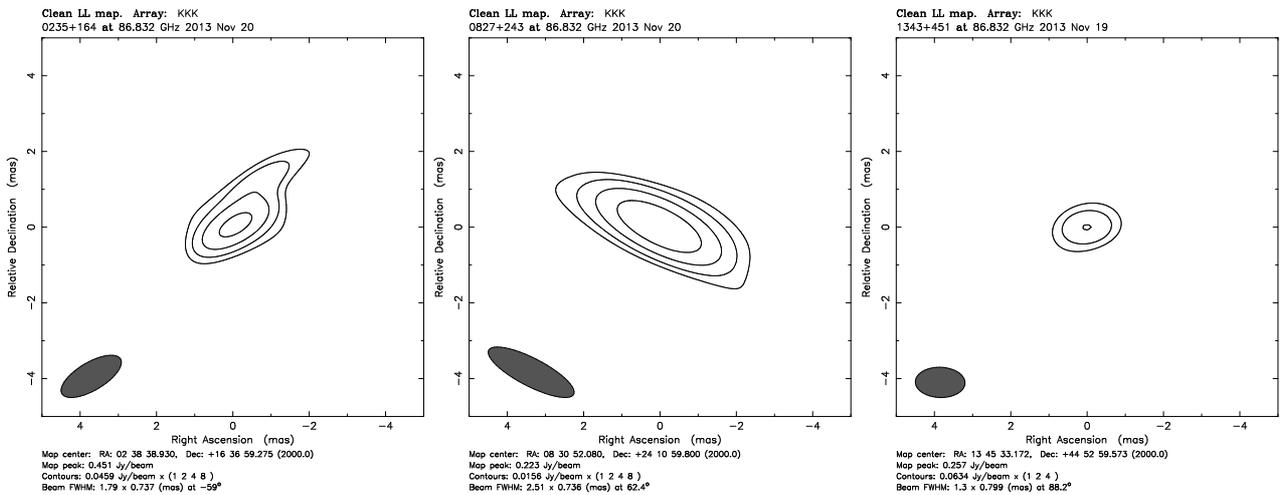

\centering
\includegraphics[angle=0,width=55mm,clip,trim=0 0cm 0 0]{iM9-0235F086PTv3.eps}
\includegraphics[angle=0,width=55mm,clip,trim=0 0cm 0 0]{iM9-0827F086PTv3.eps}
\includegraphics[angle=0,width=55mm,clip,trim=0 0cm 0 0]{iM9-1343F086PTv3.eps}
\caption{iMOGABA9 86 GHz hybrid maps of the sources that could be not imaged with standard procedures, but were successfully imaged with FPT. Contours start at 3$\times$RMS level. \label{86GHzMaps}}
\end{figure*}

All three sources could be imaged after the analysis using FPT was performed. This is in agreement with expectations from Table \ref{theoreticalflux86}. We find almost identical results for visibilities analyzed with FPT from either 22 or 43~GHz after self--calibration was performed. Maps are shown in Figure \ref{86GHzMaps}.

\subsubsection{Frequency Phase Transfer to 129~GHz}
The system equivalent flux density (SEFD) for iMOGABA9 at 129~GHz was around $5-8\times10^3$ Jy. The baseline--based sensitivity limit was then $\sigma^{theo}_{min}\sim$300--500 mJy and $\sigma^{theo}_{min}\sim$30--40 mJy for 30 and 300 seconds integration time, respectively, considering a 2-bit sampling, two antennas and 64~MHz bandwidth.  Up to twelve sources could not be imaged in iMOGABA9 with standard methods at 129~GHz.  In Table \ref{theoreticalflux129} we show these sources, their expected peak flux based on an extrapolated flux form the lower frequencies and the detection SNR for an integration time of 30 and 300 seconds.

\begin{table}[t!]
\caption{iMOGABA9 Expected Detection Limits at 129~GHz\label{theoreticalflux129}}
\centering
\begin{tabular}{ccccc}
\toprule
Source   & SEFD (Jy) & $S^{129}$ (mJy)  & SNR$^{129}_{30s}$ & SNR$^{129}_{300s}$ \\
\midrule
0235+164 & 5600 & 260 & 3 & 8 \\
0528+134 & 5300 & 300 & 3 & 10 \\
0735+178 & 5900 & 380 & 4 & 11 \\
0827+243 & 8000 & 100 & 1 & 2 \\
0836+710 & 6100 & 270 & 2 & 8 \\
1127--145 & 8000 & 750 & 5 & 16 \\
1156+295 & 5100 &   500 & 5 & 17 \\
1222+216 & 5300 & 340 & 3 & 11 \\
1308+326 & 5900 & 360 & 3 & 11 \\
1343+451 & 5300 & 90 & 1 & 3 \\
1611+343 & 5900 & 540 & 5 & 16 \\
3C446      & 5100 &   250 & 3 & 8 \\
\bottomrule
\end{tabular}
\end{table}

Based on the expected values, it seems clear that, in general, no images, or marginal detections would be obtained using standard calibration methods with typical integration times of $\sim$30~seconds for these sources. Moreover, it appears that  in some cases SNR$^{129}_{300s}<10$ for several sources. In particular, SNR$^{129}_{300s}$ is $<5$ for 0827+243 and 1343+451. This has very clear implications: the maximum achievable solution interval, ultimately given by the scan duration, which is about $\sim$5~minutes in these observations, is not enough for a detection. This implies that there is a priori no possibility to image these sources, even using the FPT technique. Improved weather and system conditions or a brightening of the targets would be needed in order to image these two sources up to 129~GHz.

\begin{figure*}[h]
\centering
\includegraphics[angle=-90,width=57mm]{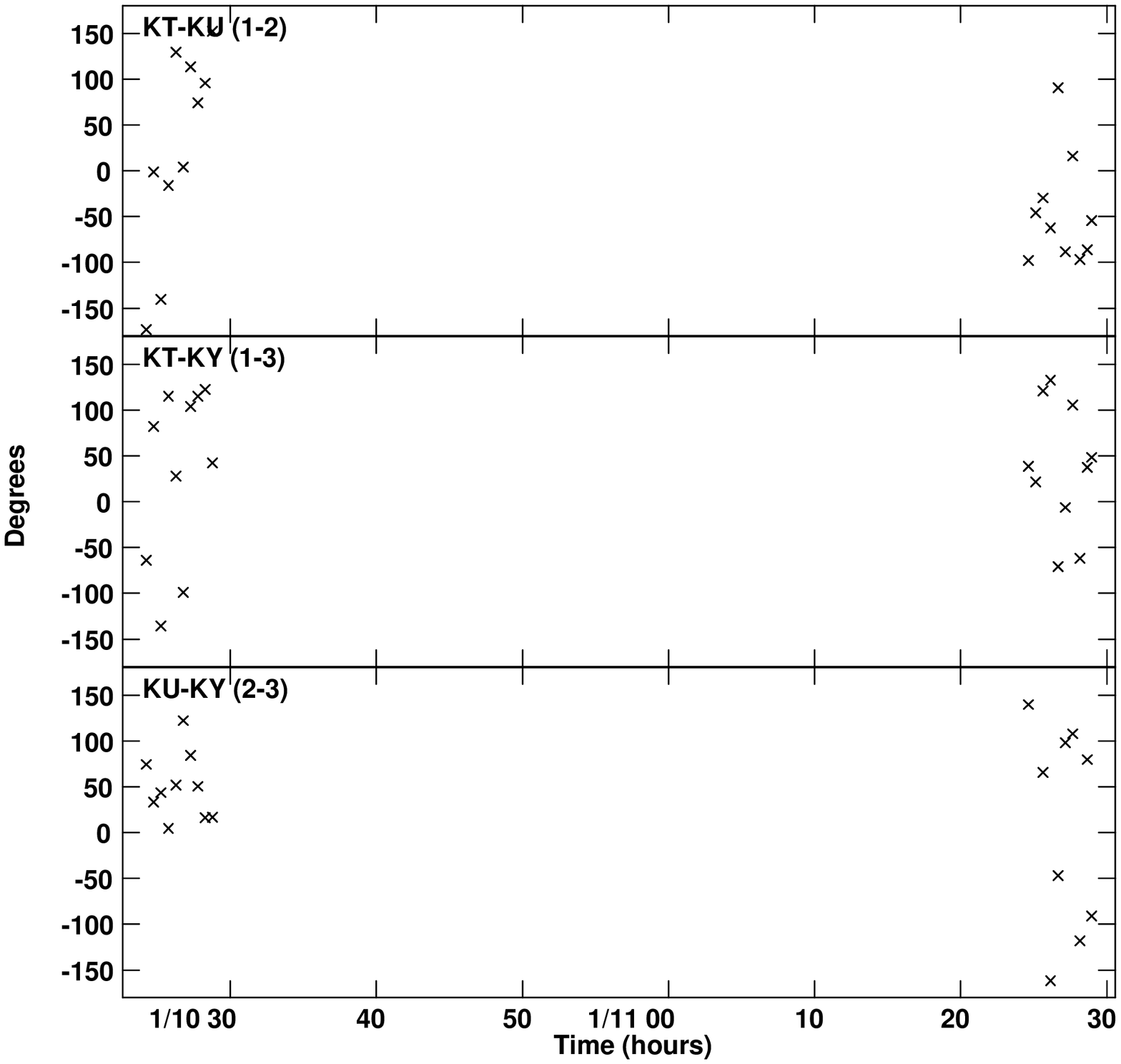}
\includegraphics[angle=-90,width=57mm]{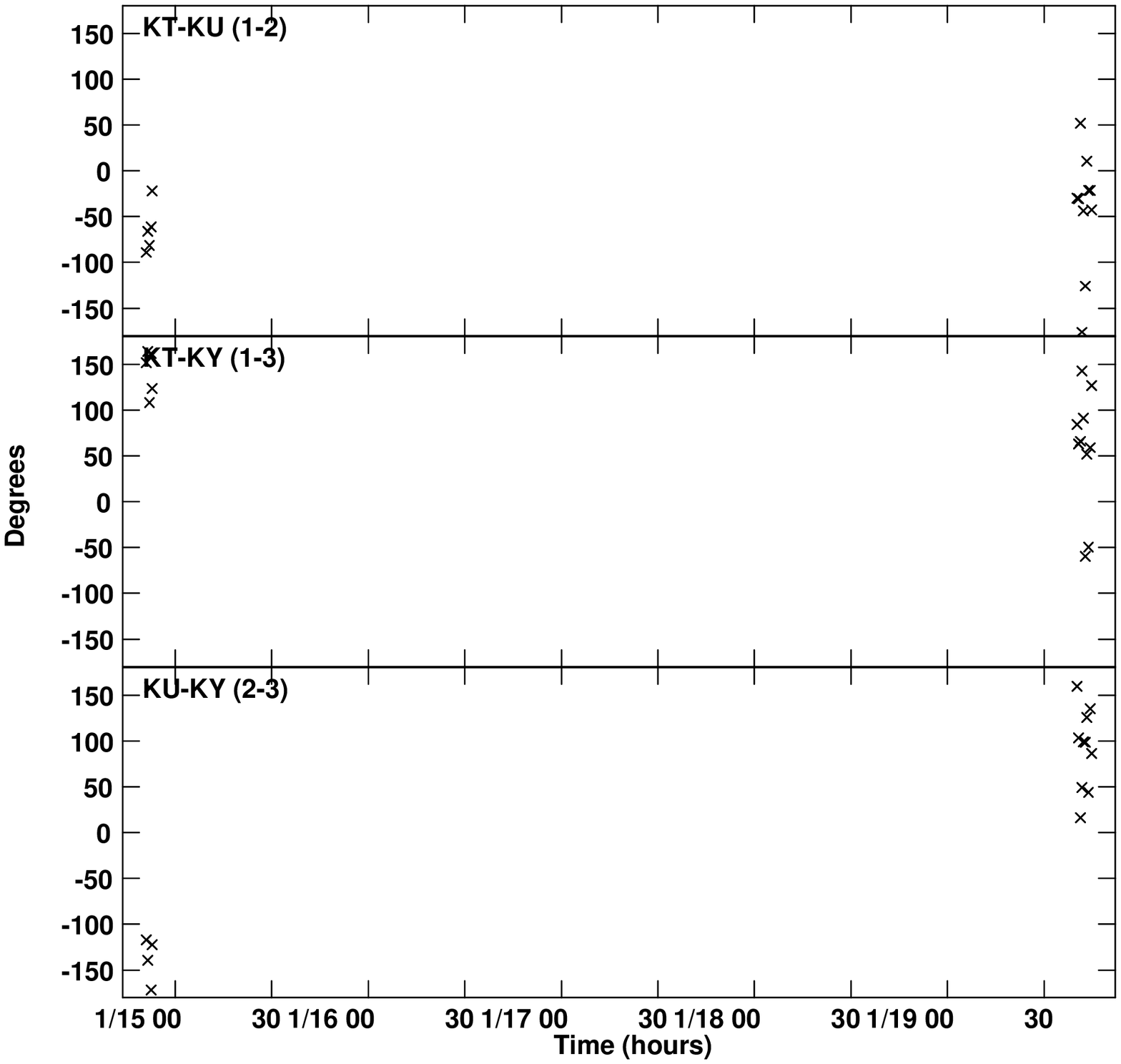}
\includegraphics[angle=-90,width=57mm]{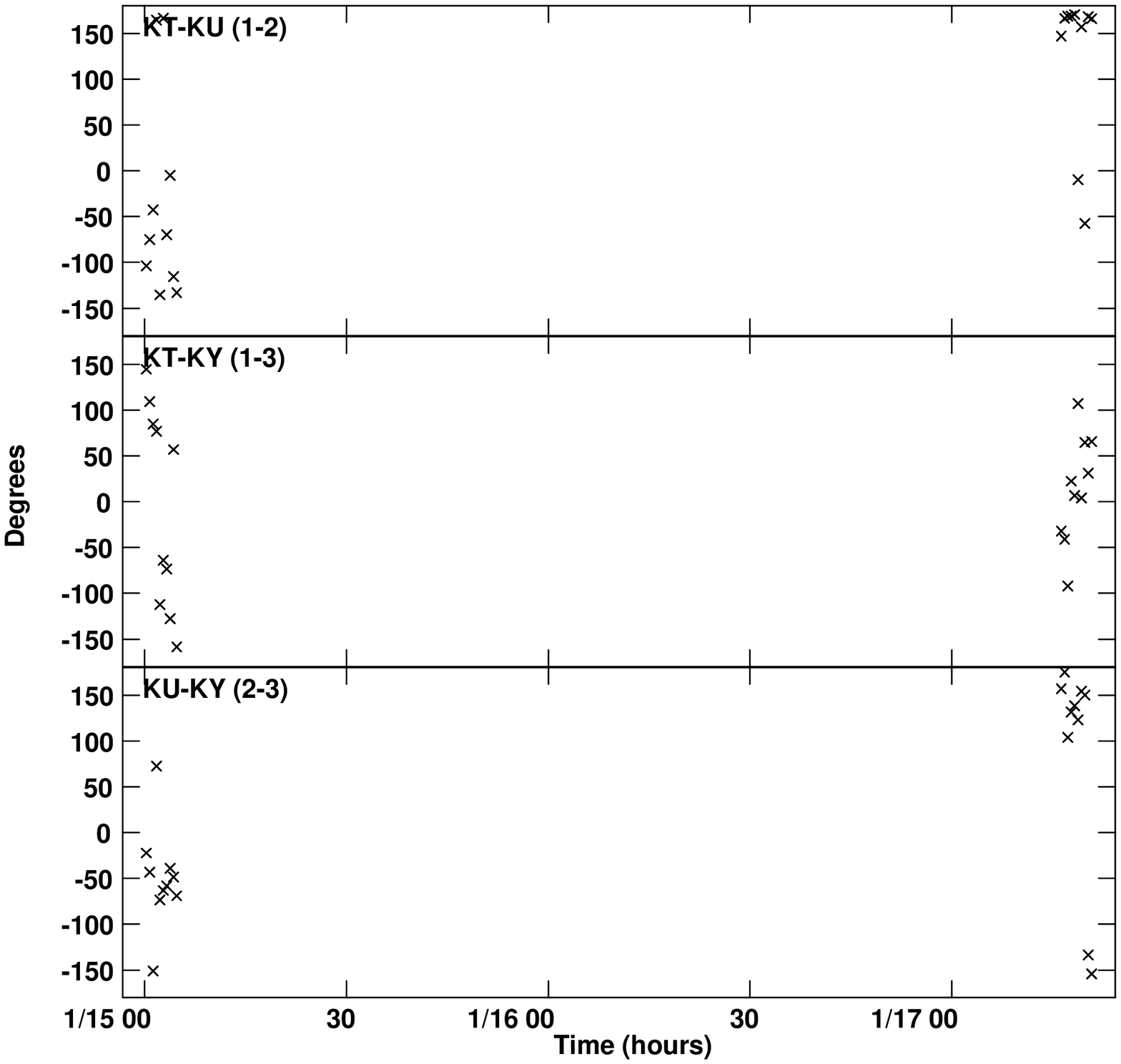}
\begin{tabular}{lll}
\hspace{0.8cm}(a) 0235+164 &\hspace{3.3cm} (b) 0528+134 &\hspace{3.3cm} (c) 0735+178\\
\end{tabular}
\includegraphics[angle=-90,width=57mm]{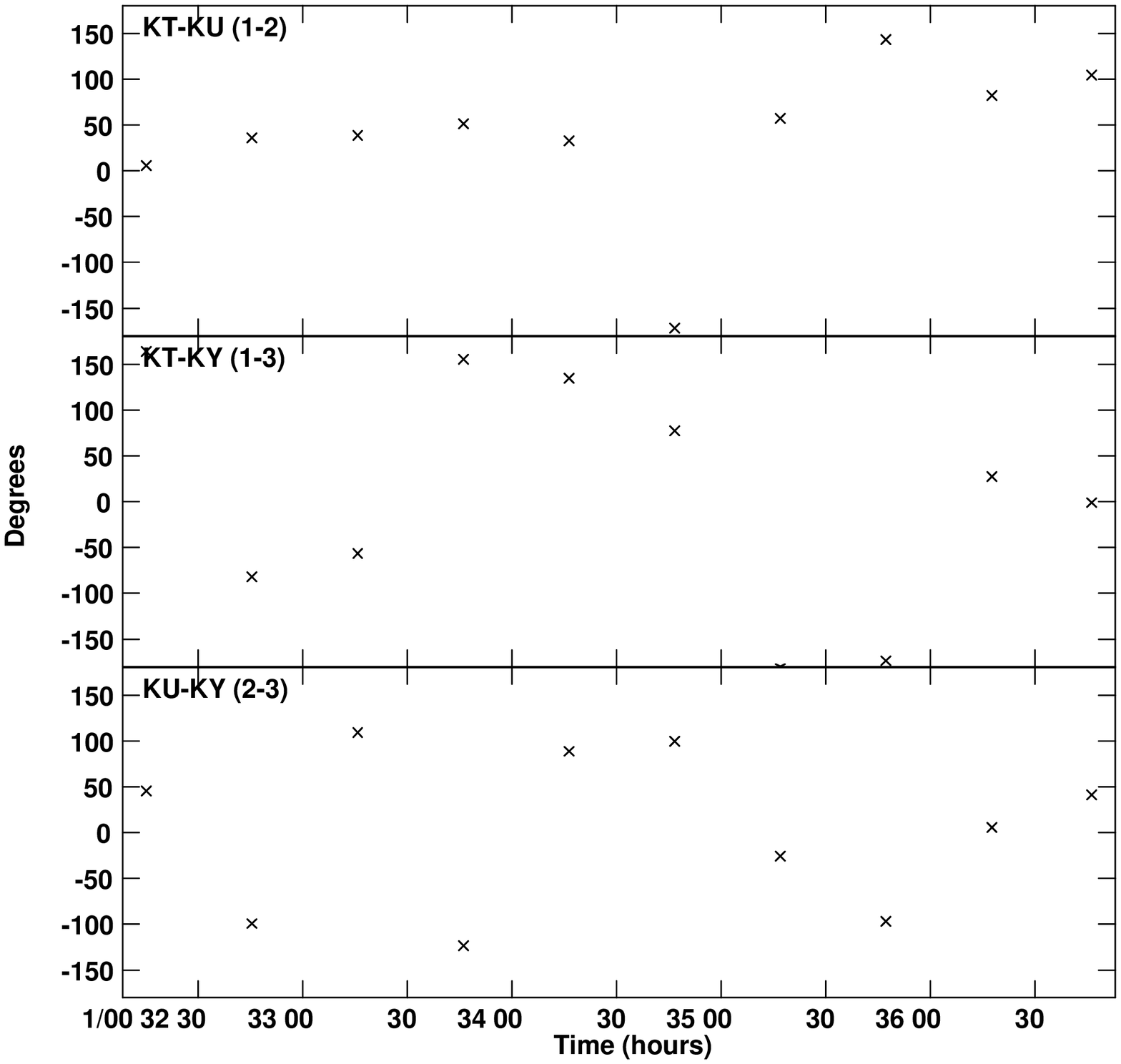}
\includegraphics[angle=-90,width=57mm]{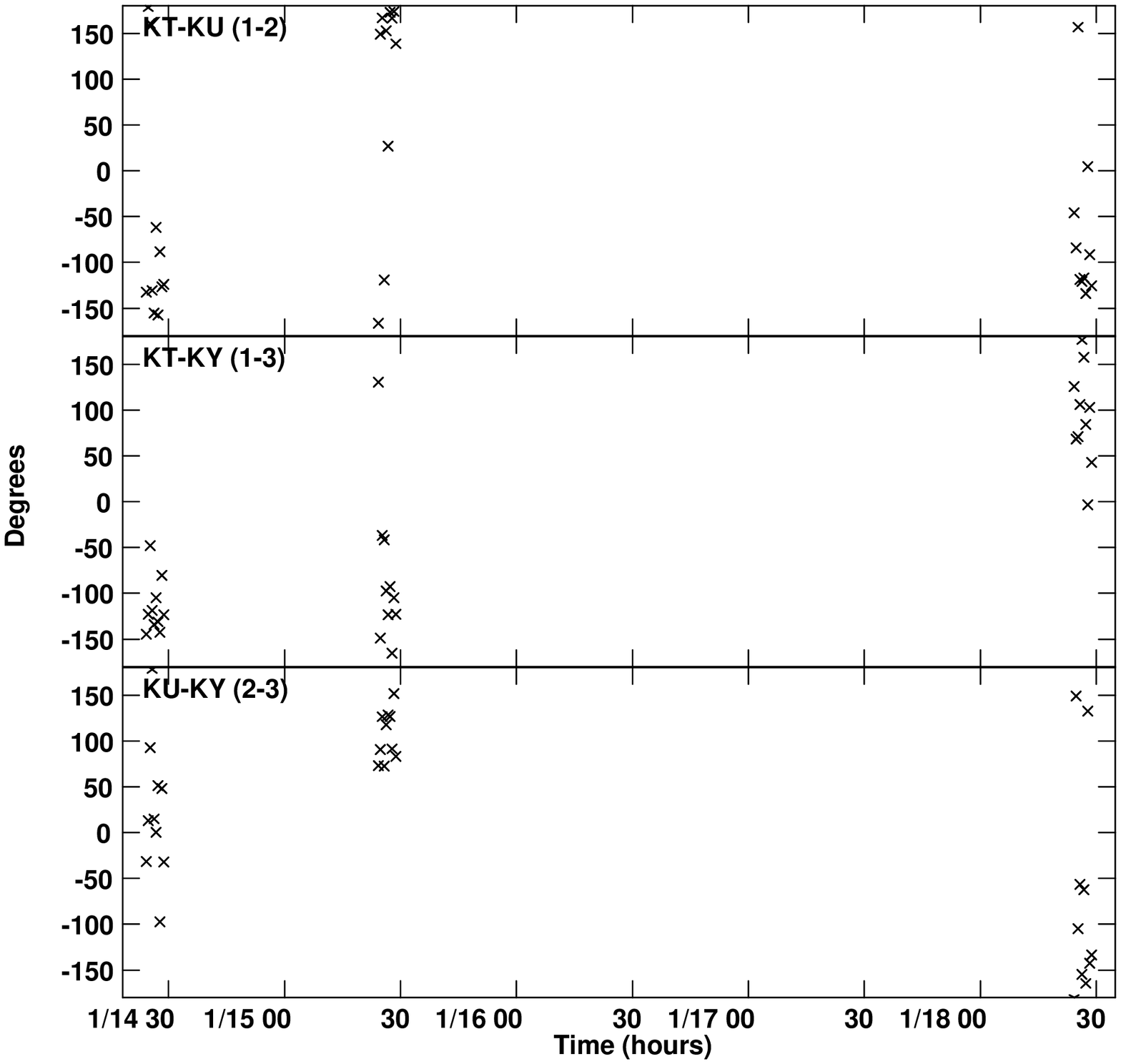}
\includegraphics[angle=-90,width=57mm]{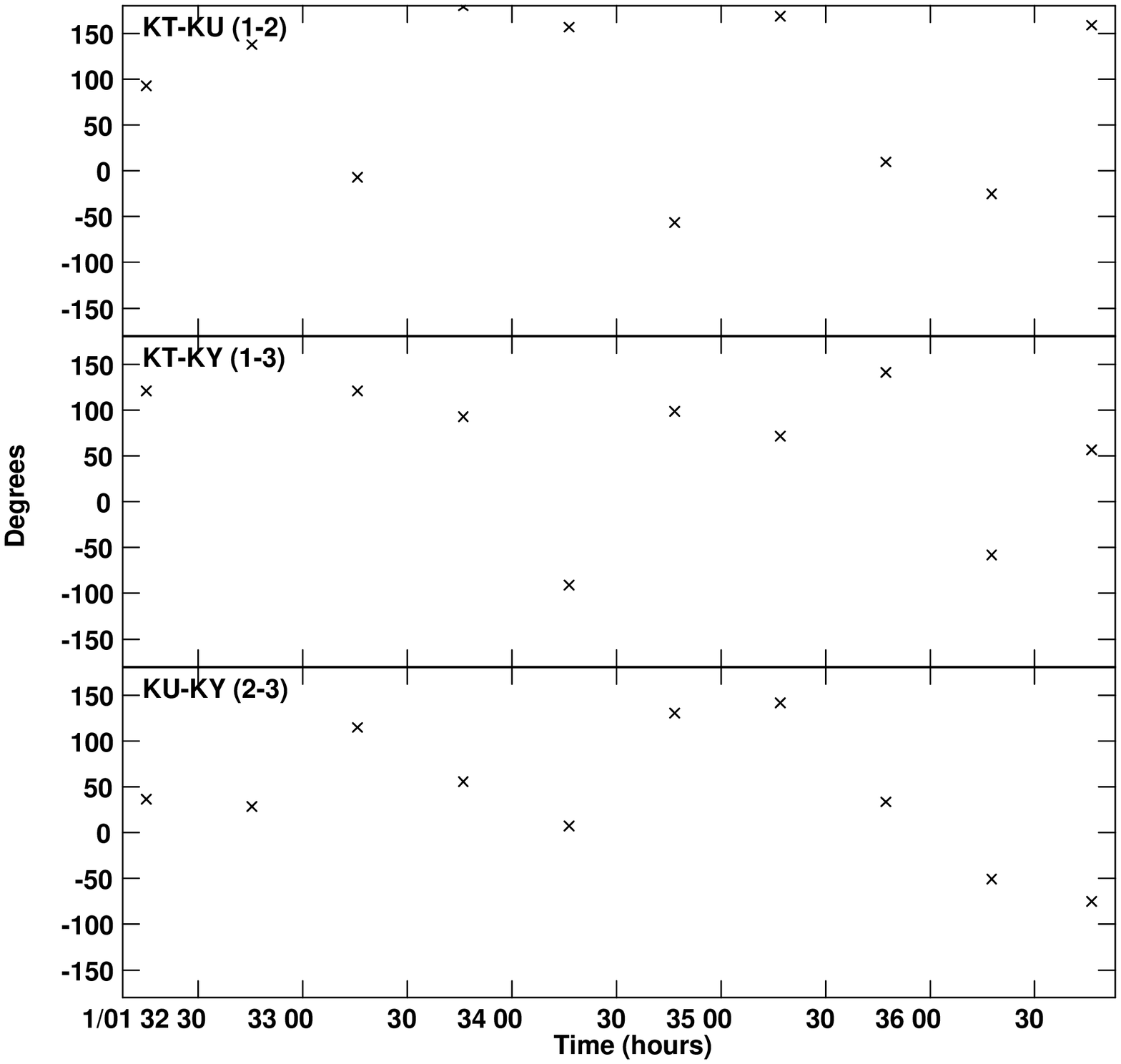}
\begin{tabular}{lll}
\hspace{0.8cm}(d) 0827+243 &\hspace{3.3cm} (e) 0836+710  &\hspace{3.3cm} (f) 1127-145\\
\end{tabular}
\includegraphics[angle=-90,width=57mm]{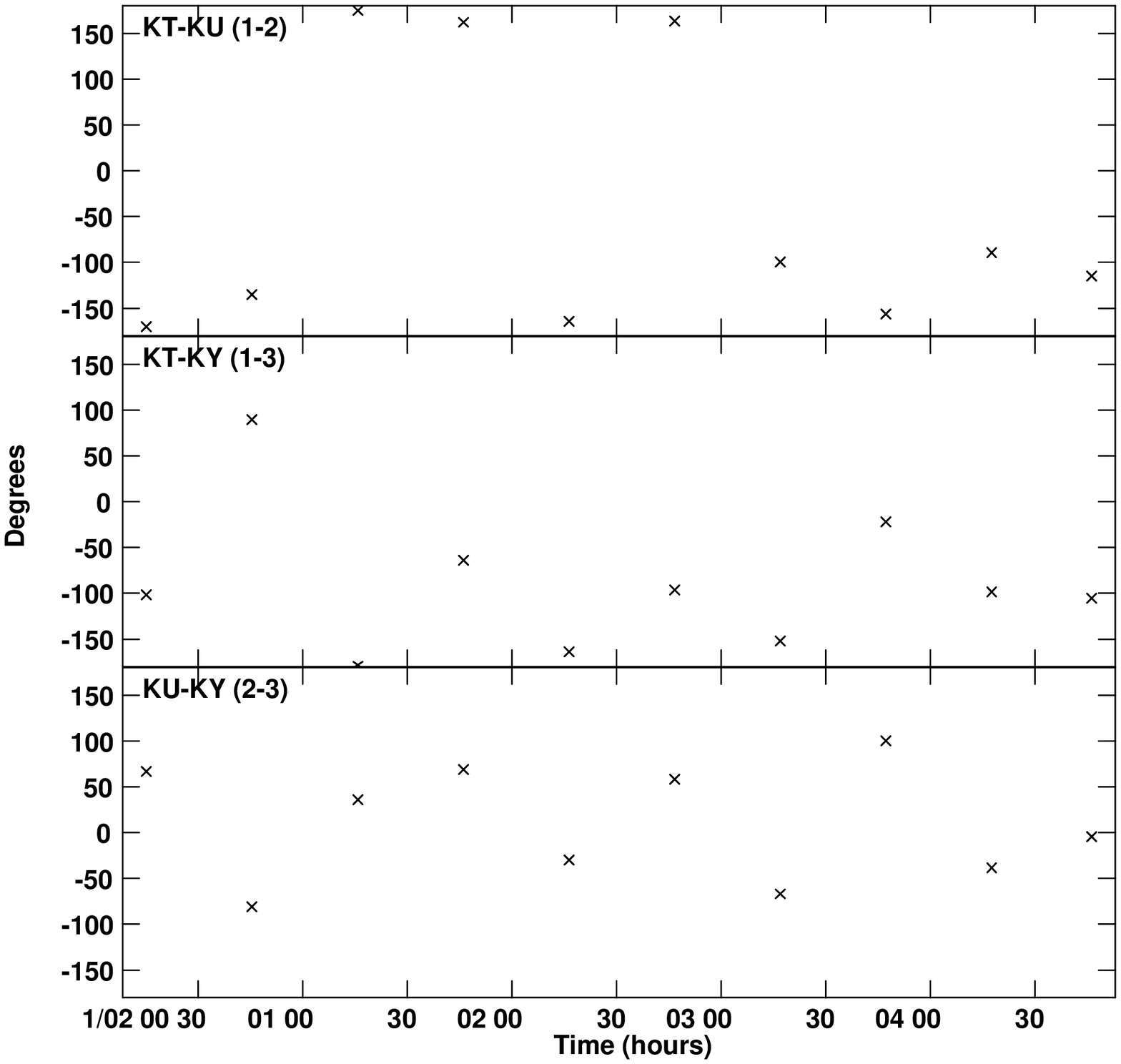}
\includegraphics[angle=-90,width=57mm]{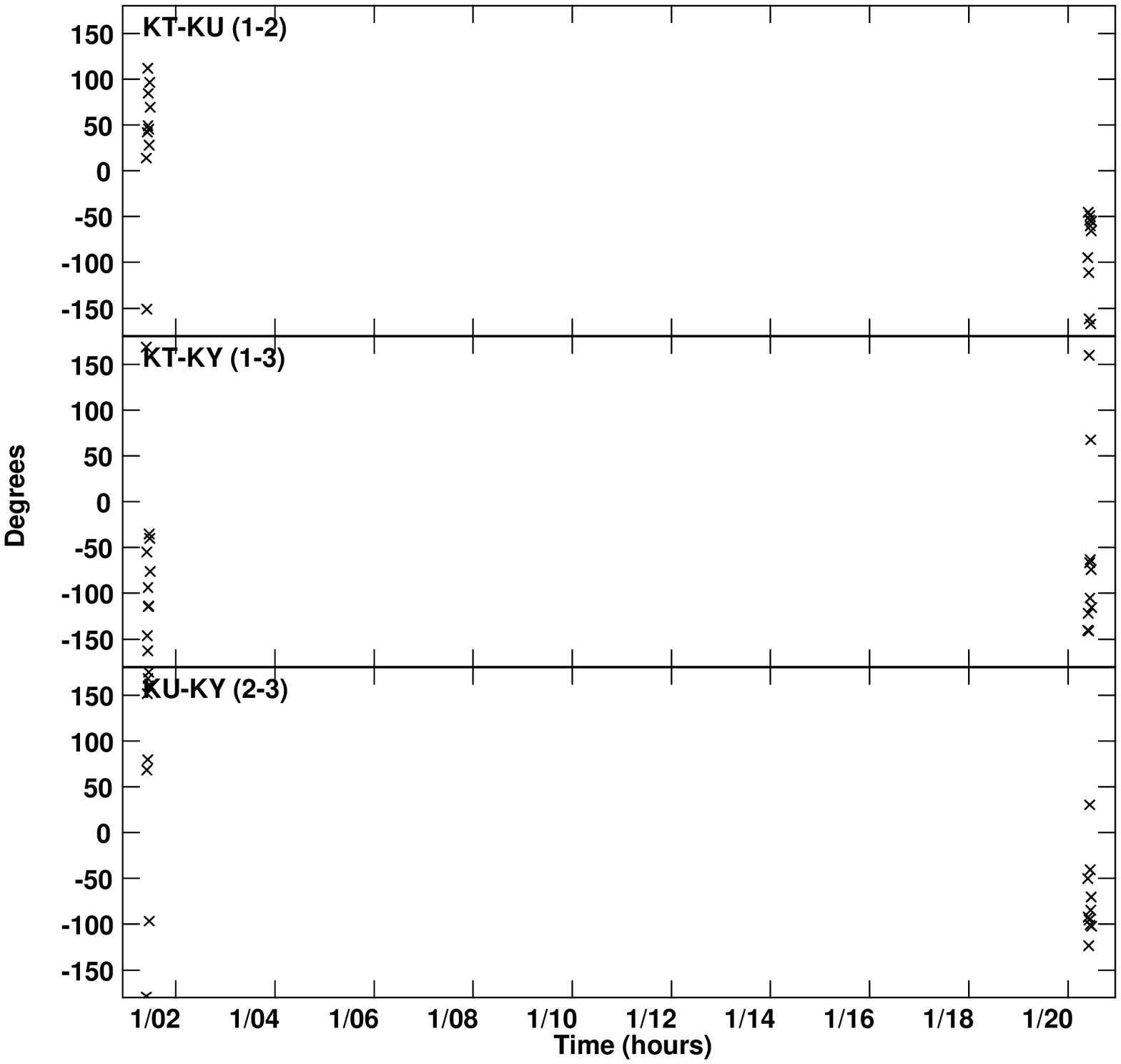}
\includegraphics[angle=-90,width=57mm]{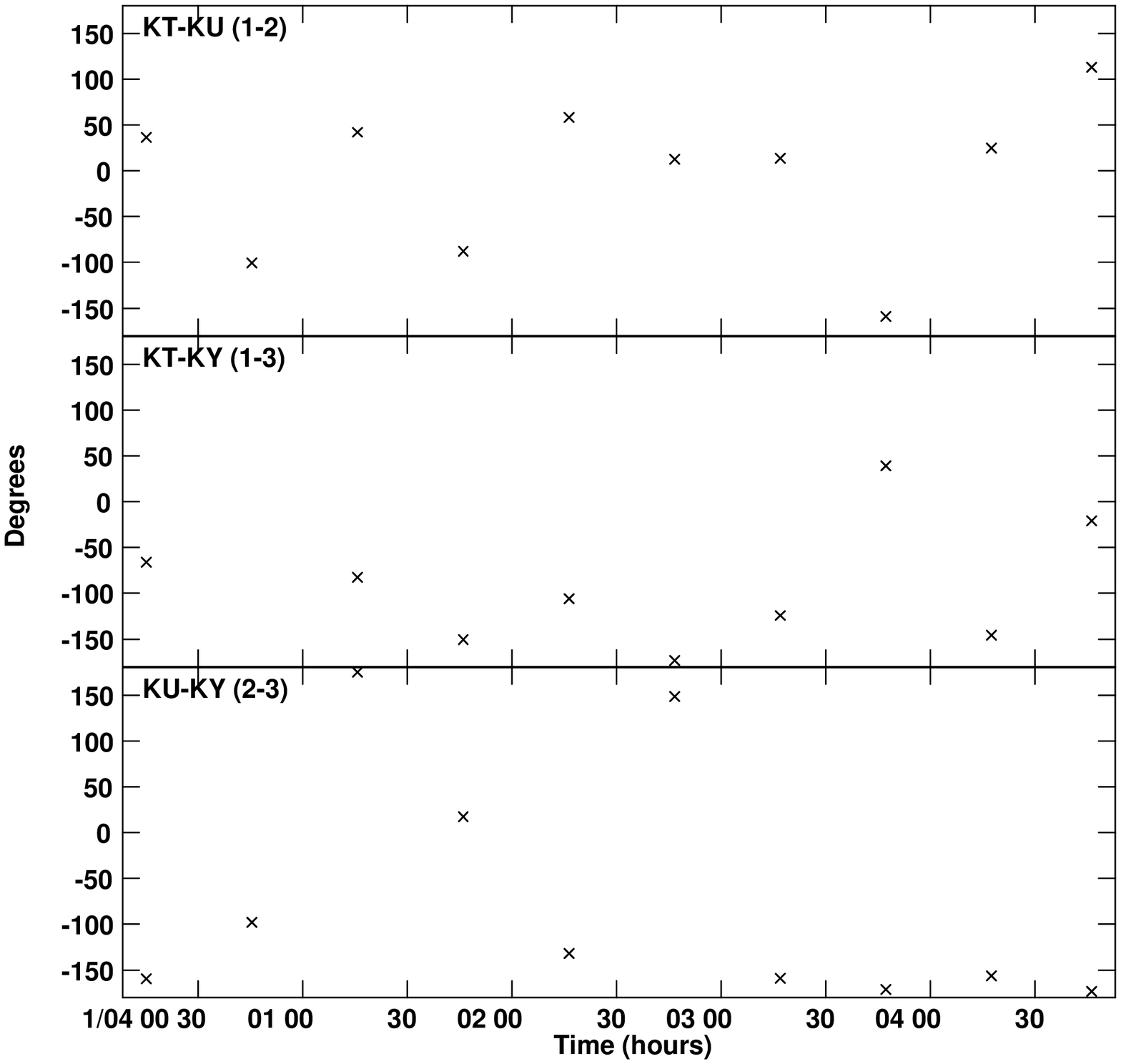}
\begin{tabular}{lll}
\hspace{0.8cm}(g) 1156+295 &\hspace{3.3cm} (h) 1222+216 &\hspace{3.3cm} (i) 1308+326\\
\end{tabular}
\includegraphics[angle=-90,width=57mm]{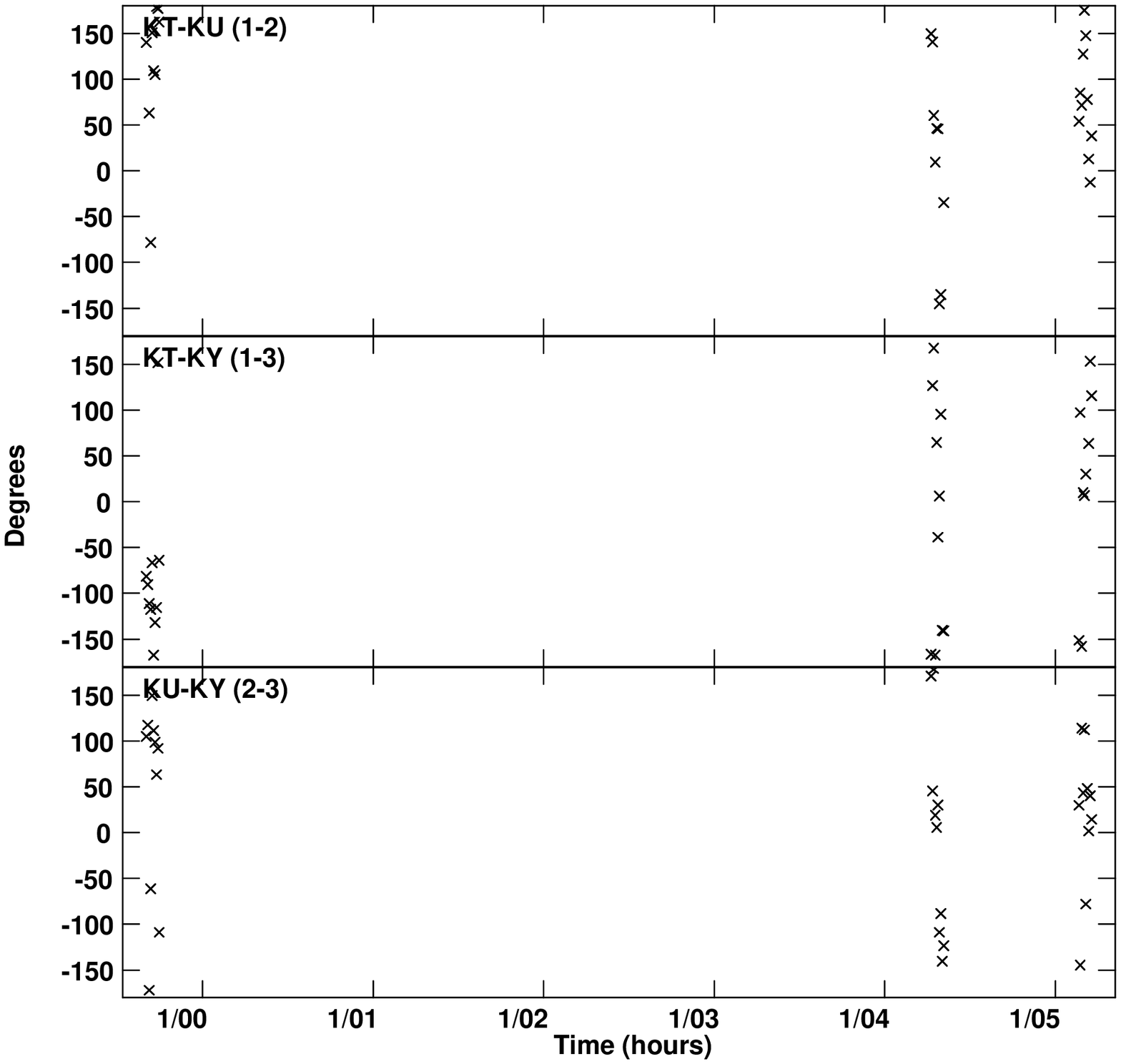}
\includegraphics[angle=-90,width=57mm]{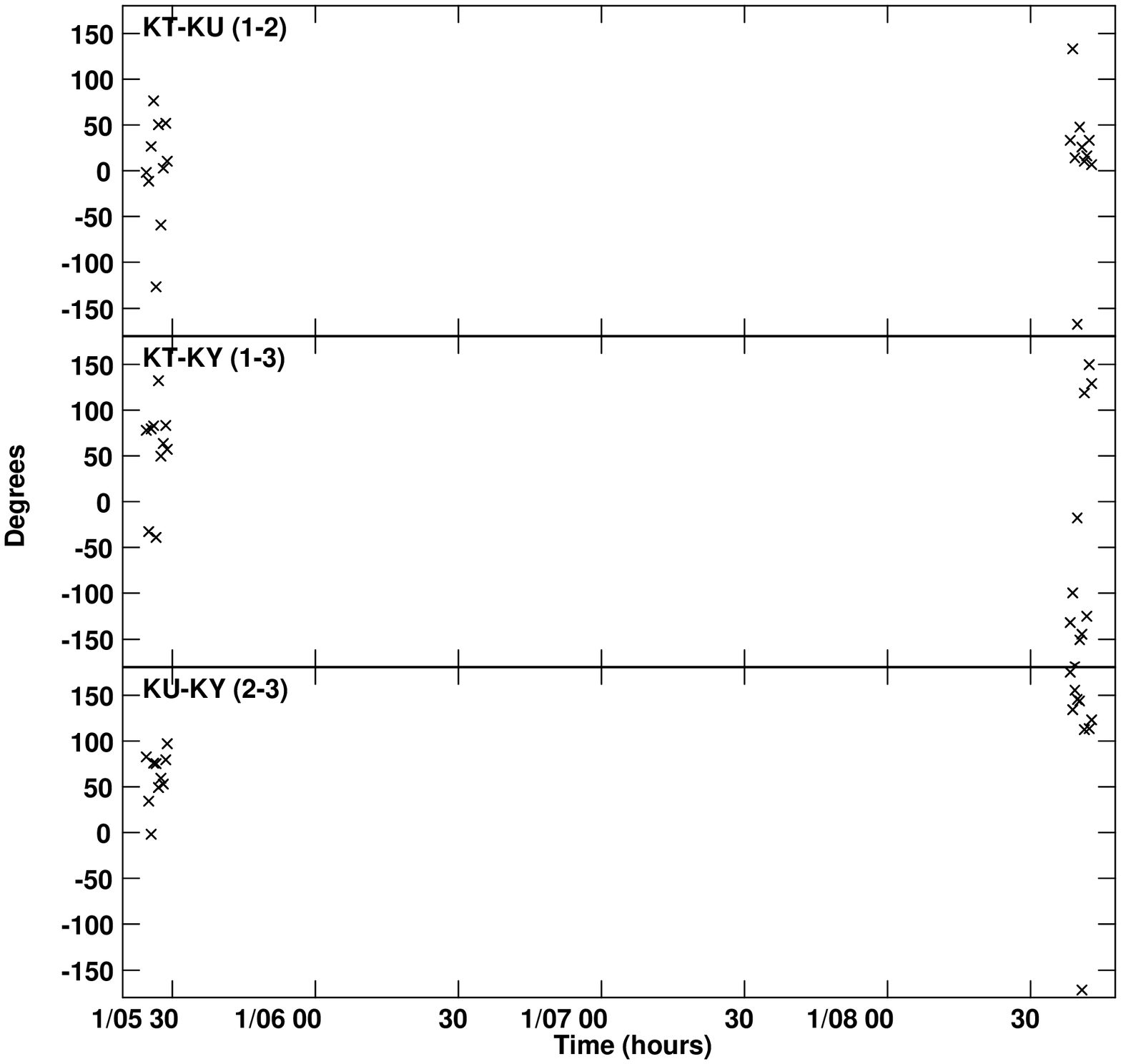}
\includegraphics[angle=-90,width=57mm]{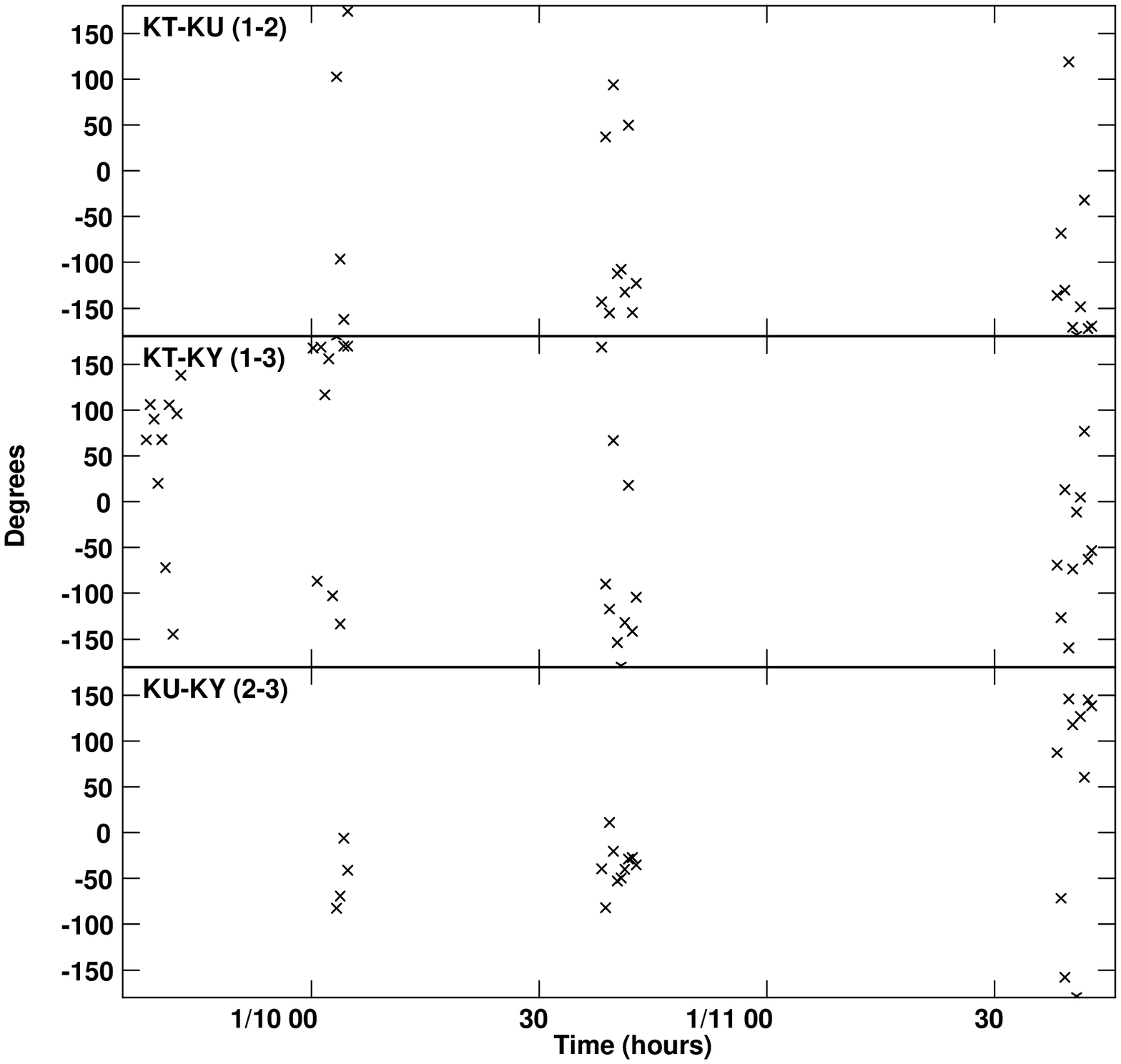}
%\begin{tabular*}{\textwidth}{c @{\extracolsep{\fill}} ccc}
\begin{tabular}{lll}
\hspace{0.8cm}(j) 1343+451 &\hspace{3.3cm} (k) 1611+343 &\hspace{3.3cm} (l) 3C446\\
\end{tabular}
\caption{FPT visibility phases for iMOGABA9 observations at 129~GHz. Each data set was calibrated using the scaled solutions from the analysis  of the same source at 22~GHz. Plots show temporal average of 30 seconds.\label{FPTphase22-129}}
%\vspace{5mm} %% add extra space ONLY when figures/tables are "colliding"!
\end{figure*}

\begin{figure*}[h]
\centering
\includegraphics[angle=-90,width=57mm]{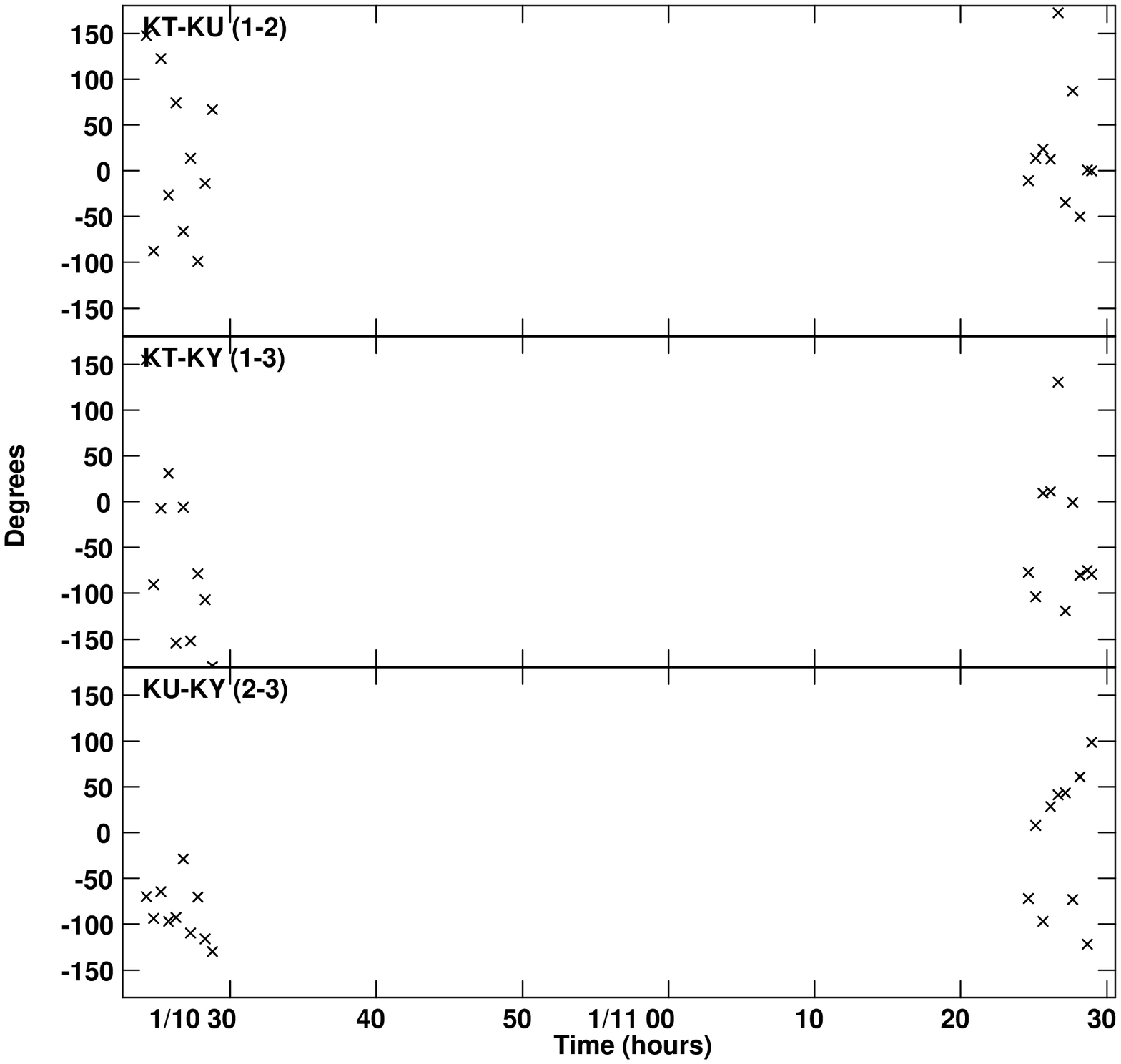}
\includegraphics[angle=-90,width=57mm]{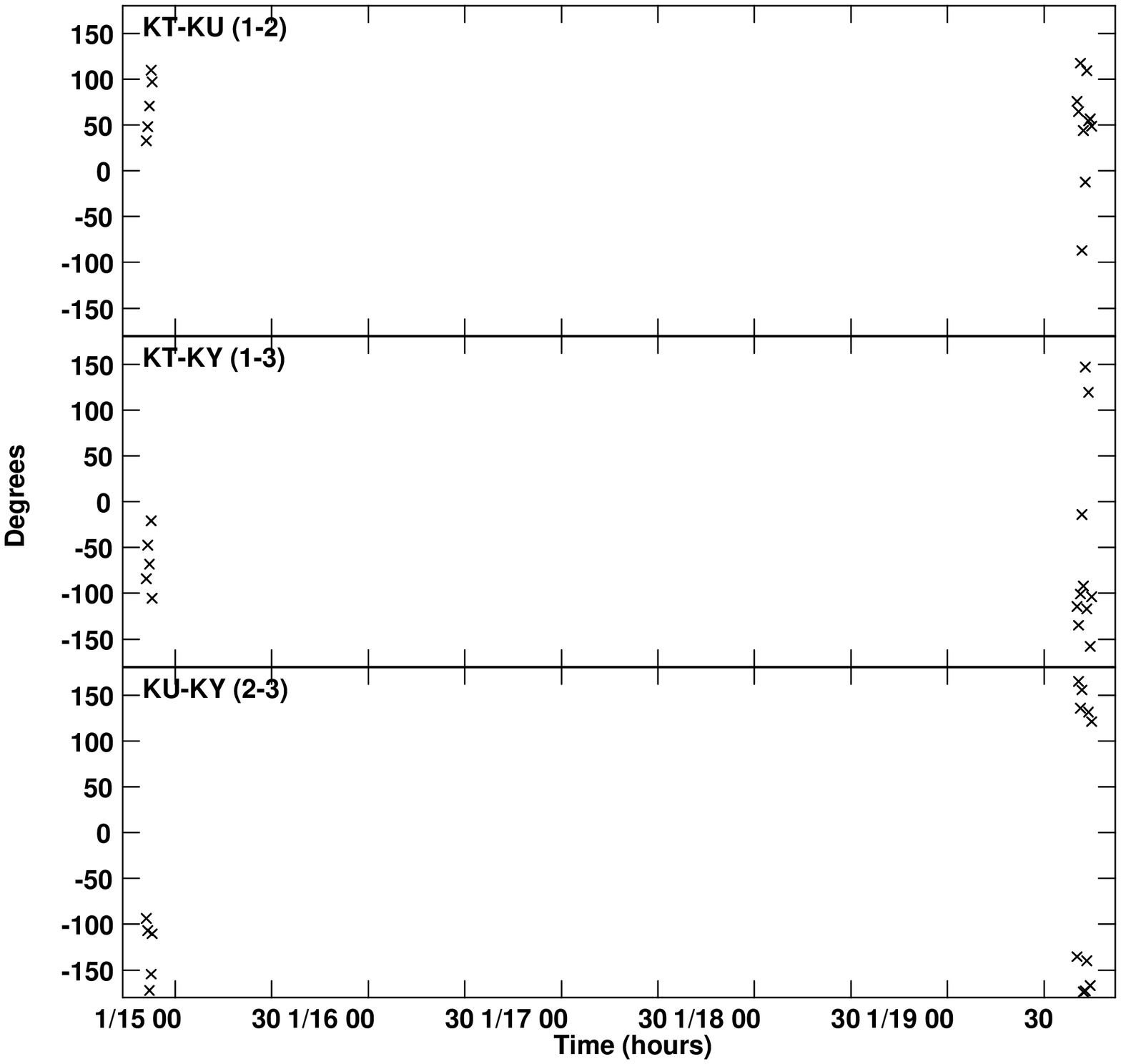}
\includegraphics[angle=-90,width=57mm]{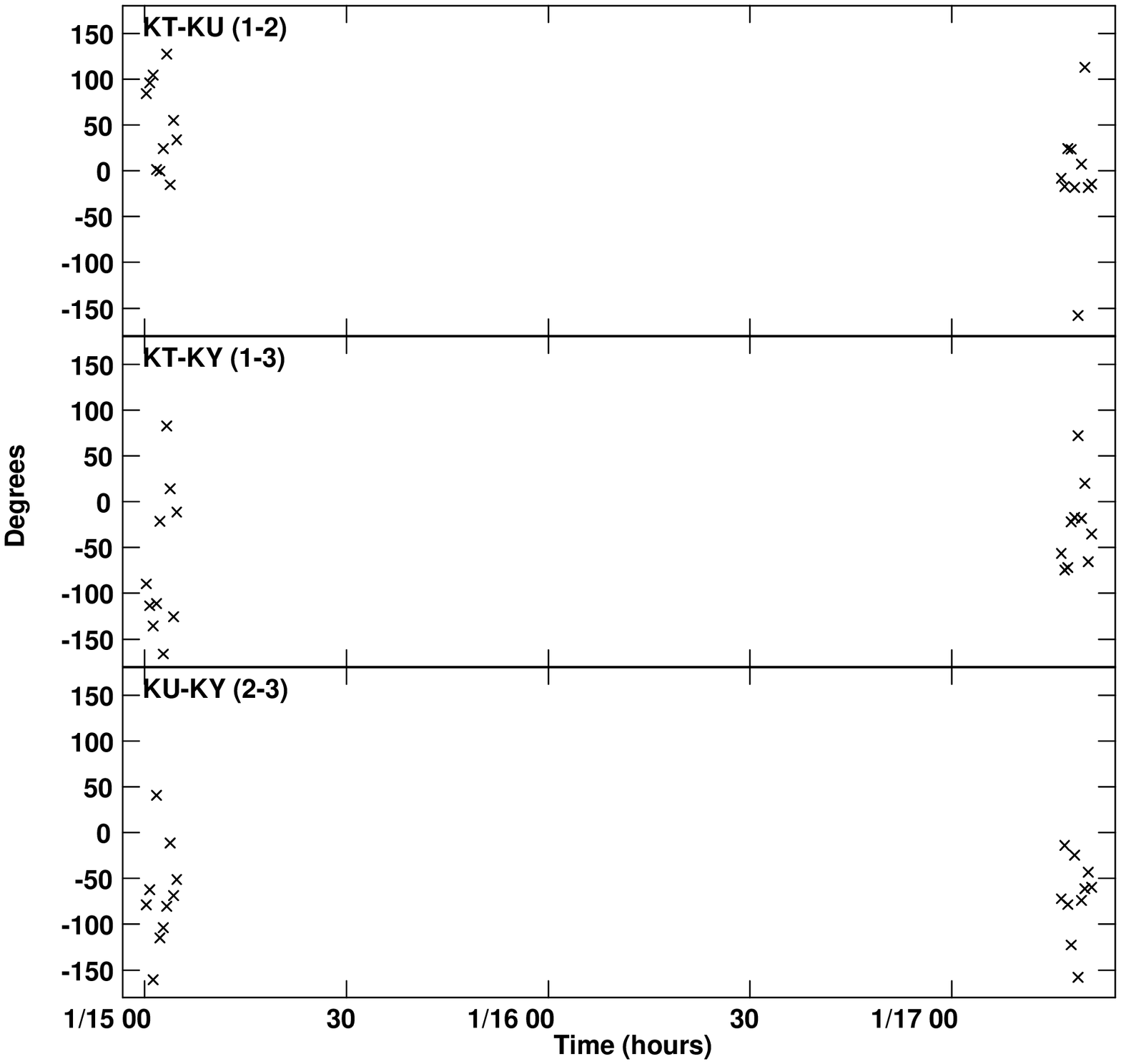}
\begin{tabular}{lll}
\hspace{0.8cm}(a) 0235+164 &\hspace{3.3cm} (b) 0528+134 &\hspace{3.3cm} (c) 0735+178\\
\end{tabular}
\includegraphics[angle=-90,width=57mm]{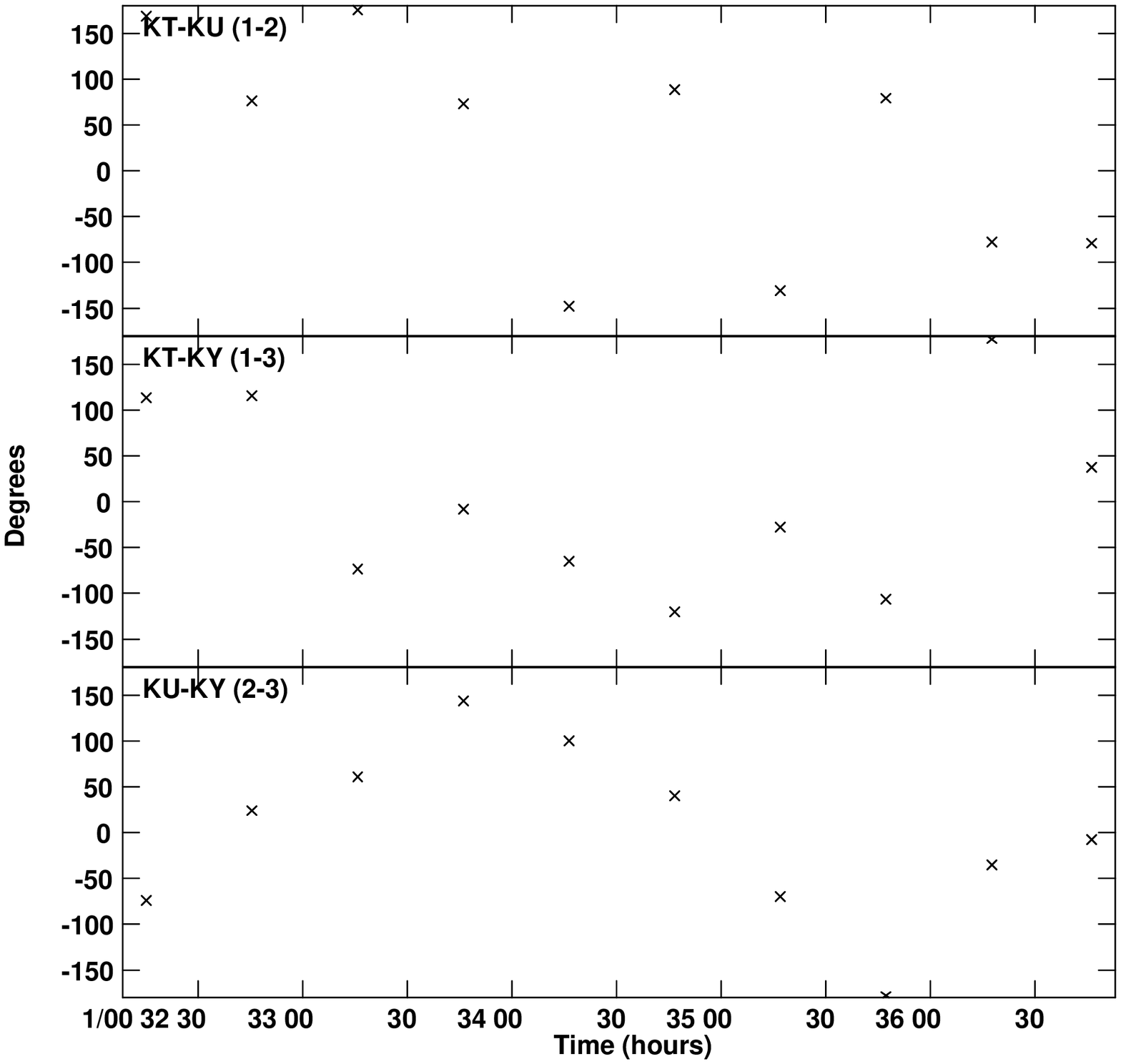}
\includegraphics[angle=-90,width=57mm]{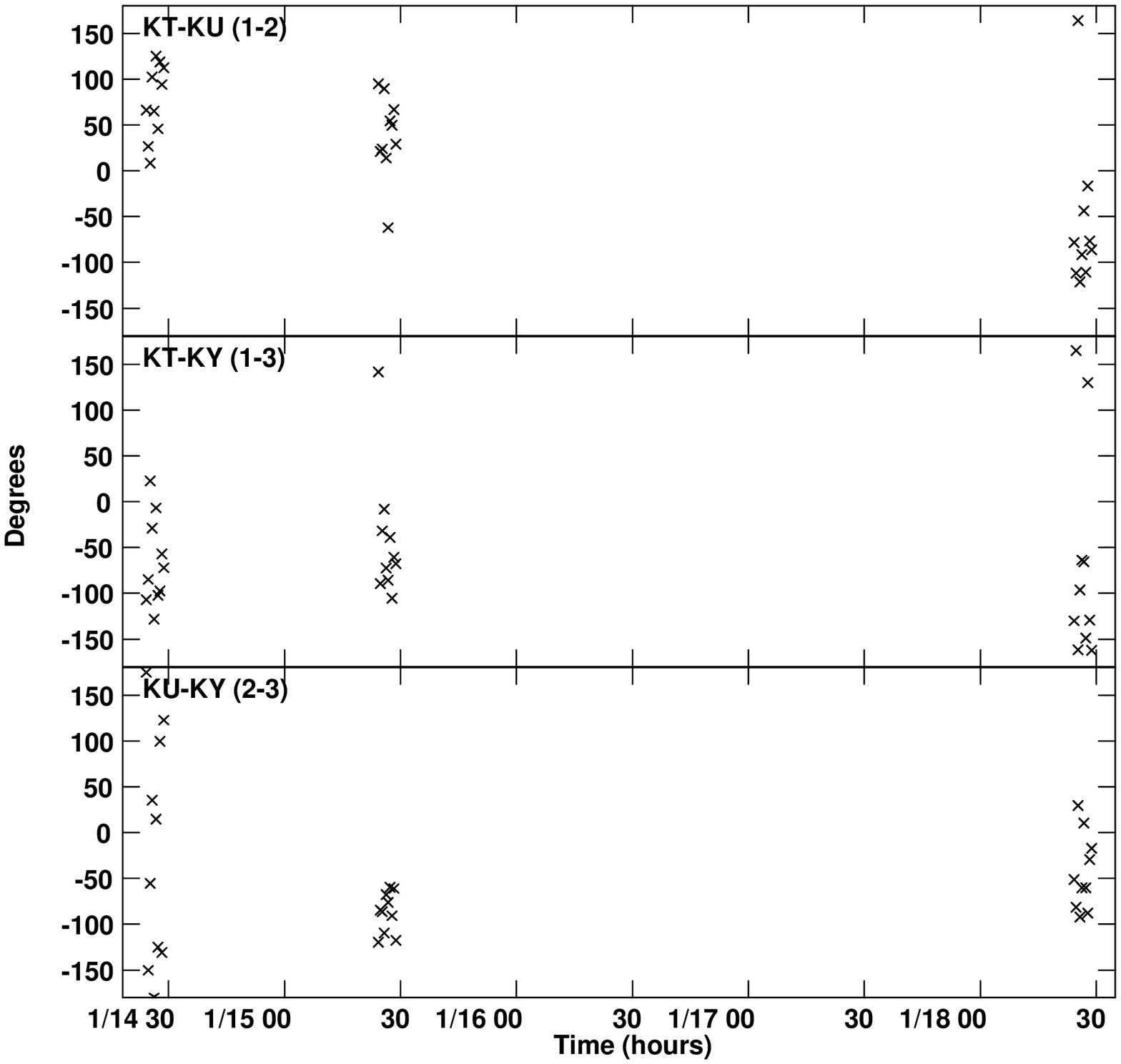}
\includegraphics[angle=-90,width=57mm]{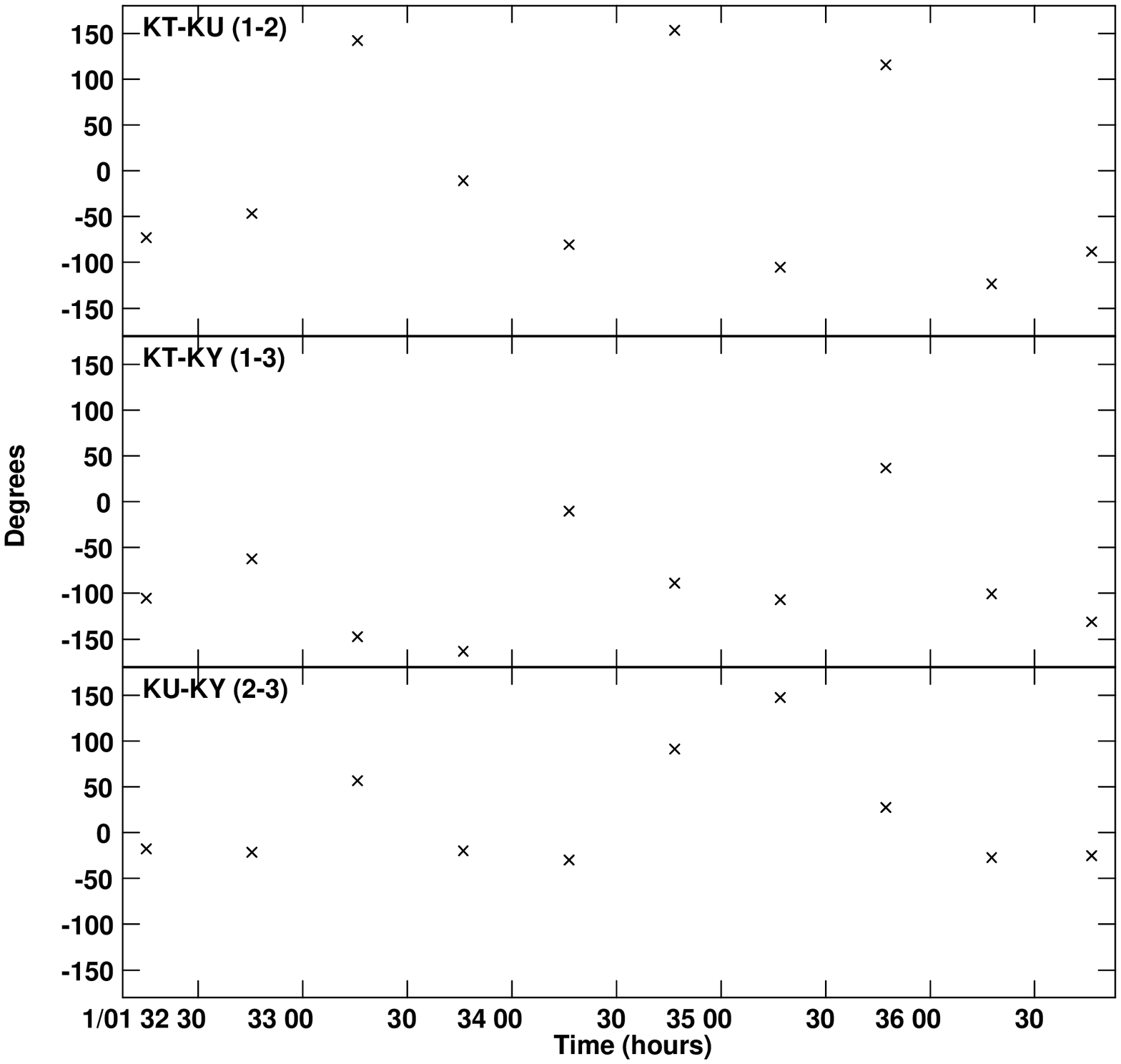}
\begin{tabular}{lll}
\hspace{0.8cm}(d) 0827+243 &\hspace{3.3cm} (e) 0836+710  &\hspace{3.3cm} (f) 1127-145\\
\end{tabular}
\includegraphics[angle=-90,width=57mm]{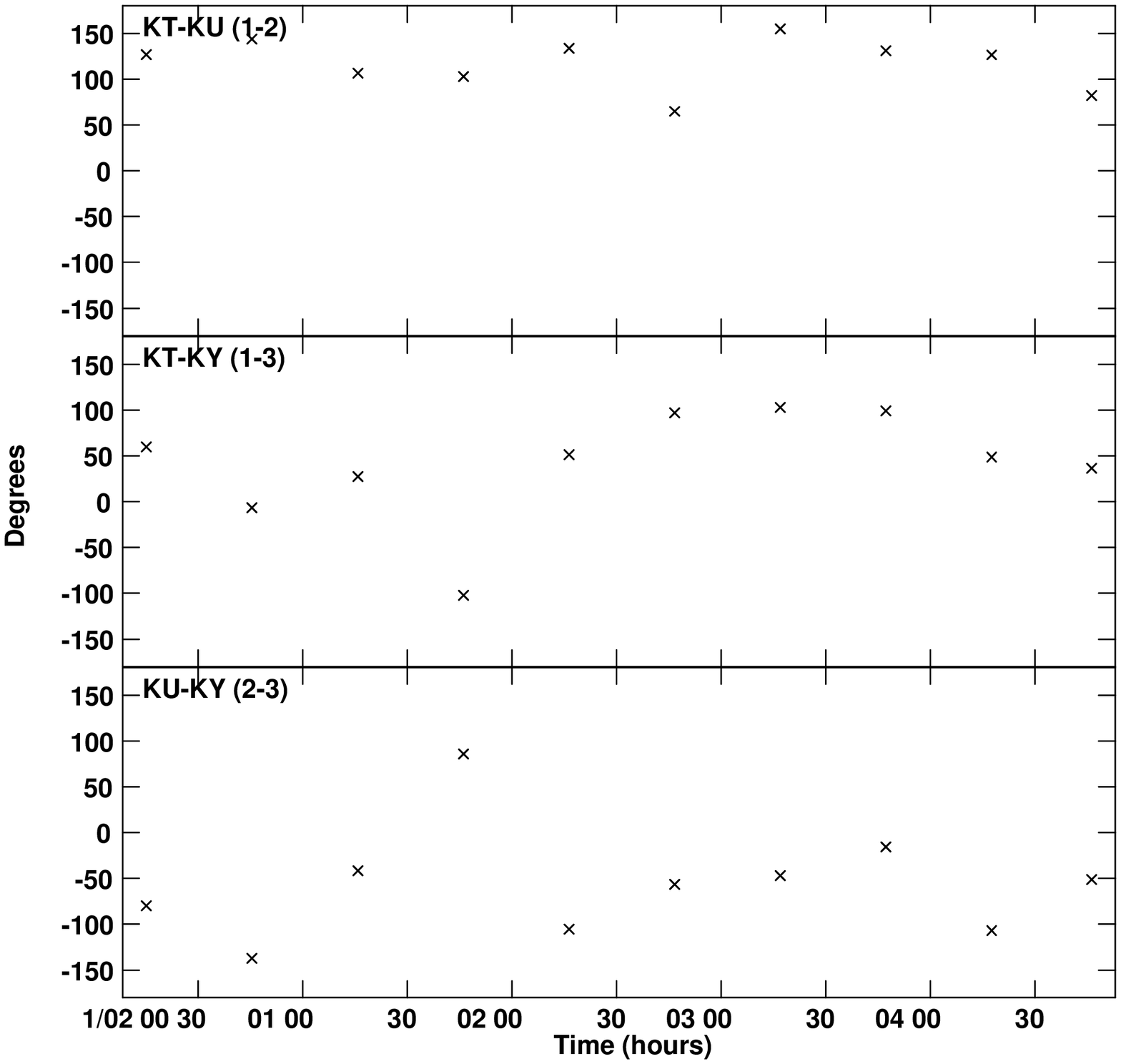}
\includegraphics[angle=-90,width=57mm]{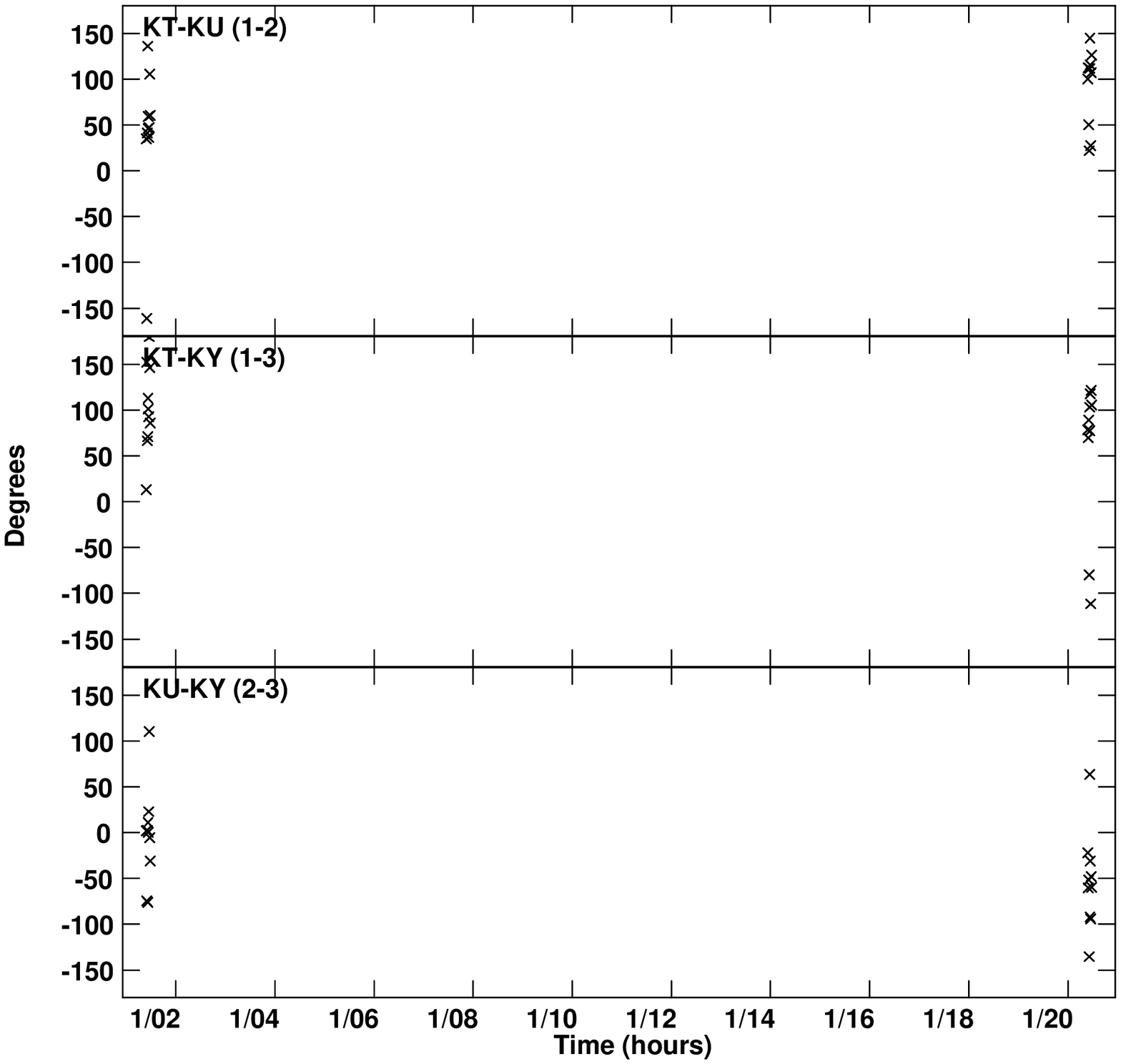}
\includegraphics[angle=-90,width=57mm]{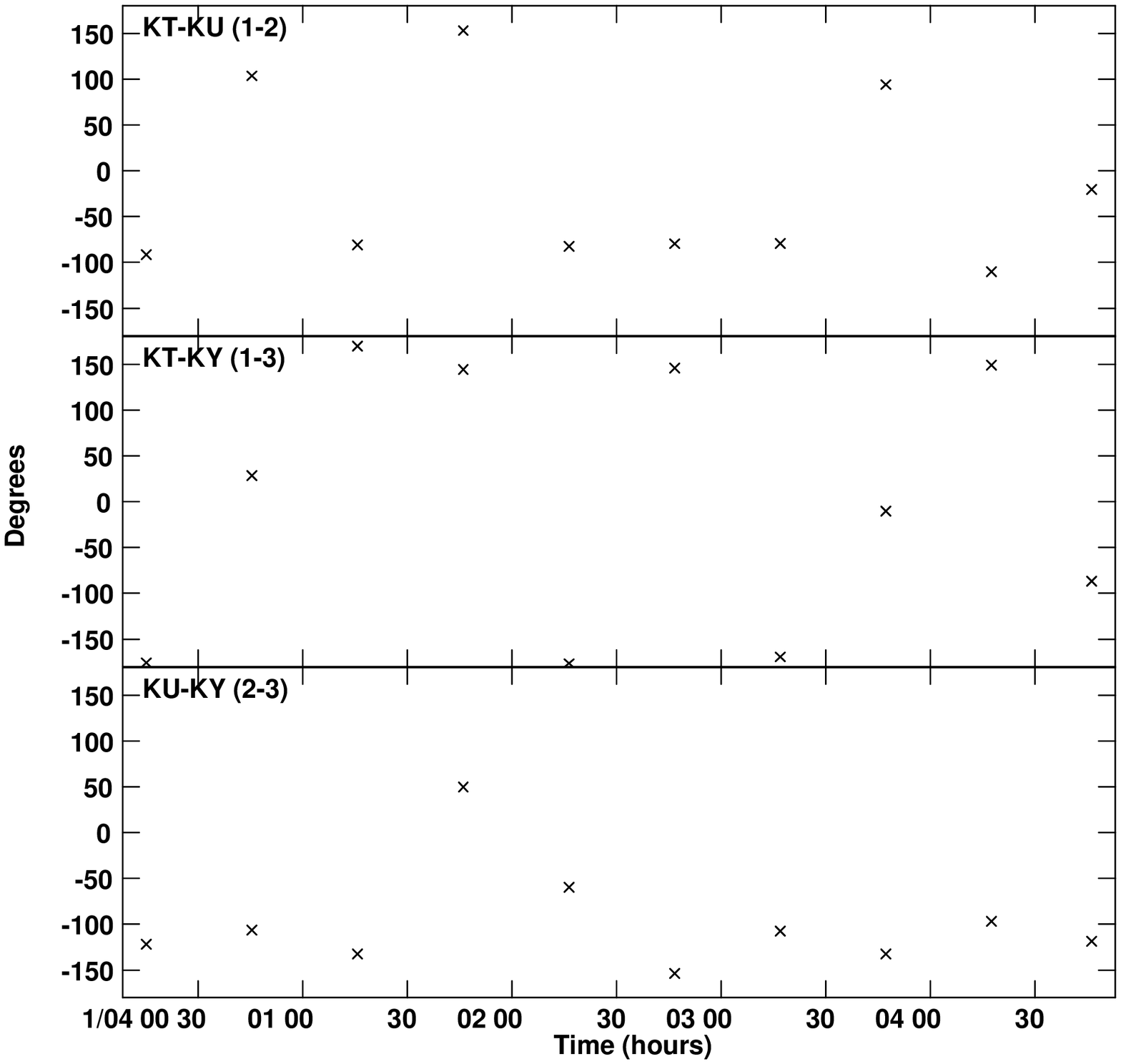}
\begin{tabular}{lll}
\hspace{0.8cm}(g) 1156+295 &\hspace{3.3cm} (h) 1222+216 &\hspace{3.3cm} (i) 1308+326\\
\end{tabular}
\includegraphics[angle=-90,width=57mm]{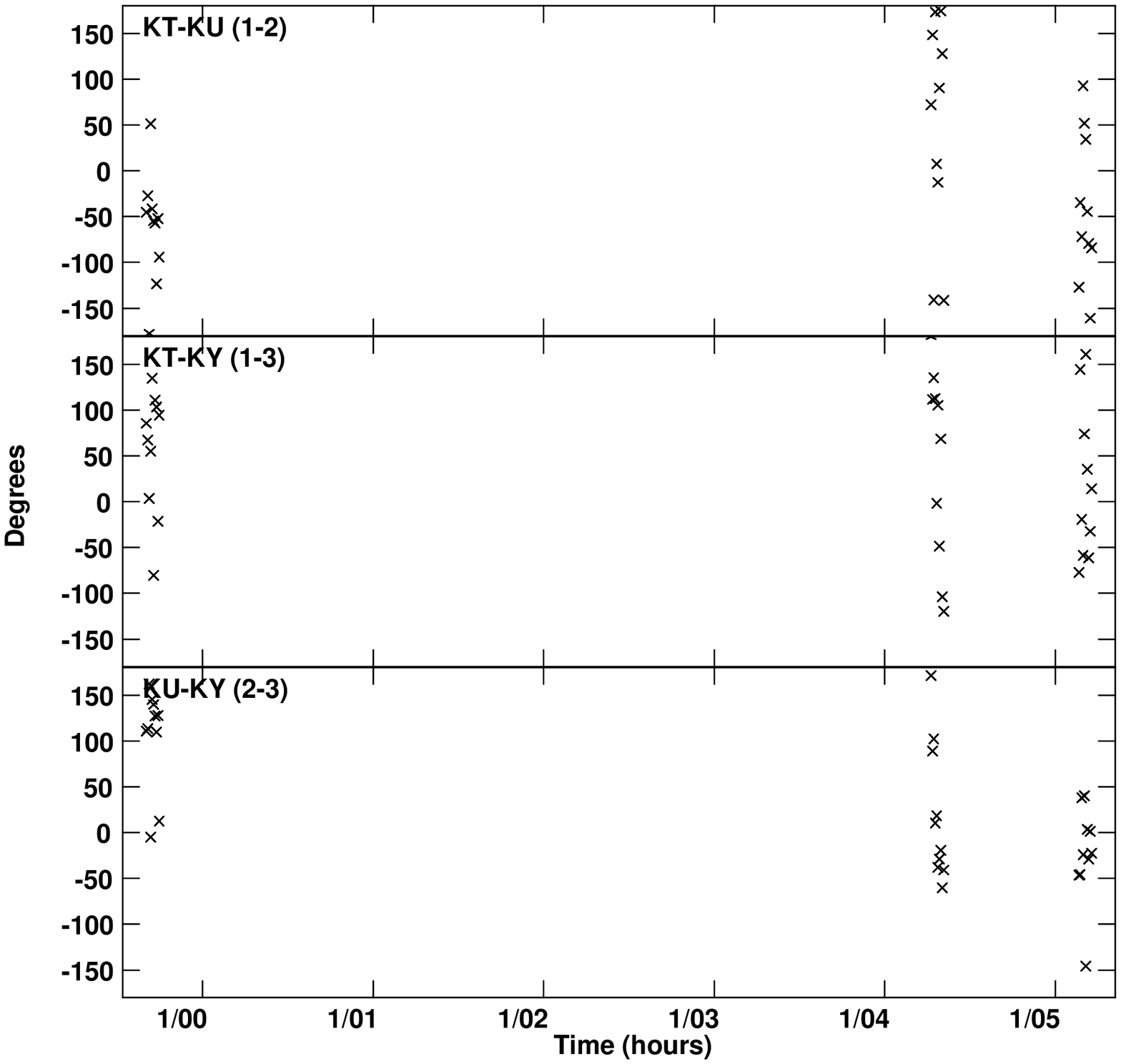}
\includegraphics[angle=-90,width=57mm]{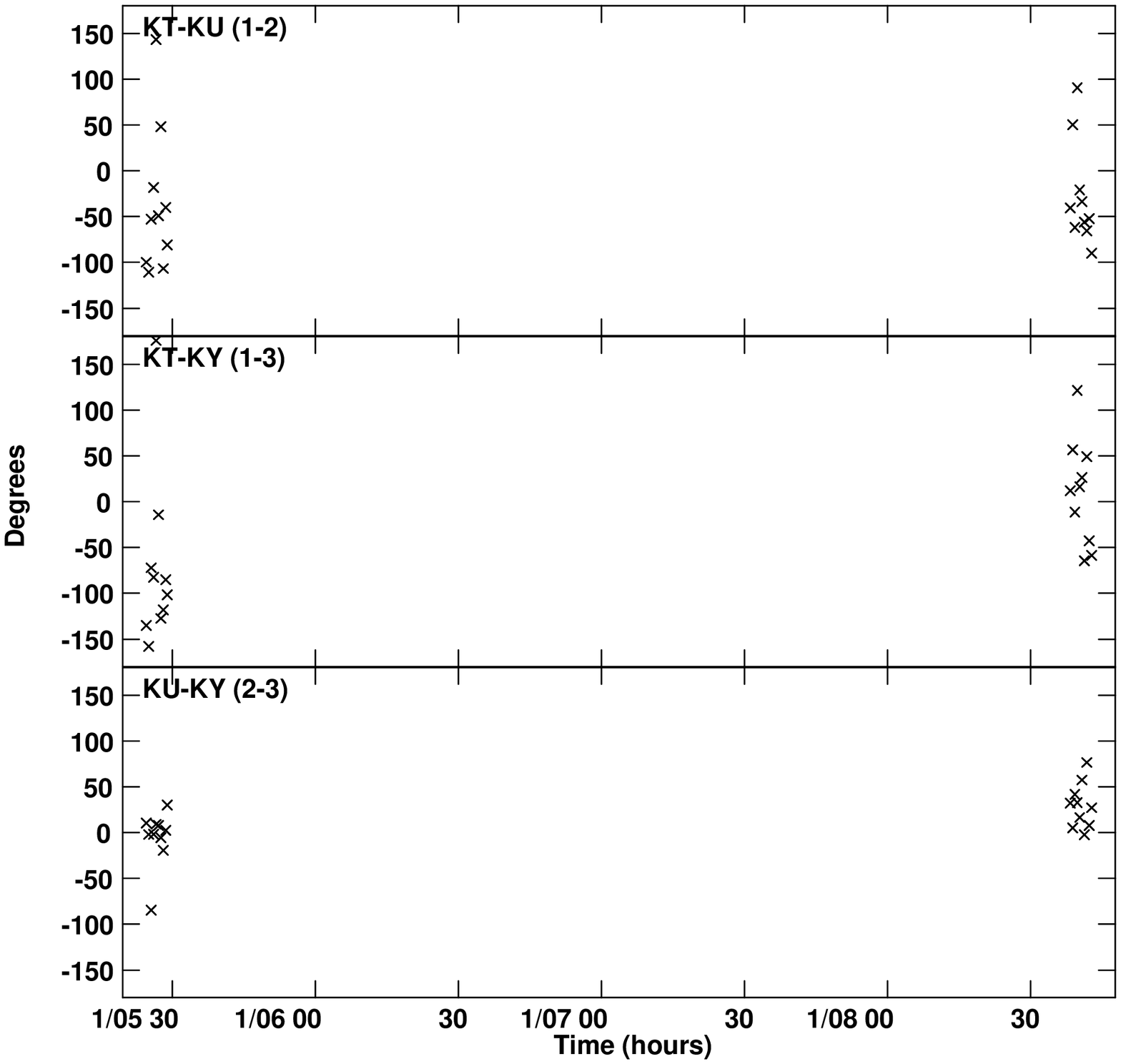}
\includegraphics[angle=-90,width=57mm]{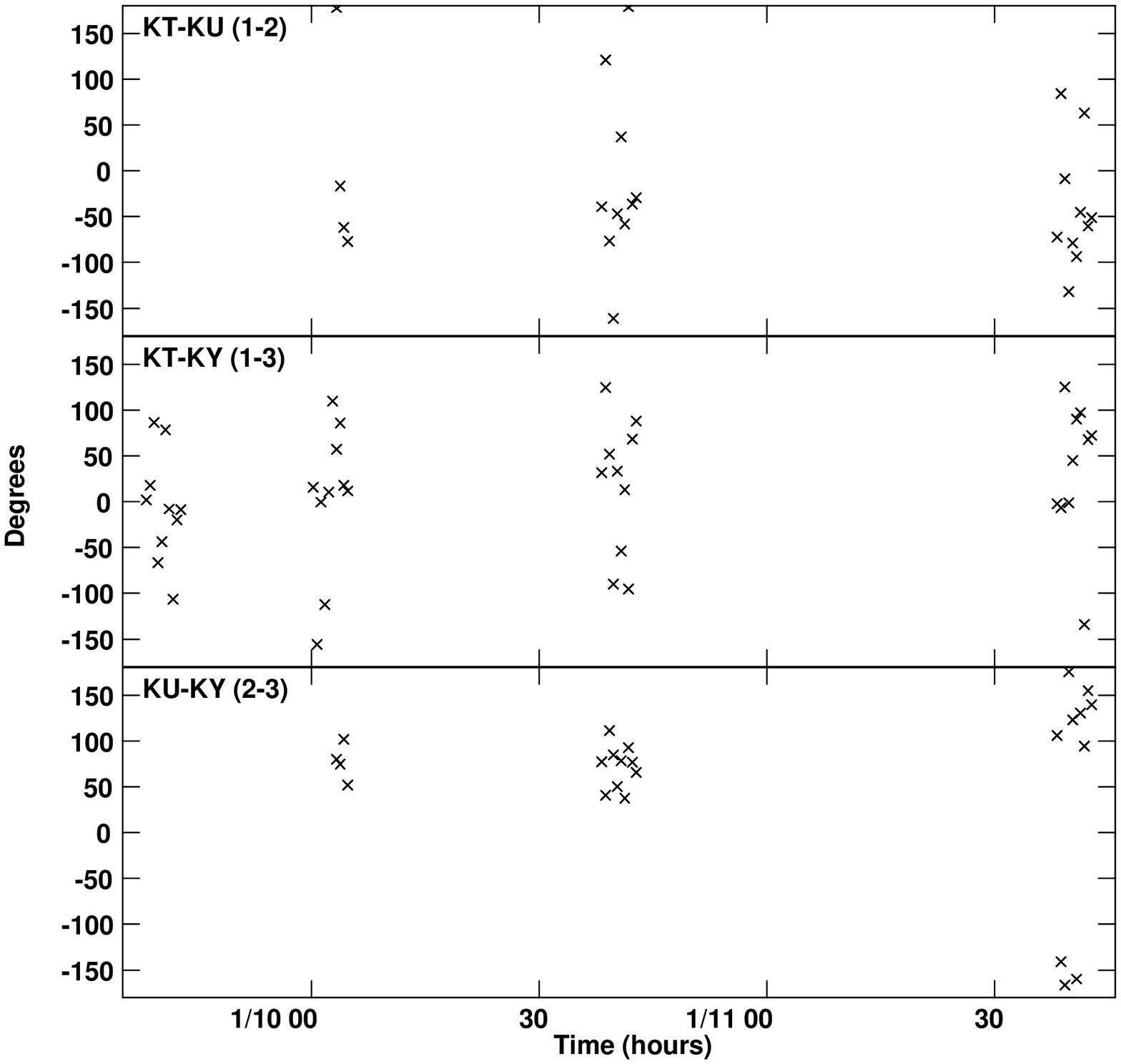}
%\begin{tabular*}{\textwidth}{c @{\extracolsep{\fill}} ccc}
\begin{tabular}{lll}
\hspace{0.8cm}(j) 1343+451 &\hspace{3.3cm} (k) 1611+343 &\hspace{3.3cm} (l) 3C446\\
\end{tabular}
\caption{Same as Figure \ref{FPTphase22-129}, but using the scaled solutions from the analysis of the same source at 43 GHz.\label{FPTphase43-129}}
%\vspace{5mm} %% add extra space ONLY when figures/tables are "colliding"!
\end{figure*}

Following identical procedure as in Section \ref{sec:FPT86},  we transferred the solutions from 22 and 43~GHz to 129~GHz (frequency ratios R=6 and 3, respectively). We did not attempt FPT from 86~GHz given that the ratio would not be an integer number. Plots of the FPT calibrated visibilities are shown in Figure \ref{FPTphase22-129} for a transfer from 22~GHz and Figure \ref{FPTphase43-129} for a transfer from 43~GHz, respectively.

As seen for the transfer to 86~GHz, comparison of Figures \ref{FPTphase22-129} and \ref{FPTphase43-129} indicate that both transfers to 129~GHz also lead to similar results. Further inspection shows that for these sources with a larger estimated  SNR$^{129}$, the scatter in the transferred phases is smaller. This is the case for example for 0528+134, 0735+178, and 1222+216. On the other hand, sources with smaller  SNR$^{129}$, such as 1343+451, have a comparatively larger scatter.

\begin{figure*}[h]
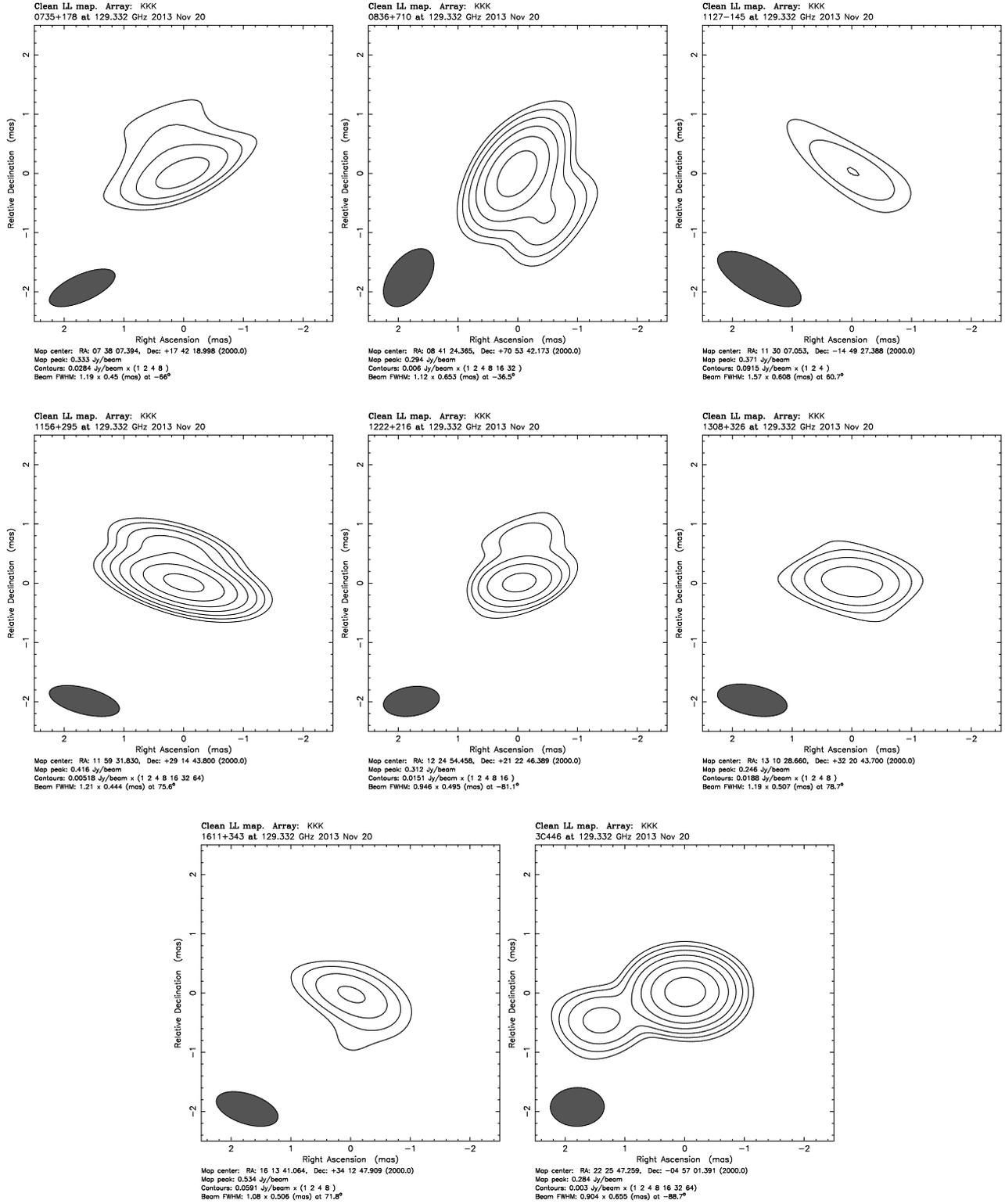

\centering
\includegraphics[angle=0,width=55mm,clip,trim=0 0cm 0 0]{iM9-0735F129PTv3.eps}
\includegraphics[angle=0,width=55mm,clip,trim=0 0cm 0 0]{iM9-0836F129PTv3.eps}
\includegraphics[angle=0,width=55mm,clip,trim=0 0cm 0 0]{iM9-1127F129PTv3.eps}\\
\vspace{0.5cm}
\includegraphics[angle=0,width=55mm,clip,trim=0 0cm 0 0]{iM9-1156F129PTv3.eps}
\includegraphics[angle=0,width=55mm,clip,trim=0 0cm 0 0]{iM9-1222F129PTv3.eps}
\includegraphics[angle=0,width=55mm,clip,trim=0 0cm 0 0]{iM9-1308F129PTv3.eps}\\
\vspace{0.5cm}
\includegraphics[angle=0,width=55mm,clip,trim=0 0cm 0 0]{iM9-1611F129PTv3.eps}
\includegraphics[angle=0,width=55mm,clip,trim=0 0cm 0 0]{iM9-3C446F129PTv3.eps}
\caption{iMOGABA9 129 GHz hybrid maps of the sources that could be not imaged with standard procedures, but were successfully imaged with FPT. \label{129GHzMaps}}
\end{figure*}

After re-fringing the phase, we found no solutions for 0235+164, 0827+243, and 1343+451. Solutions were found at all baselines for at least one scan for the other sources. These results are in agreement with the a priori expectation based on Table \ref{theoreticalflux129}. Maps of sources that could be imaged are shown in Figure \ref{129GHzMaps}. Obtained peak fluxes are in good agreement with the expected values (c.f. Table \ref{theoreticalflux129}), except for \mbox{1127--145}, which seems to have a lower flux than expected. In terms of detection limits, this implies that SNR$^{129}_{300s}$ for this source has been overestimated, and the real value should be also smaller, but still larger than our cutoff value.

\subsection{iMOGABA15}
\subsubsection{Frequency Phase Transfer to 86~GHz}\label{sec:FPT86v15}
Given the poor weather conditions during this epoch, representative system equivalent flux densities (SEFDs) for iMOGABA15 at 86~GHz were in the range $5-28\times10^3$ Jy\footnote{We considered here SEFDs from Ulsan and Yonsei antennas, due to the amplitude issue on Tamna, which may lead to incorrect values for its SEFD}. This leads to a baseline--based sensitivity limit $\sigma^{theo}_{min}\sim$100--500~mJy and $\sigma^{theo}_{min}\sim$30--150~mJy for 30 and 300 seconds integration times, respectively, considering a 2-bit sampling, two antennas and 64~MHz bandwidth. Up to eleven sources could not be imaged in iMOGABA15 with standard methods at 86~GHz.  In Table \ref{theoreticalflux86b} we show these sources, their expected peak flux based on an extrapolation from lower frequencies and the detection SNR for an integration times of 30 and 300 seconds.

\begin{table}[t!]
\caption{iMOGABA15 Expected Detection Limits at 86~GHz\label{theoreticalflux86b}}
\centering
\begin{tabular}{ccccc}
\toprule
Source   & SEFD (Jy) & $S^{86}$ (mJy)  & SNR$^{86}_{30s}$ & SNR$^{86}_{300s}$ \\
\midrule
0235+164  & 4700    &     630 & 7 & 23 \\
0420--014 & 5600    &    880 & 9 & 27 \\
0727--115 & 7500    &    1000 & 8 & 24 \\
0735+178 & 5600    &     500 & 5 & 16 \\
0827+243 & 9400    &     440 & 3 & 8 \\
0836+710 & 4700    &     580 & 7 & 21 \\
1510--089 & 7500    &    1300 & 10 & 30 \\
1611+343 & 22500  &   1200 & 3 & 9 \\
4C39.25   & 22500   &   1300 & 3 & 10 \\
CTA102    & 22500  &   1000 & 2 & 8 \\
NRAO530 & 28000  &   1400 & 3 & 9 \\
\bottomrule
\end{tabular}
\end{table}

Based on the values in Table \ref{theoreticalflux86b}, we expect to detect some of these sources with standard methods. Note however that the amplitude offset problem caused by the Tamna antenna decreased the flux in related baselines, thus making it more difficult in practice to get fringes with the expected SNR and achieve a proper self-calibration. With an amplitude loss of about 30\%, all sources with the exception of 0420--014 and 1510--089 have SNR$\lesssim5$, which falls below the detection limits.

\begin{figure*}[h]
\centering
\includegraphics[angle=-90,width=57mm]{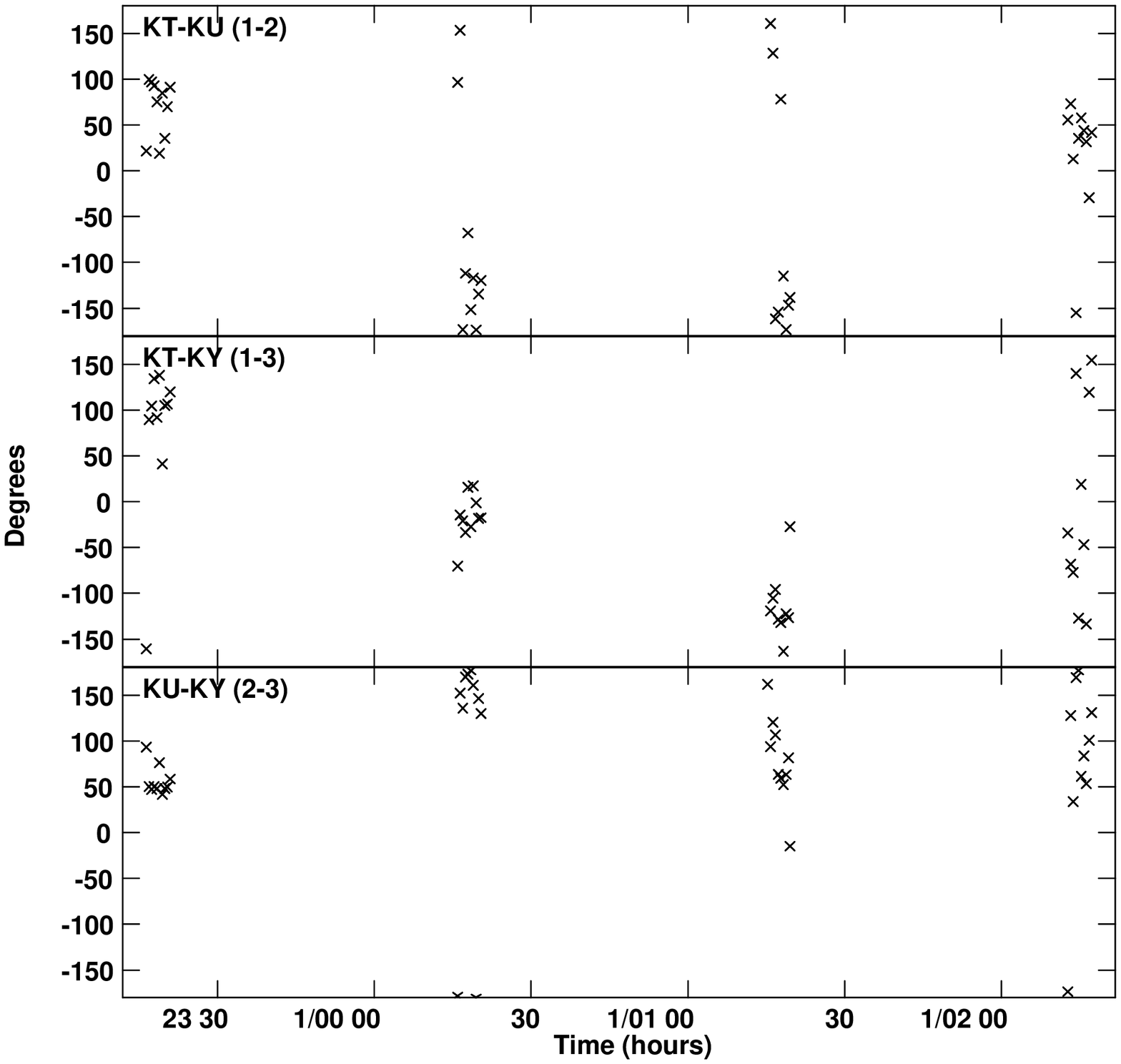}
\includegraphics[angle=-90,width=57mm]{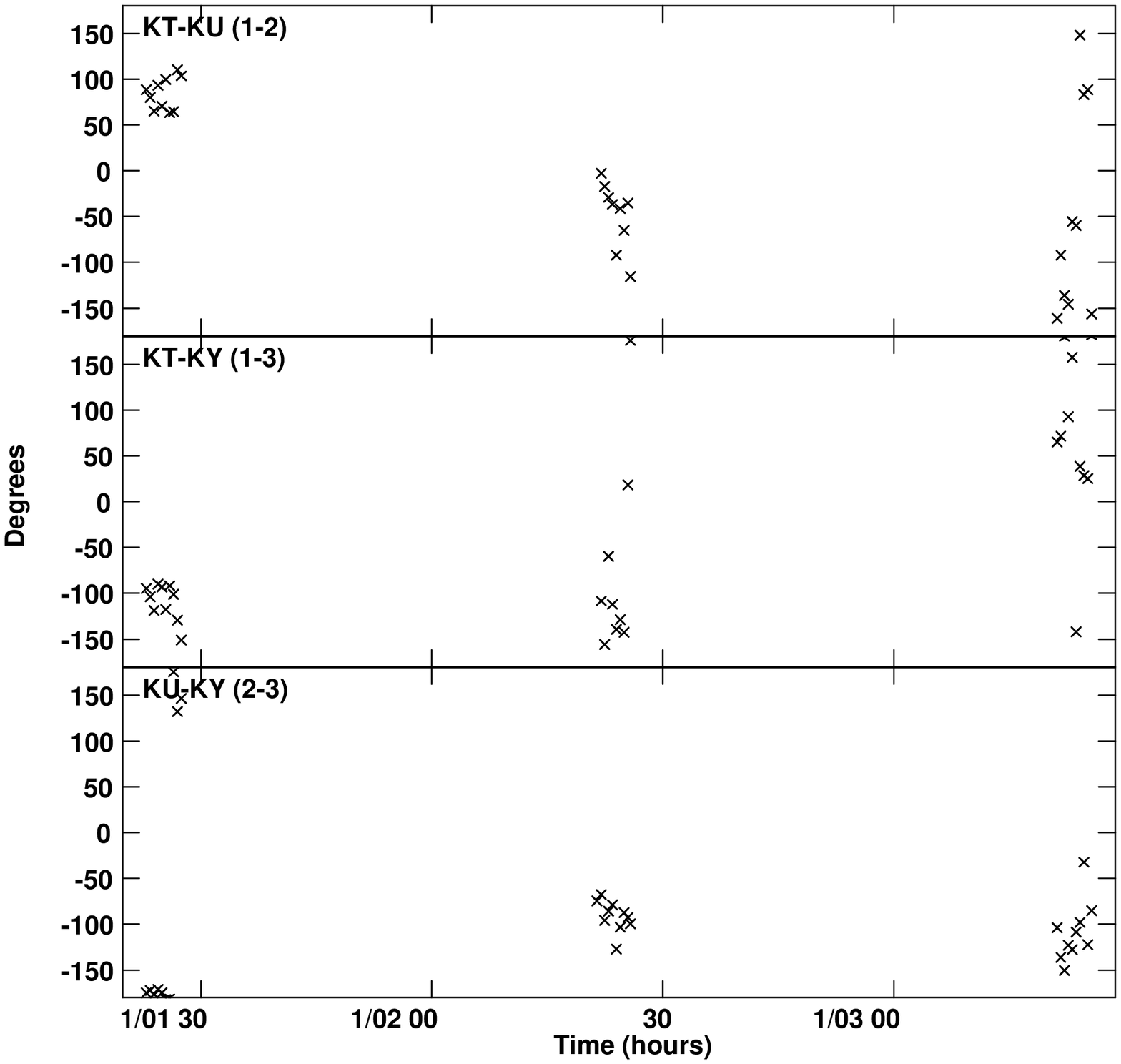}
\includegraphics[angle=-90,width=57mm]{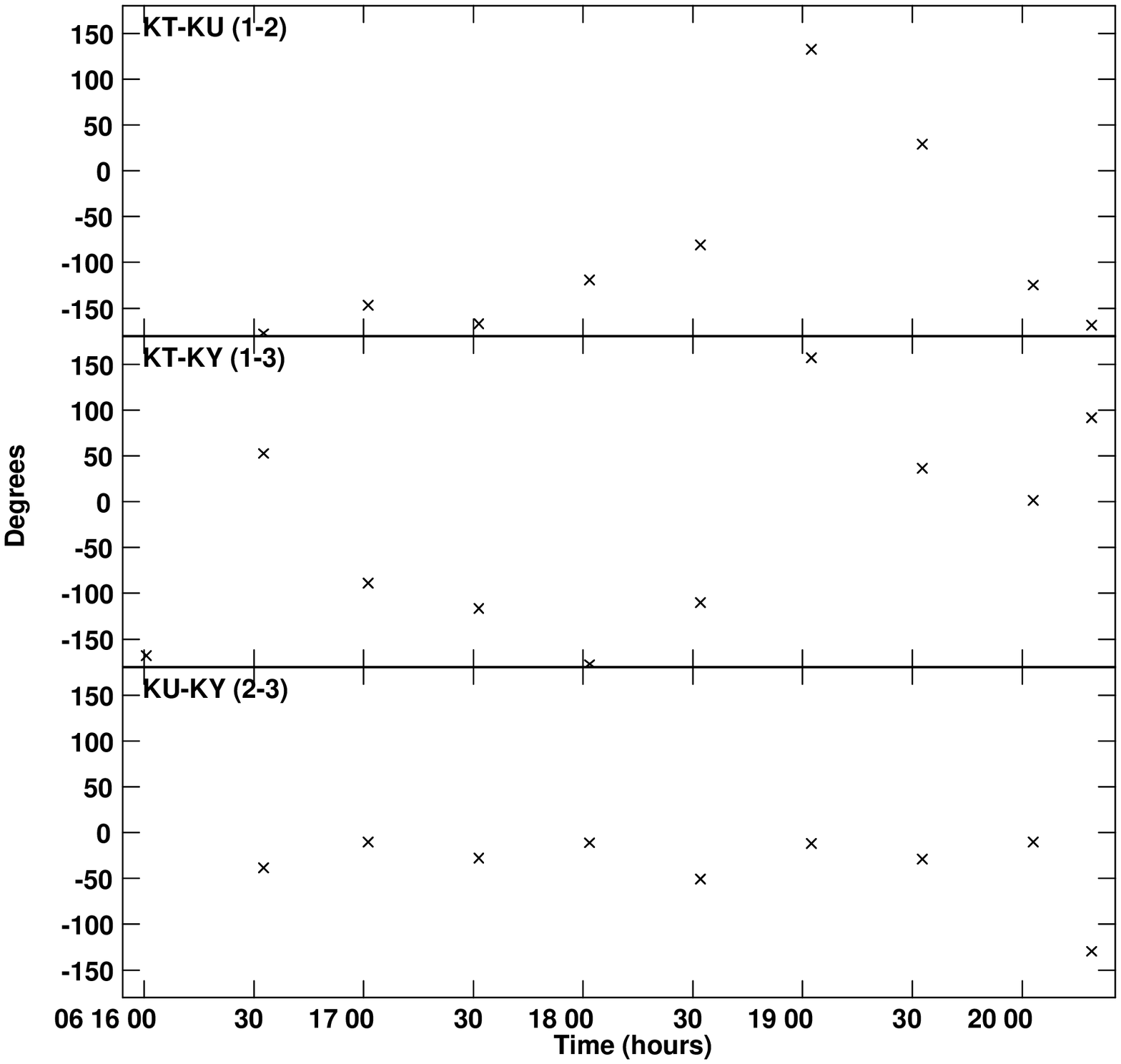}
\begin{tabular}{lll}
\hspace{0.8cm}(a) 0235+164 &\hspace{3.3cm} (b) 0420--014 &\hspace{3.3cm} (c) 0727--115\\
\end{tabular}
\includegraphics[angle=-90,width=57mm]{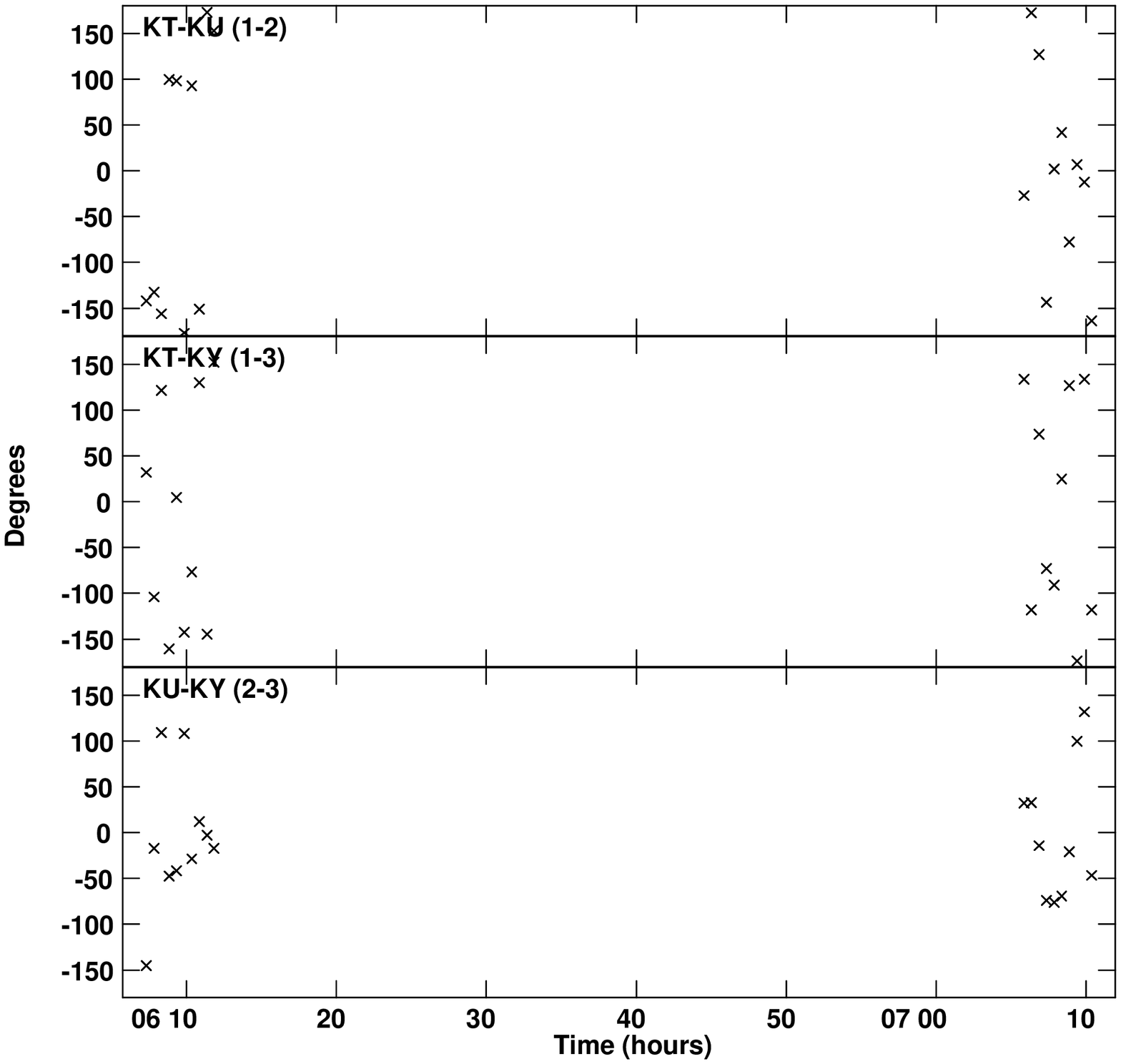}
\includegraphics[angle=-90,width=57mm]{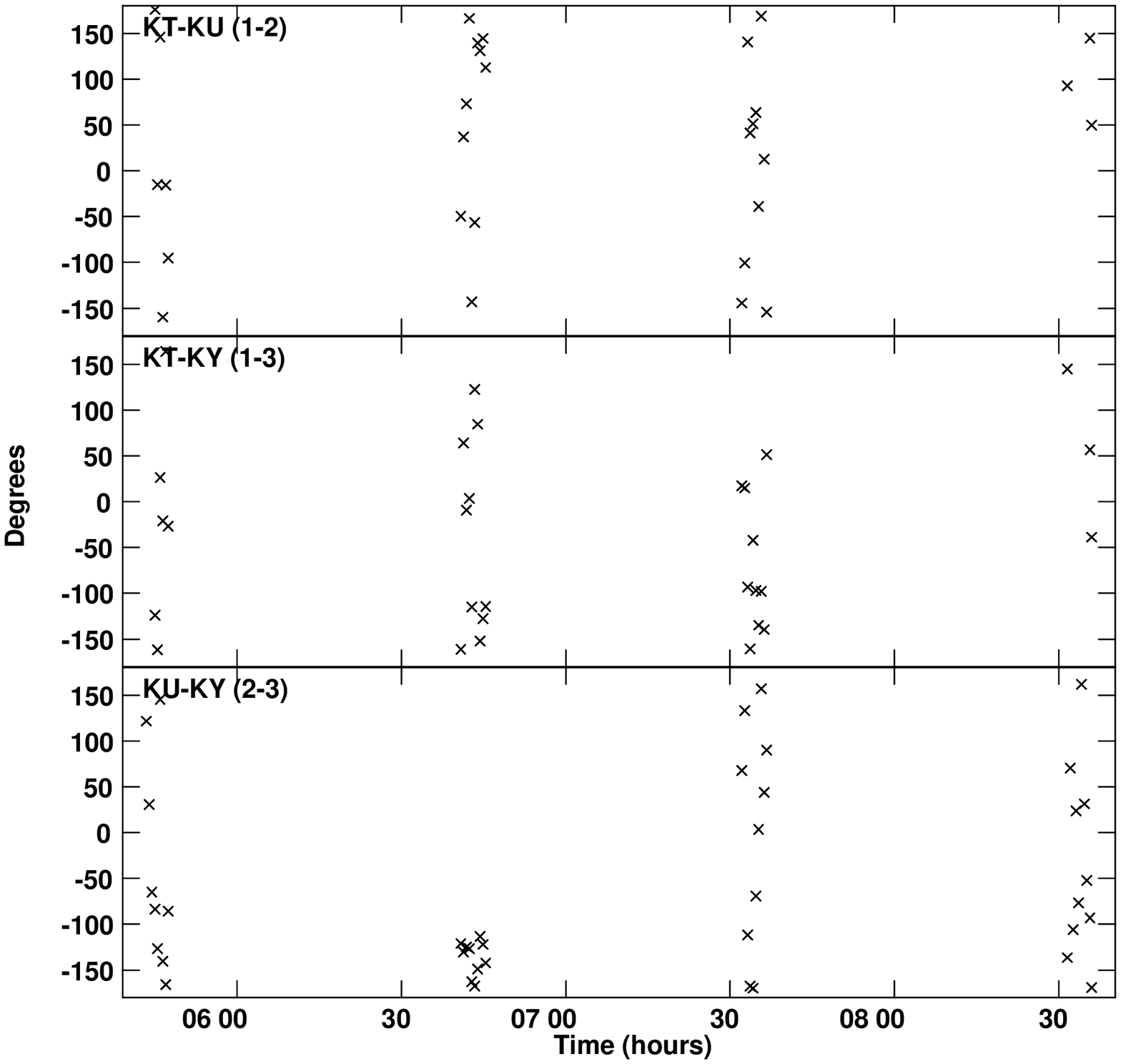}
\includegraphics[angle=-90,width=57mm]{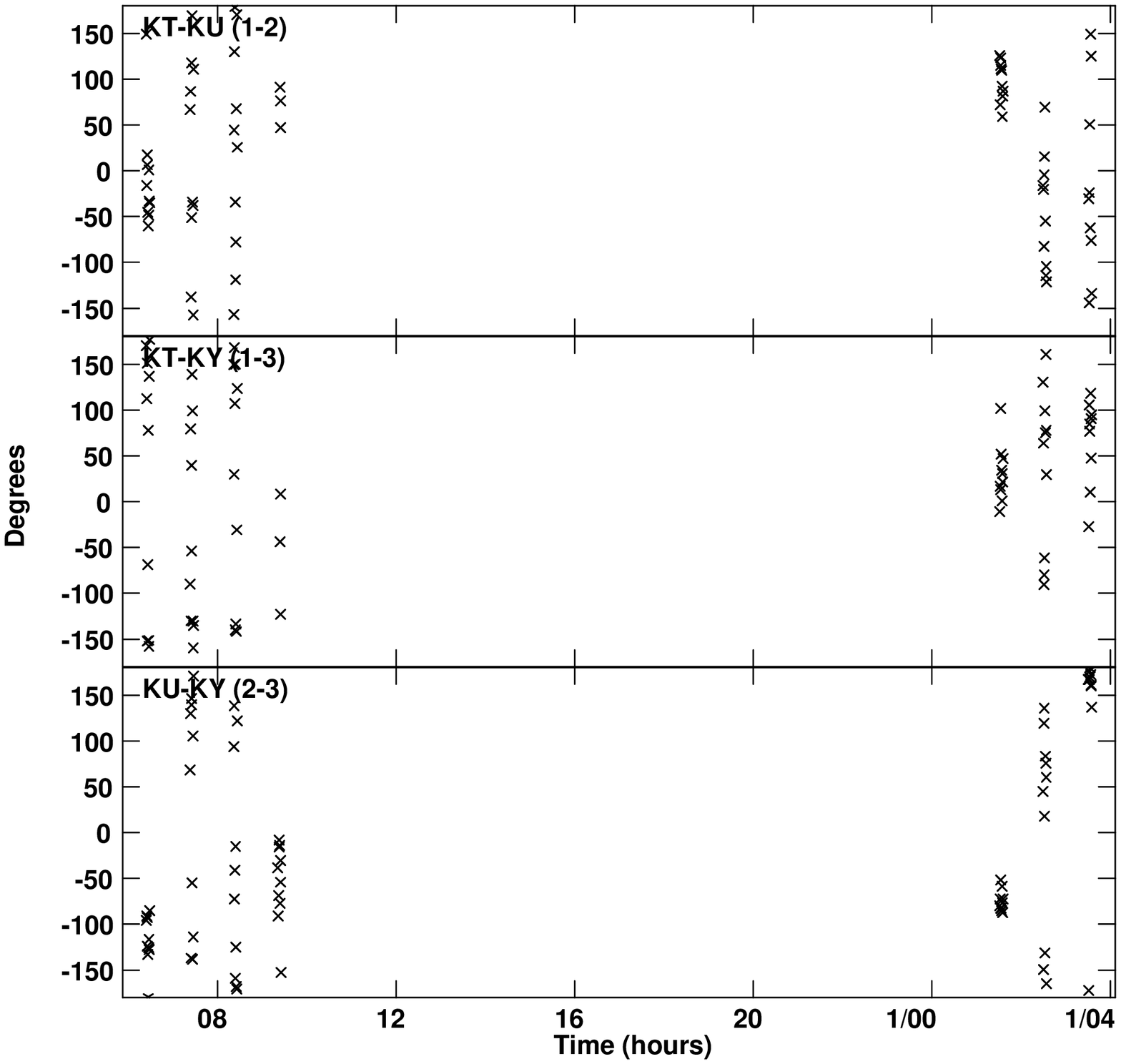}
\begin{tabular}{lll}
\hspace{0.8cm}(d) 0735+178 &\hspace{3.3cm} (e) 0827+243  &\hspace{3.3cm} (f) 0836+710\\
\end{tabular}
\includegraphics[angle=-90,width=57mm]{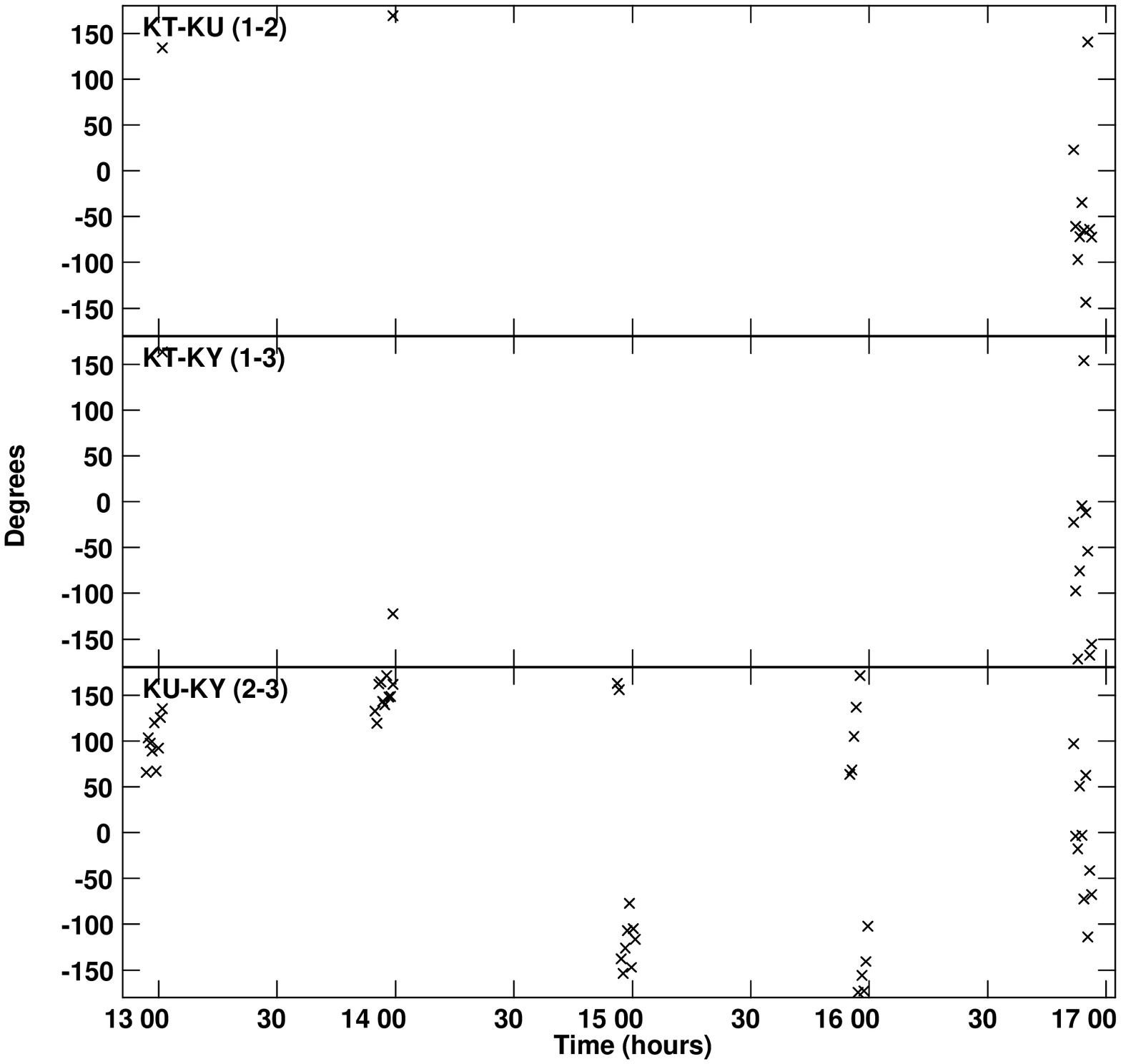}
\includegraphics[angle=-90,width=57mm]{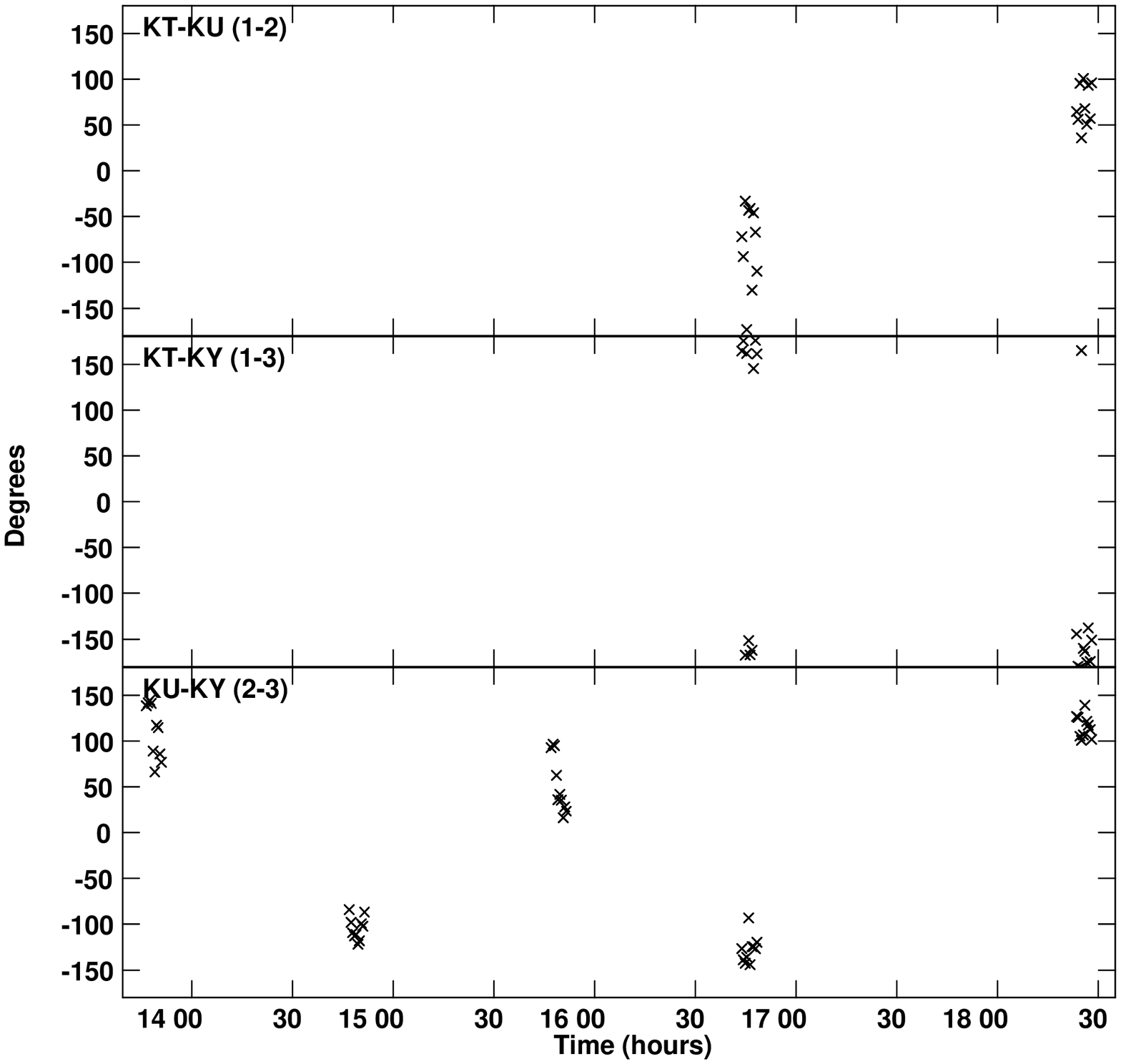}
\includegraphics[angle=-90,width=57mm]{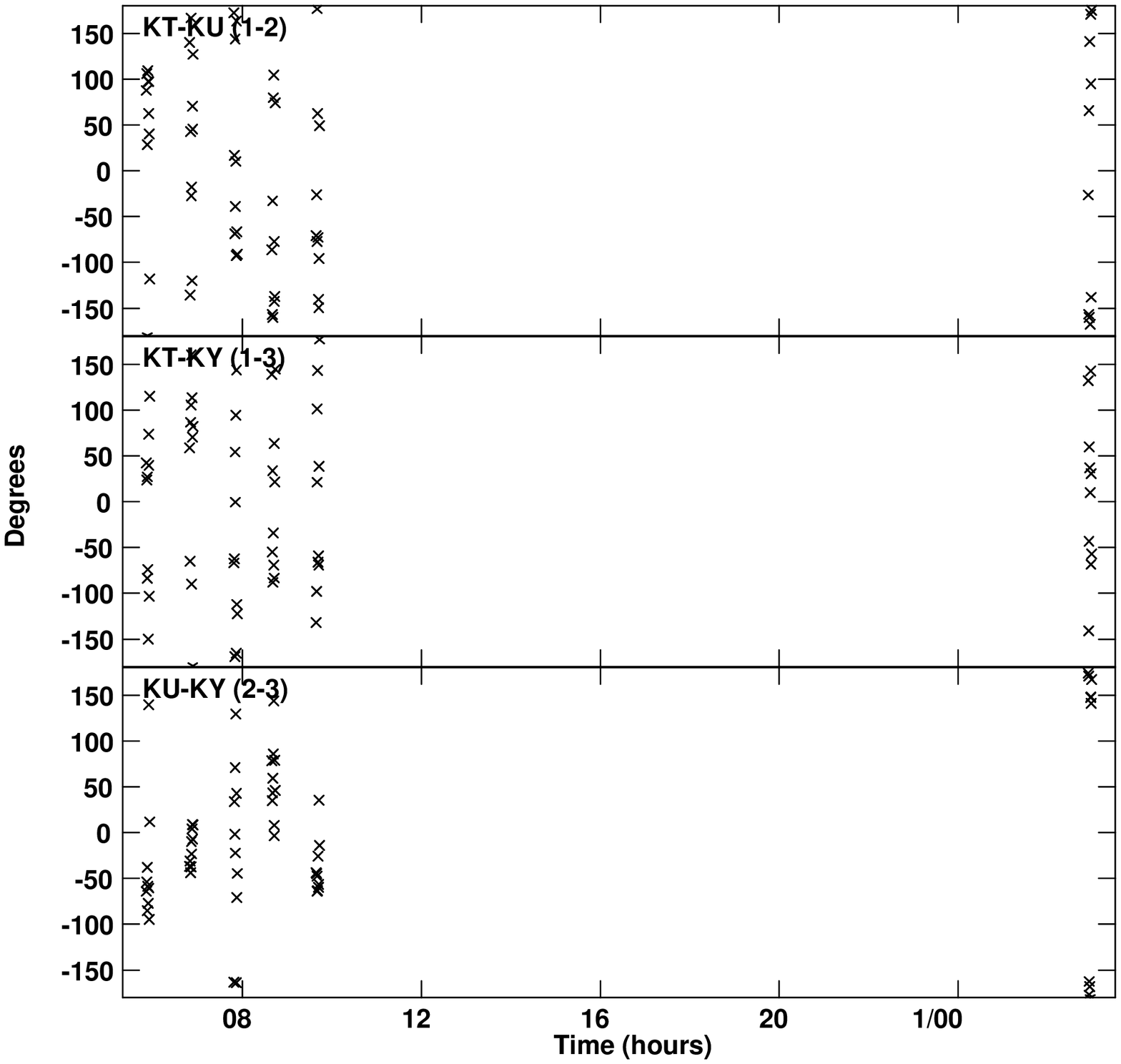}
\begin{tabular}{lll}
\hspace{0.8cm}(g) 1510--089 &\hspace{3.3cm} (h) 1611+343 &\hspace{3.3cm} (i) 4C39.25\\
\end{tabular}
\includegraphics[angle=-90,width=57mm]{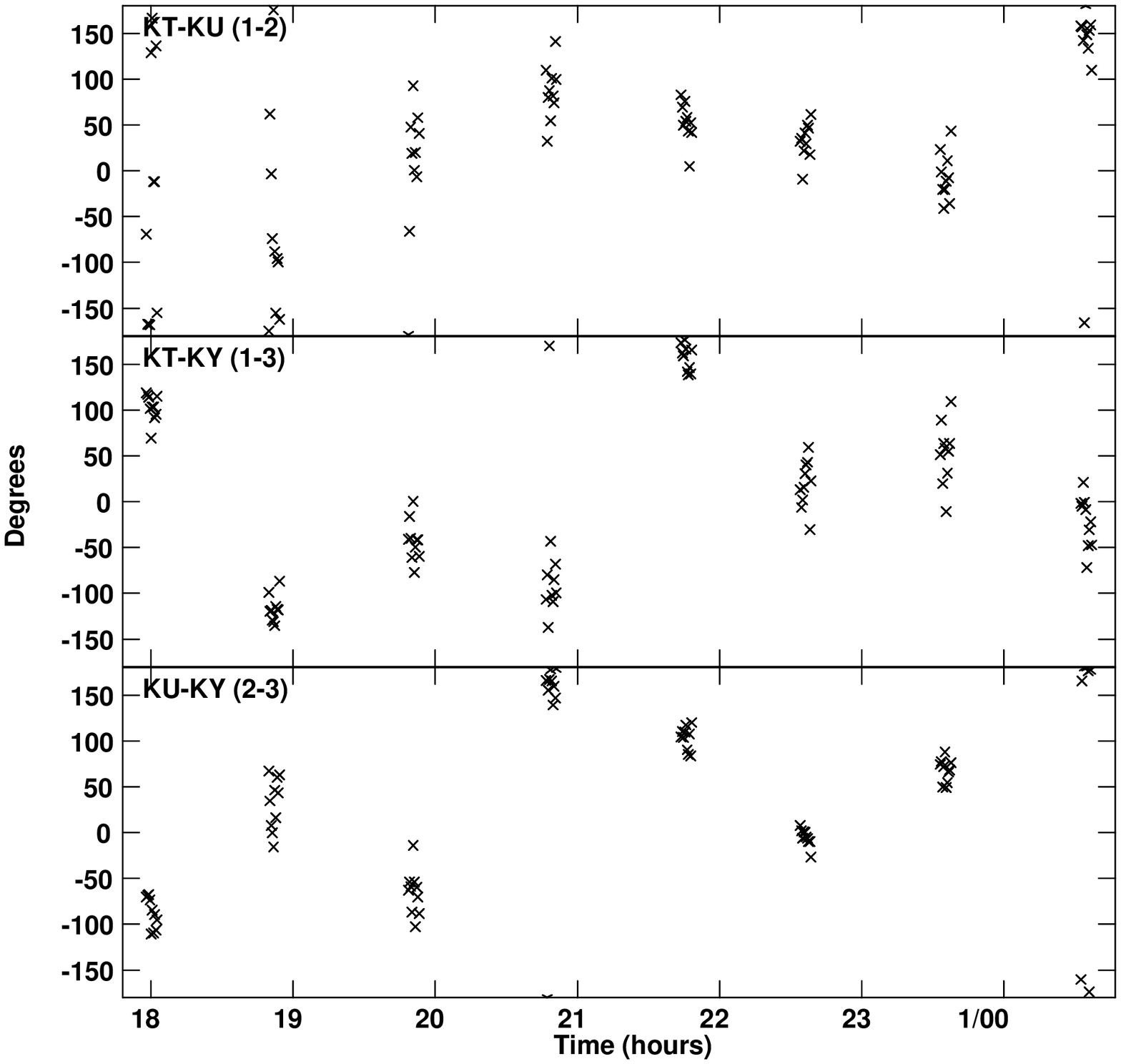}
\includegraphics[angle=-90,width=57mm]{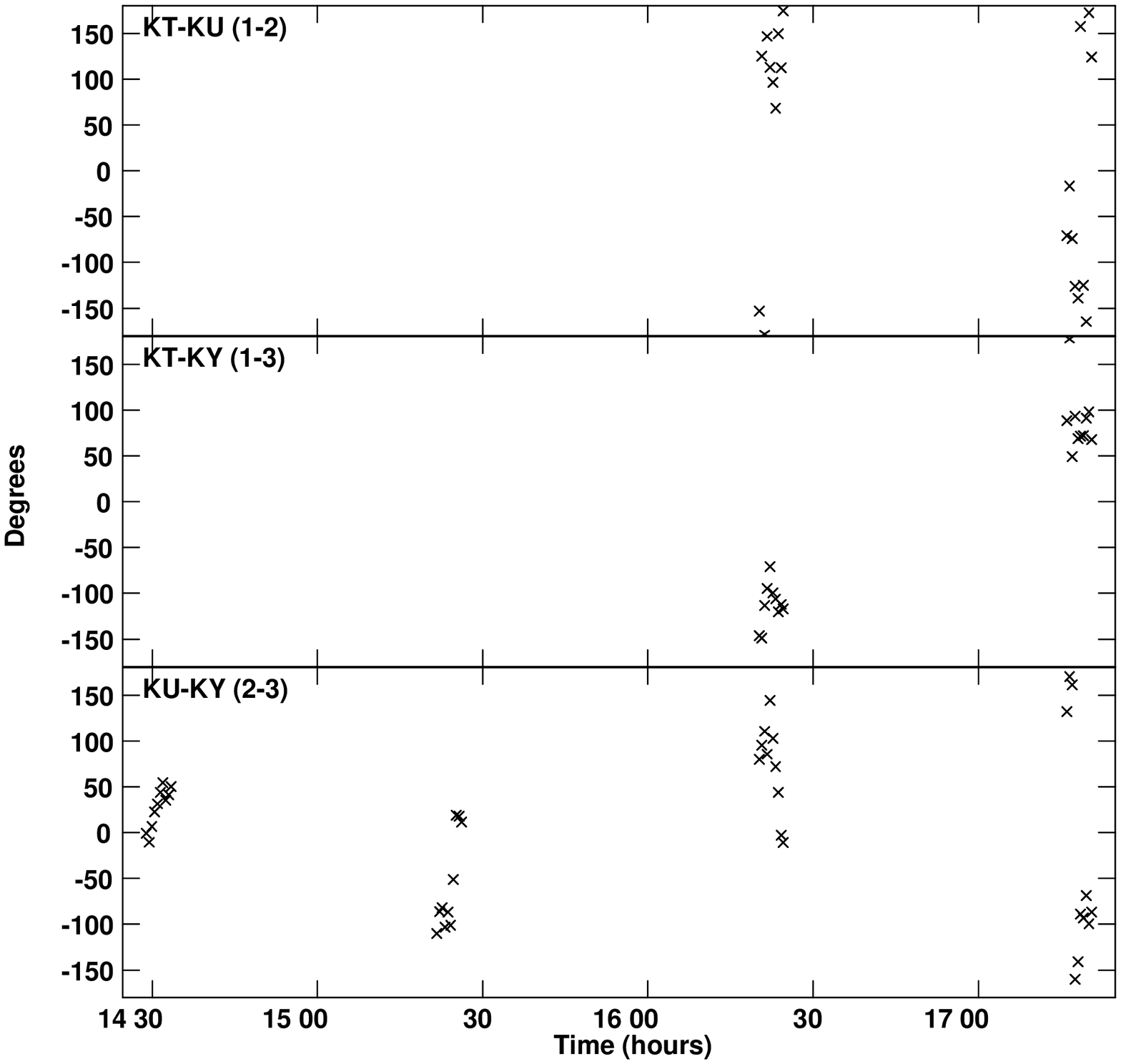}
%\begin{tabular*}{\textwidth}{c @{\extracolsep{\fill}} ccc}
\begin{tabular}{lll}
\hspace{0.8cm}(j) CTA102 &\hspace{3.3cm} (k) NRAO530\\
\end{tabular}
\caption{FPT visibility phases for iMOGABA15 observations at 86~GHz. Each data set was calibrated using the scaled solutions from the analysis  of the same source at 22~GHz. Plots show temporal average of 30 seconds.\label{iM15-FPTphase22-86}}
%\vspace{5mm} %% add extra space ONLY when figures/tables are "colliding"!
\end{figure*}

\begin{figure*}[h]
\centering
\includegraphics[angle=-90,width=57mm]{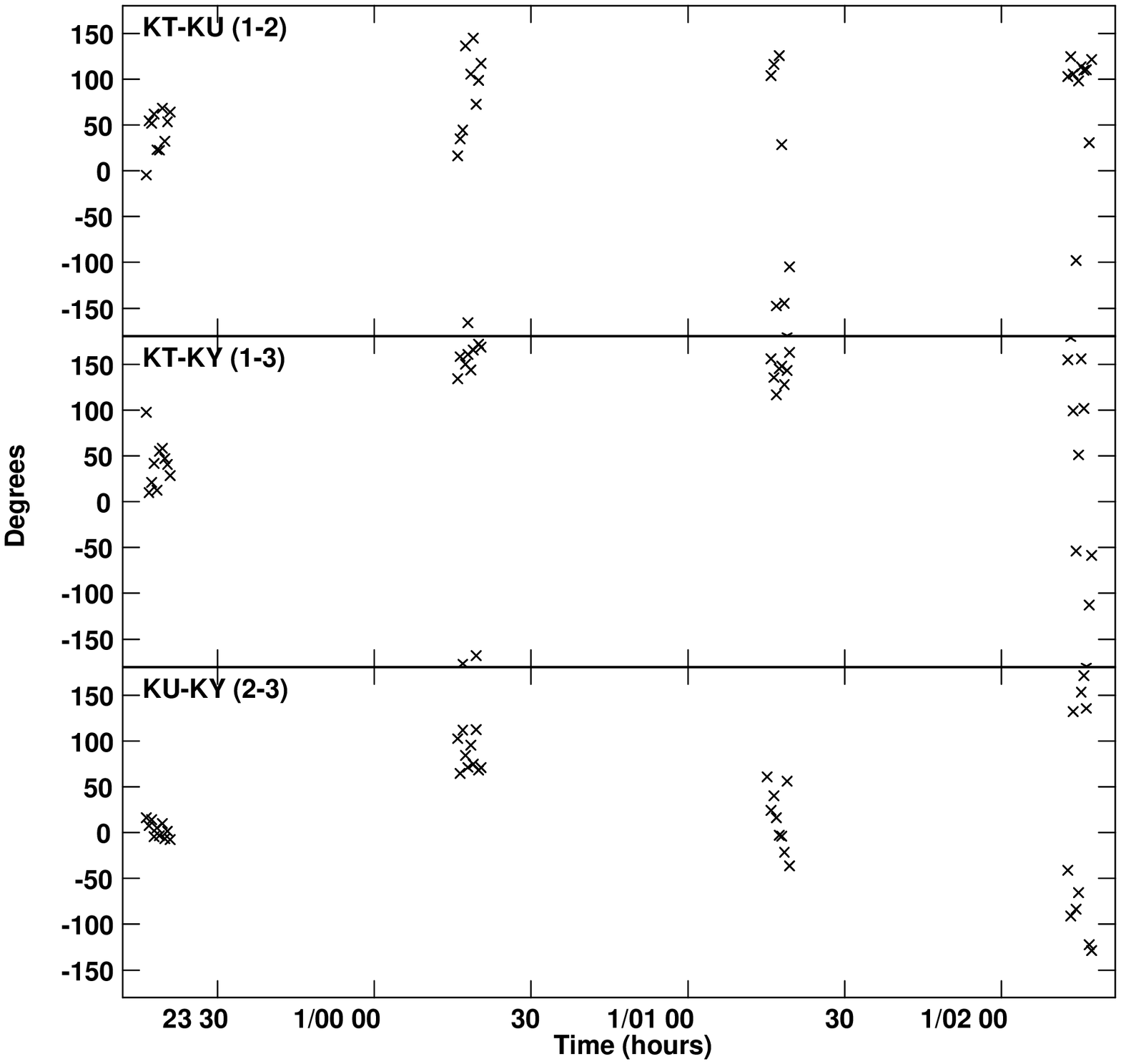}
\includegraphics[angle=-90,width=57mm]{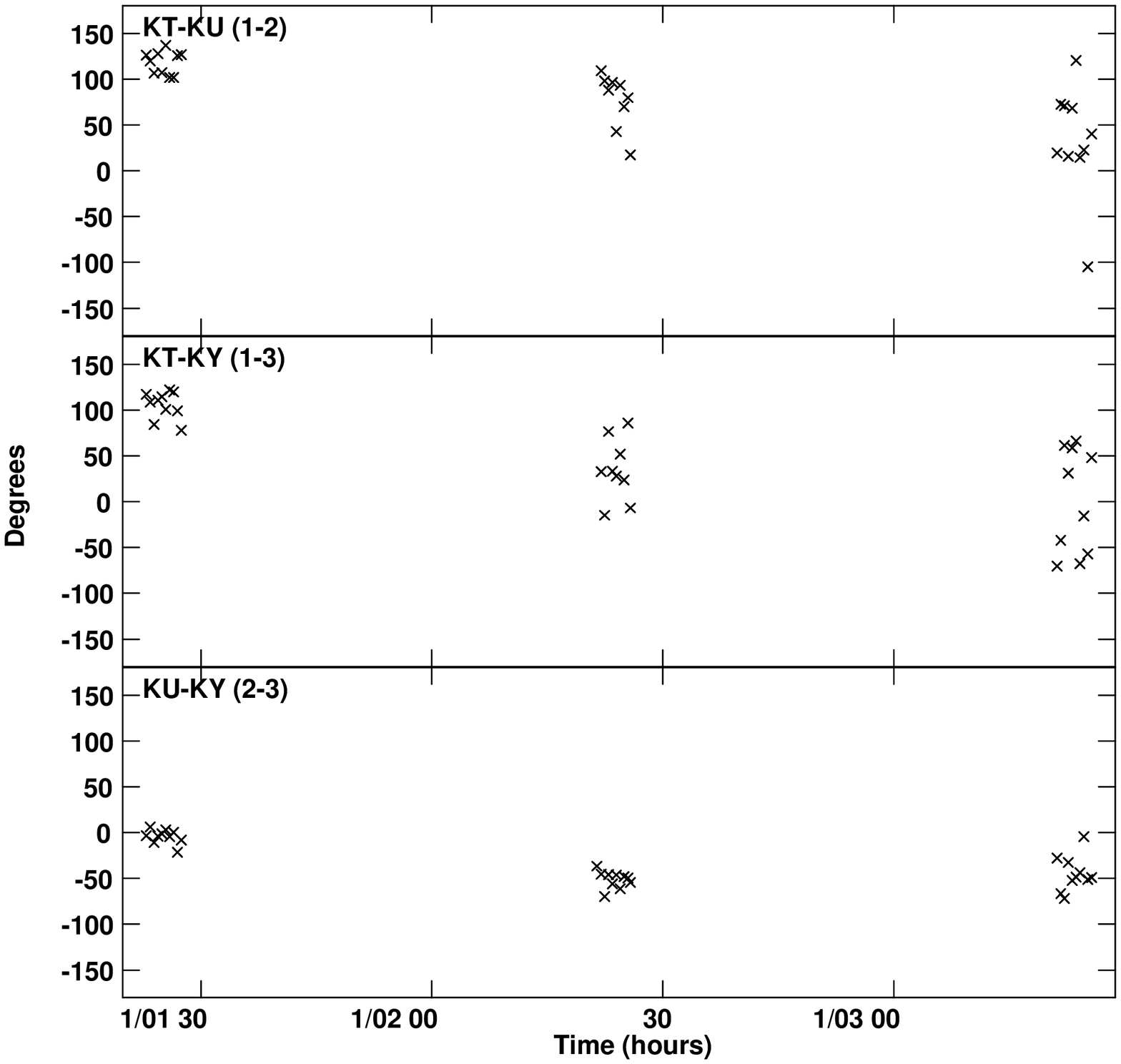}
\includegraphics[angle=-90,width=57mm]{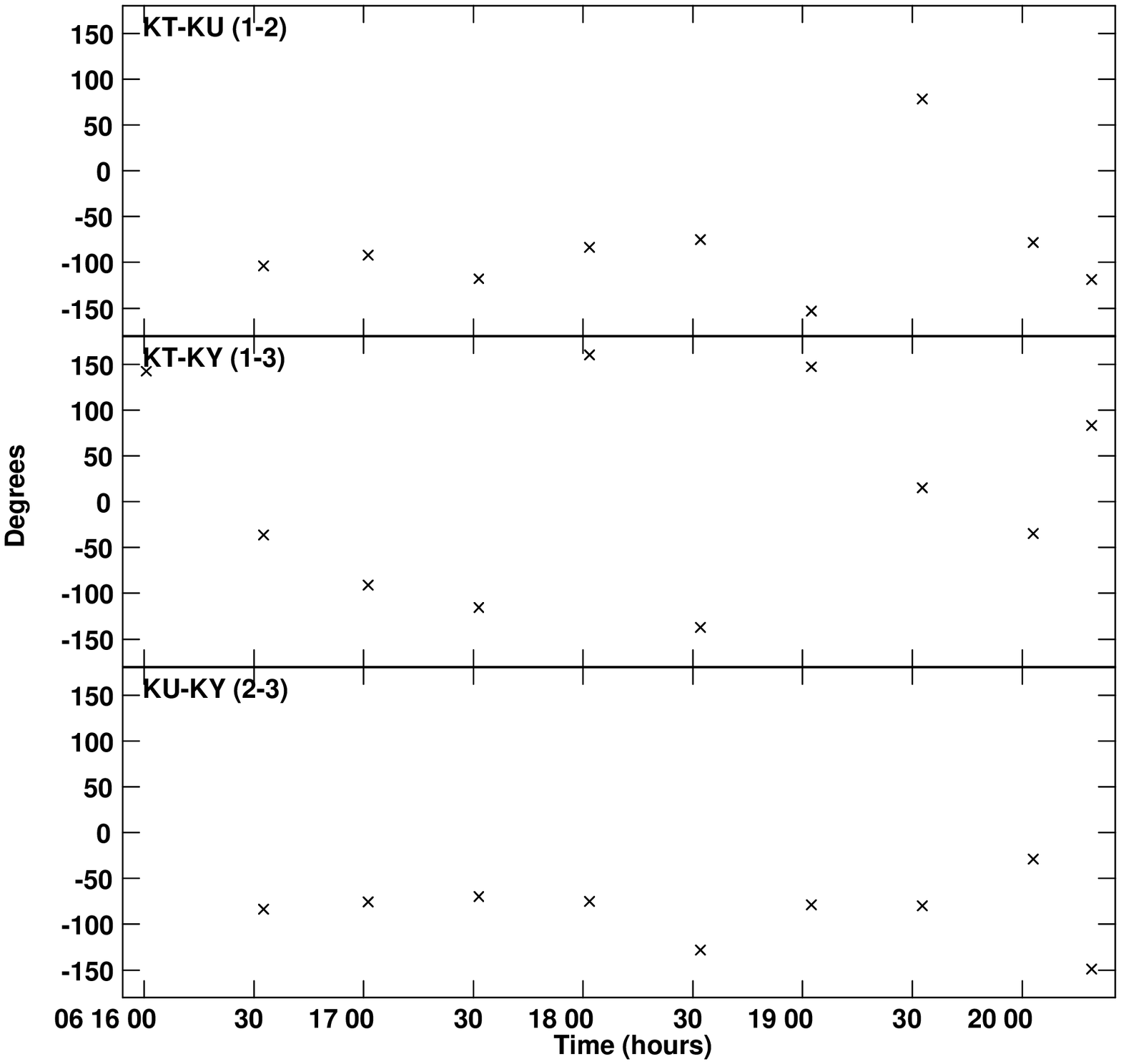}
\begin{tabular}{lll}
\hspace{0.8cm}(a) 0235+164 &\hspace{3.3cm} (b) 0420--014 &\hspace{3.3cm} (c) 0727--115\\
\end{tabular}
\includegraphics[angle=-90,width=57mm]{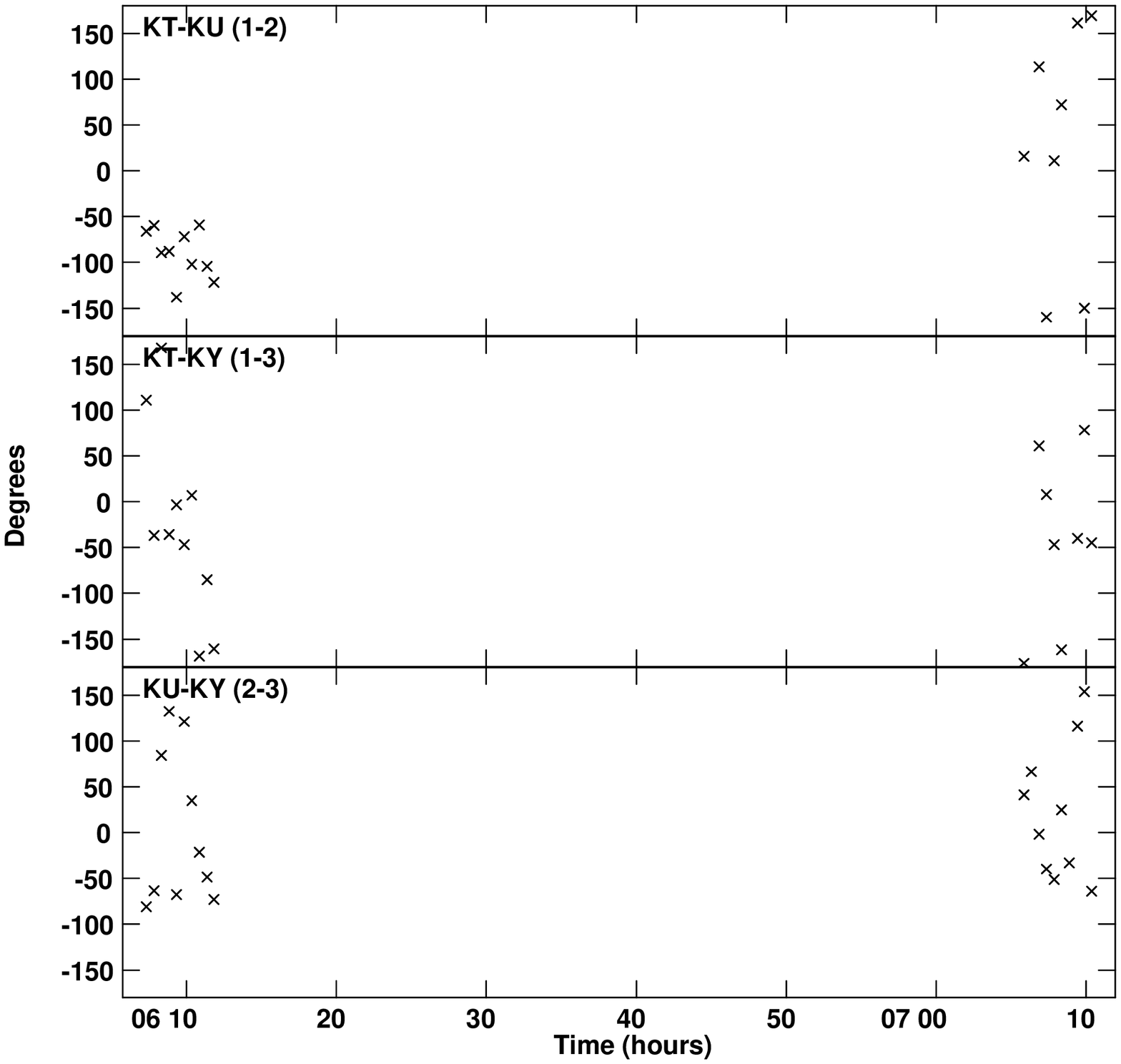}
\includegraphics[angle=-90,width=57mm]{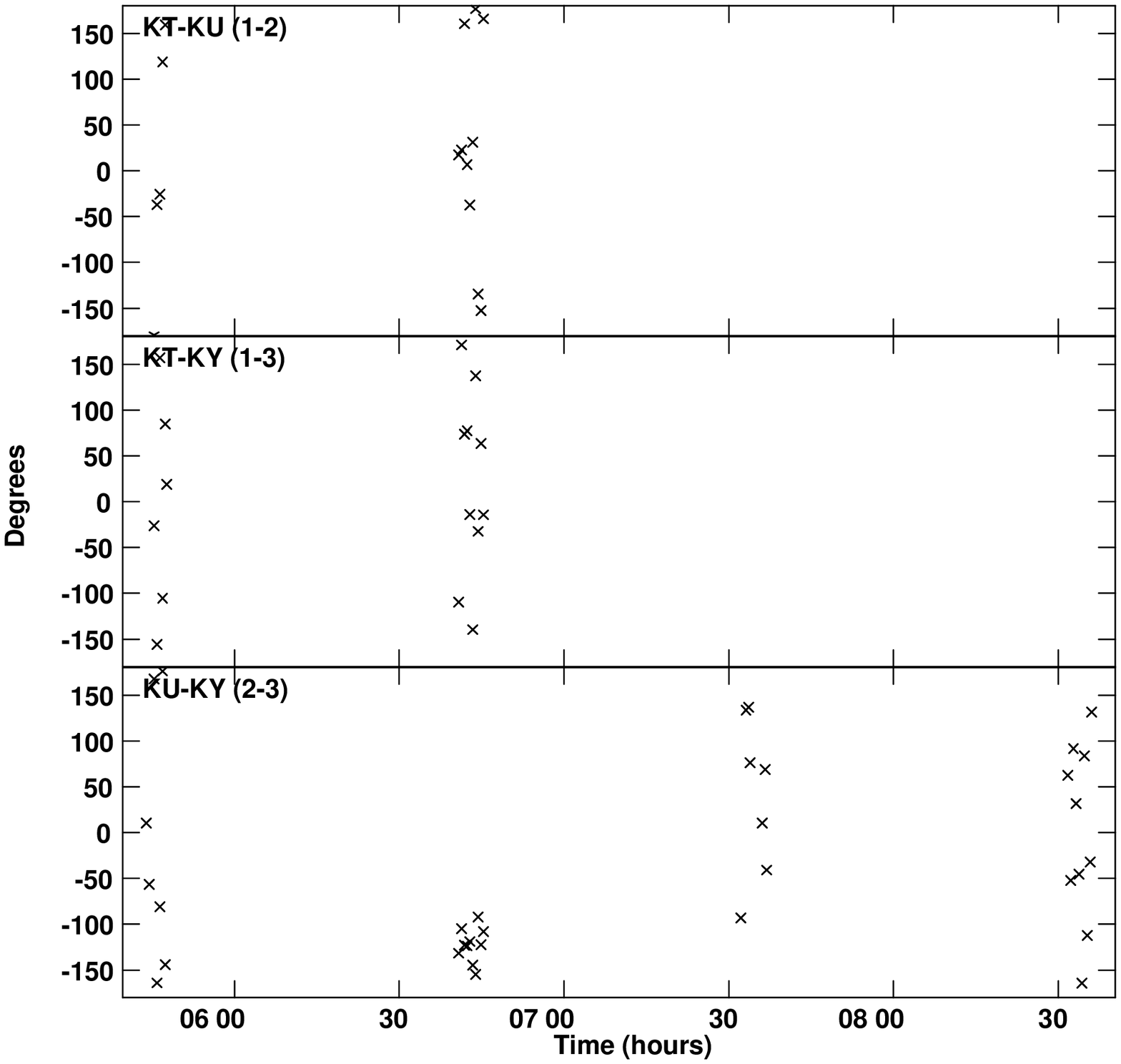}
\includegraphics[angle=-90,width=57mm]{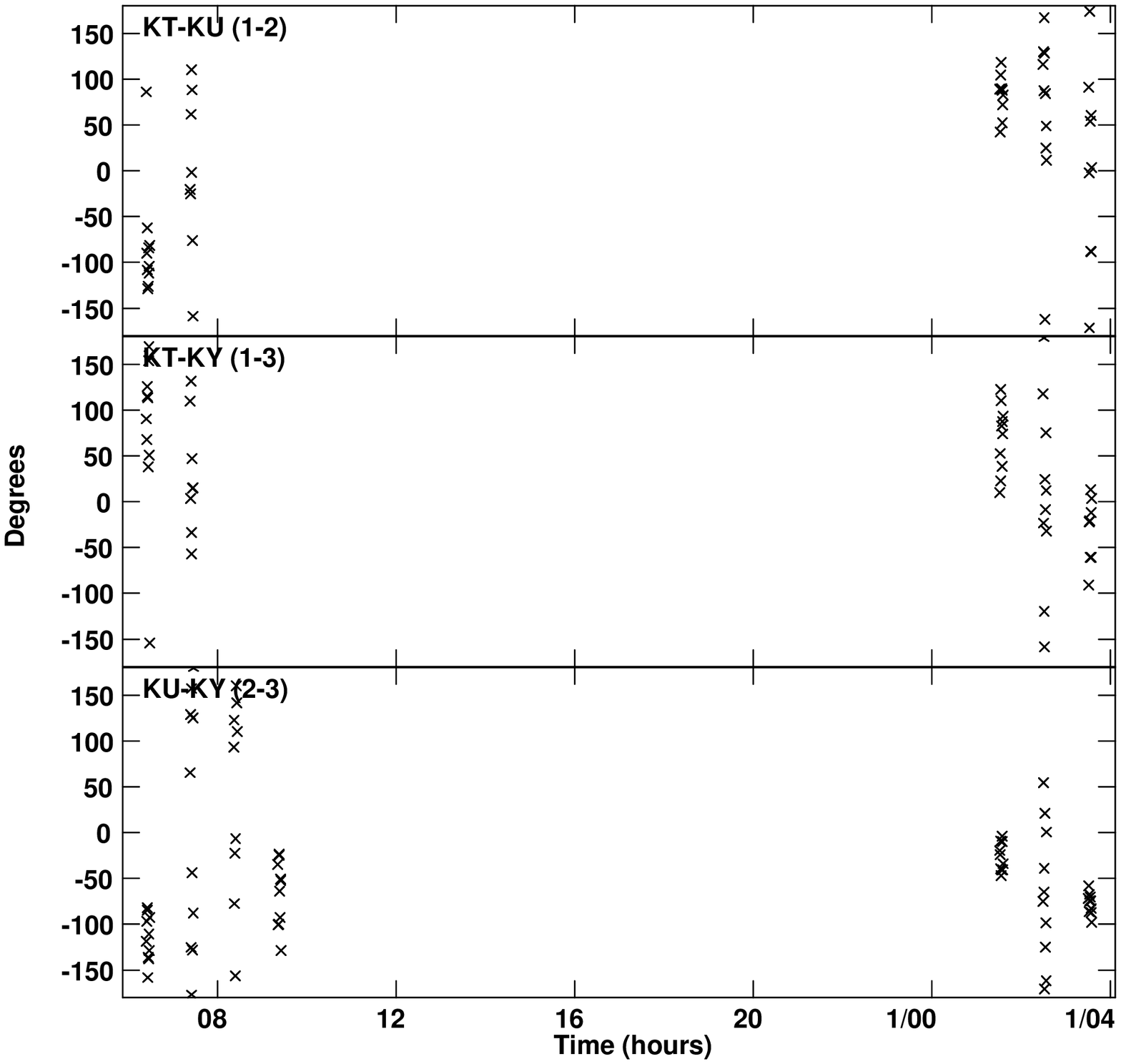}
\begin{tabular}{lll}
\hspace{0.8cm}(d) 0735+178 &\hspace{3.3cm} (e) 0827+243  &\hspace{3.3cm} (f) 0836+710\\
\end{tabular}
\includegraphics[angle=-90,width=57mm]{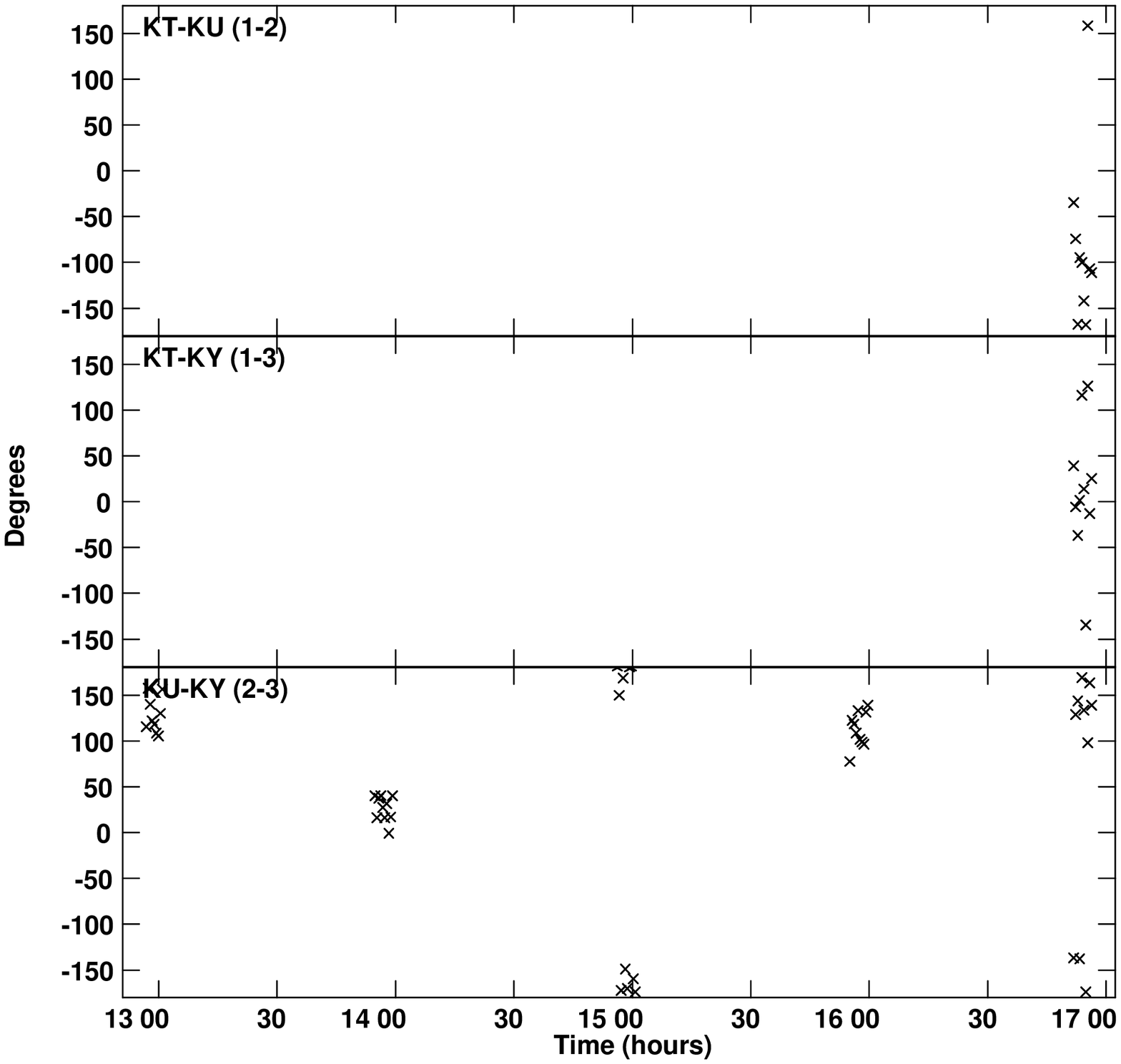}
\includegraphics[angle=-90,width=57mm]{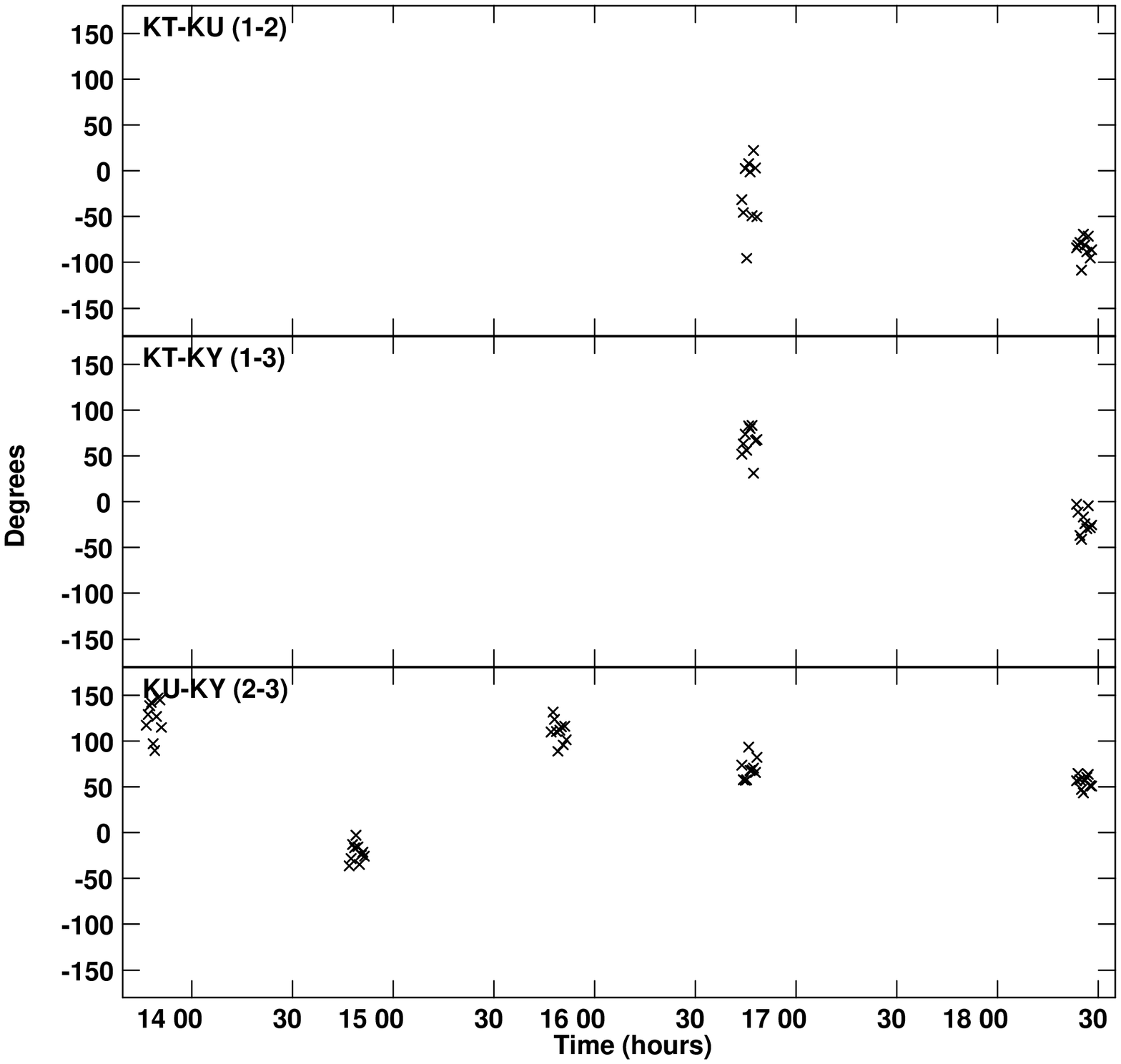}
\includegraphics[angle=-90,width=57mm]{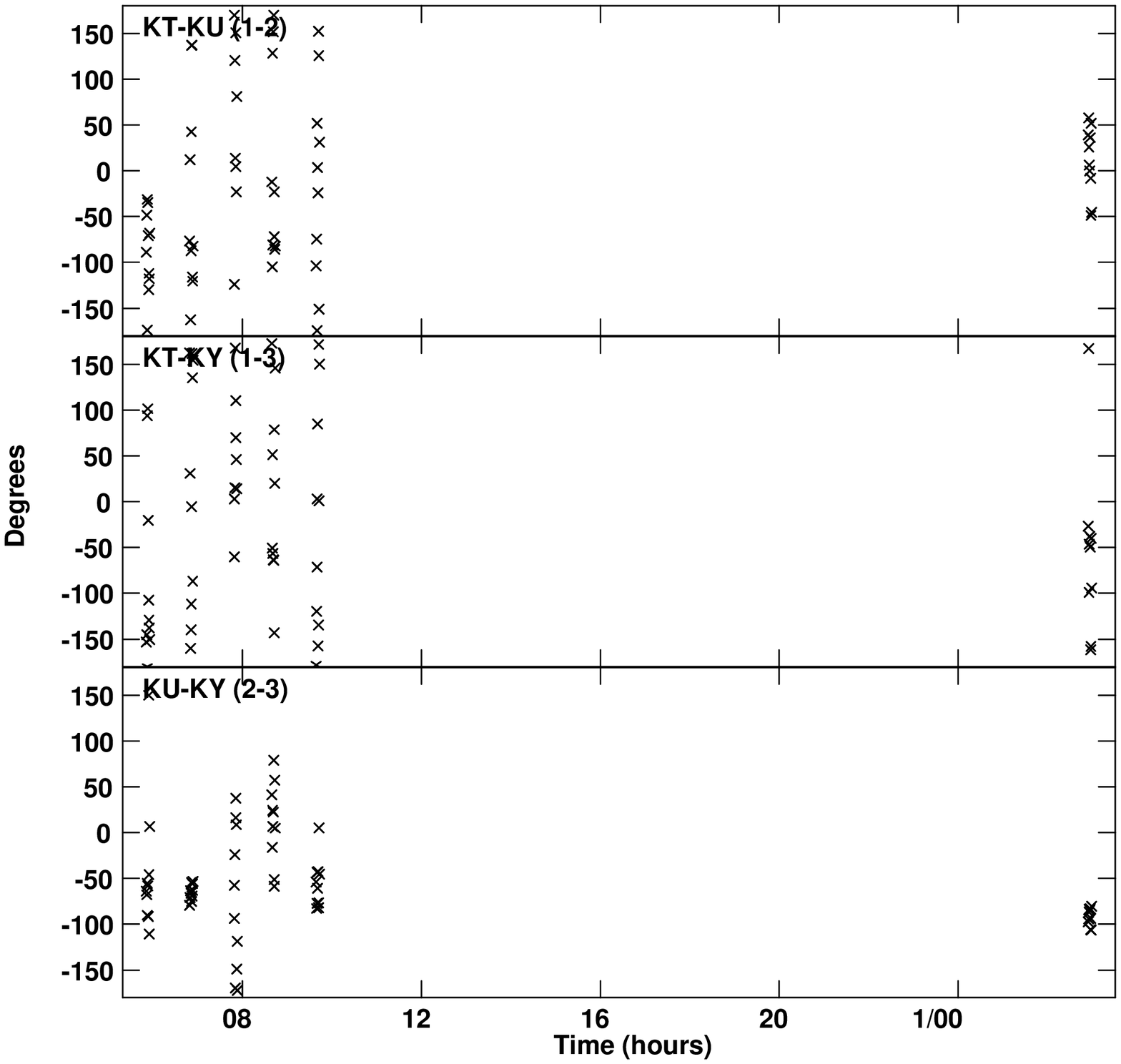}
\begin{tabular}{lll}
\hspace{0.8cm}(g) 1510--089 &\hspace{3.3cm} (h) 1611+343 &\hspace{3.3cm} (i) 4C39.25\\
\end{tabular}
\includegraphics[angle=-90,width=57mm]{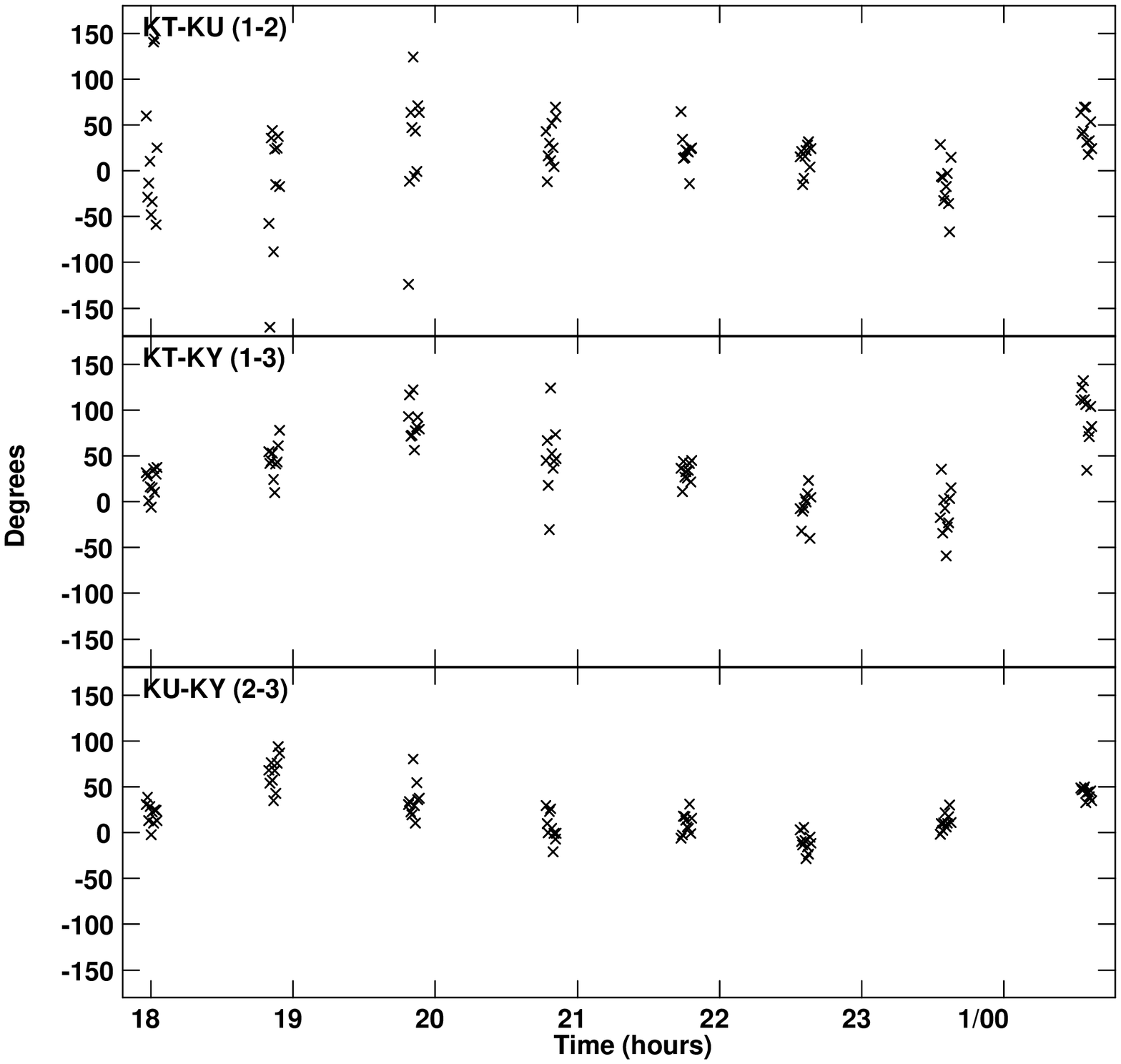}
\includegraphics[angle=-90,width=57mm]{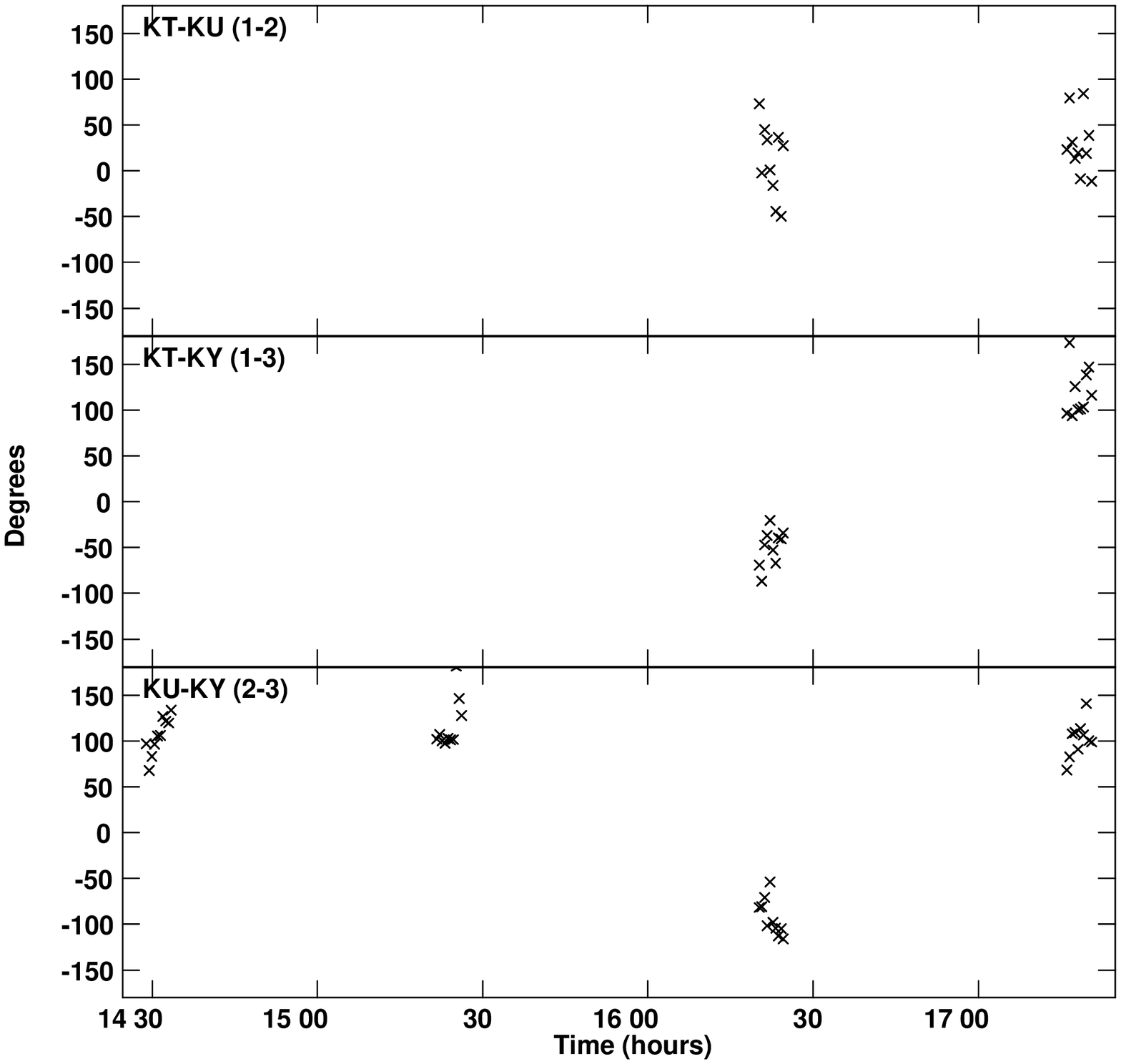}
%\begin{tabular*}{\textwidth}{c @{\extracolsep{\fill}} ccc}
\begin{tabular}{lll}
\hspace{0.8cm}(j) CTA102 &\hspace{3.3cm} (k) NRAO530\\
\end{tabular}
\caption{Same as Figure \ref{iM15-FPTphase22-86}, but using the scaled solutions from the analysis of the same source at 43 GHz.\label{iM15-FPTphase43-86}}
%\vspace{5mm} %% add extra space ONLY when figures/tables are "colliding"!
\end{figure*}

We use the FPT technique for this data, given that a five minute integration time allows us to detect all the sources in our sample. As for iMOGABA9, we have performed the transfer from both 22 and 43~GHz for consistency checking and comparison. Plots of the FPT calibrated phase visibilities from 22 and 43~GHz are shown in Figures \ref{iM15-FPTphase22-86} and \ref{iM15-FPTphase43-86} respectively. 

A relatively low scatter in the phases is obtained for a significant number of sources (e.g., 0235+164, 0420--014, 1611+343, CTA102 or NRAO530). Much larger scatter is found for some sources, including 0827+243, 0836+710 and 4C39.25, despite the fact that some of them show an a priori larger SNR$^{86}$. This effect may be due to the amplitude loss problem. Note indeed that 0836+710 and 4C39.25 were directly affected by the Tamna antenna offset issue. Furthermore, it is clear that the Ulsan--Yonsei baseline, which is free from this problem, shows on average better solutions\footnote{This can also be in combination with the fact that Ulsan--Yonsei is the shortest baseline.}.

Comparison between Figures \ref{iM15-FPTphase22-86} and \ref{iM15-FPTphase43-86} shows an agreement between the results from the FPT from 22 and 43~GHz. In some sources (e.g., CTA102), the phase dispersion seems to be larger based on the transfer from 22~GHz. This is expected as this source was optically thick during our observations, with a peak flux of $S^{22}\sim100$ mJy and $S^{43}\sim150$ mJy. This leads to a higher SNR at 43 than 22~GHz which, combined with the smaller frequency ratio scaling $R$, naturally leads to better phase stability when performing the transfer.

We inspected the data after the successful frequency transfer and proceeded to re--fringe fitting with a solution interval of 5 minutes, and applied the solutions to the data. 
However, due to the various amplitude offset problems, we considered the data is not robust enough to consider a reliable imaging with accurate flux and structure. We thus do not show the related frequency phase transferred maps here.

\subsubsection{Frequency Phase Transfer to 129~GHz}\label{sec:iM15FPT86v15}
No sources were imaged for iMOGABA15 at 129~GHz using standard VLBI reduction methods. SEFDs for this frequency are in the range $10-80\times10^3$ Jy, which implies baseline based sensitivity limits $\sigma^{theo}_{min}\sim$200--1500~mJy and $\sigma^{theo}_{min}\sim$60--500~mJy for 30 and 300 seconds integration times, respectively. In Table \ref{theoreticalflux129b} we summarize the SEFDs, expected flux at 129~GHz and detection limits for integration times of 30 and 300 seconds. Note that not all the sources in iMOGABA15 are included because a proper estimate for the extrapolated flux at 129~GHz could not be obtained due to the amplitude offset problem at lower frequencies.

Inspection of Table \ref{theoreticalflux129b} allows us to quantitatively understand the effects of the bad weather on the iMOGABA targets at 129~GHz. Due to the large SEFDs, it is challenging to get significant detections for the majority of our sources. Even with integration times of about 5 minutes  there are 13 sources that do not reach $SNR^{129}_{300s}>5$ and only 8 of the brightest sources reach larger values.

\begin{figure*}[h]
\centering
\includegraphics[angle=-90,width=57mm]{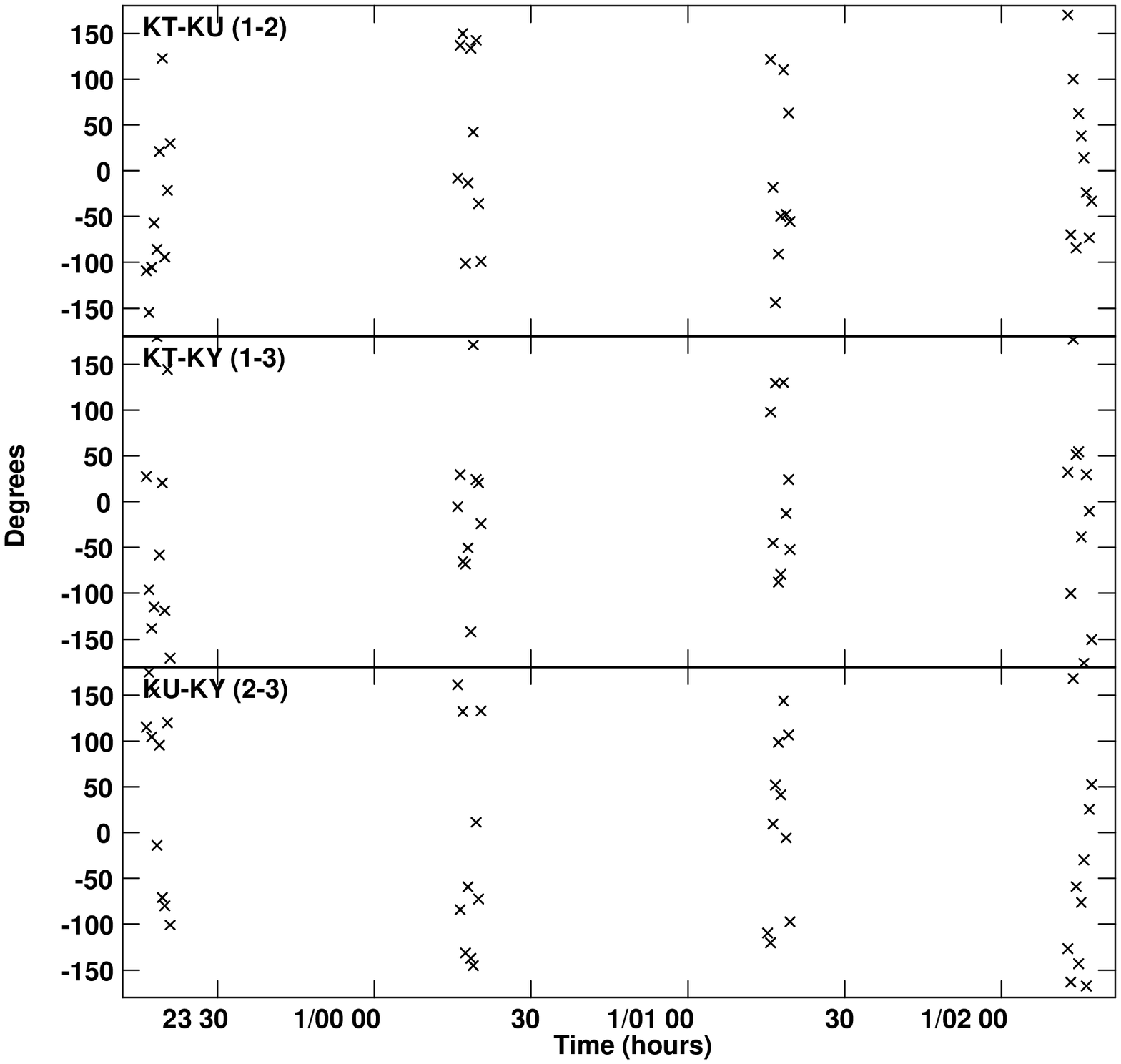}
\includegraphics[angle=-90,width=57mm]{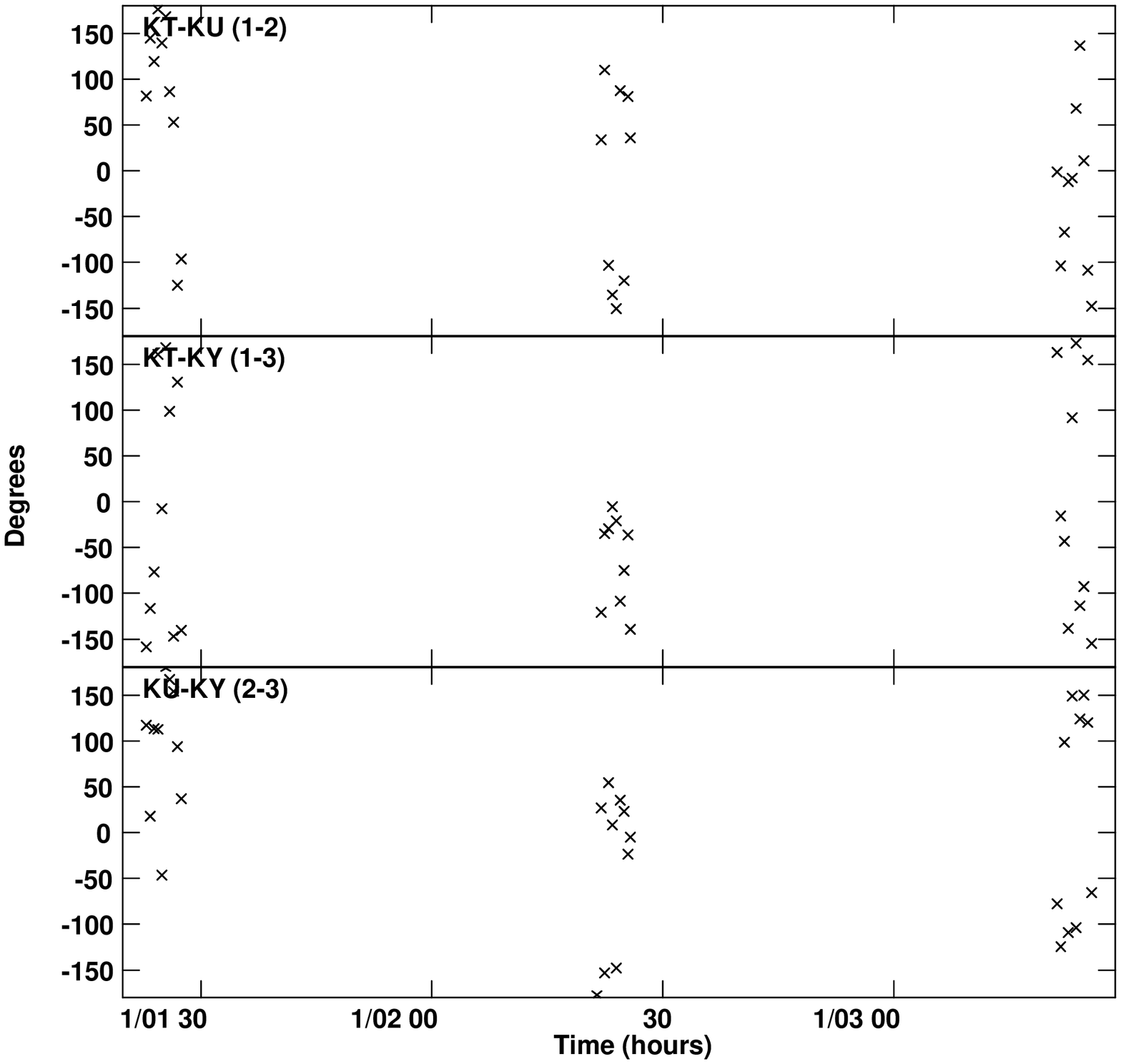}
\includegraphics[angle=-90,width=57mm]{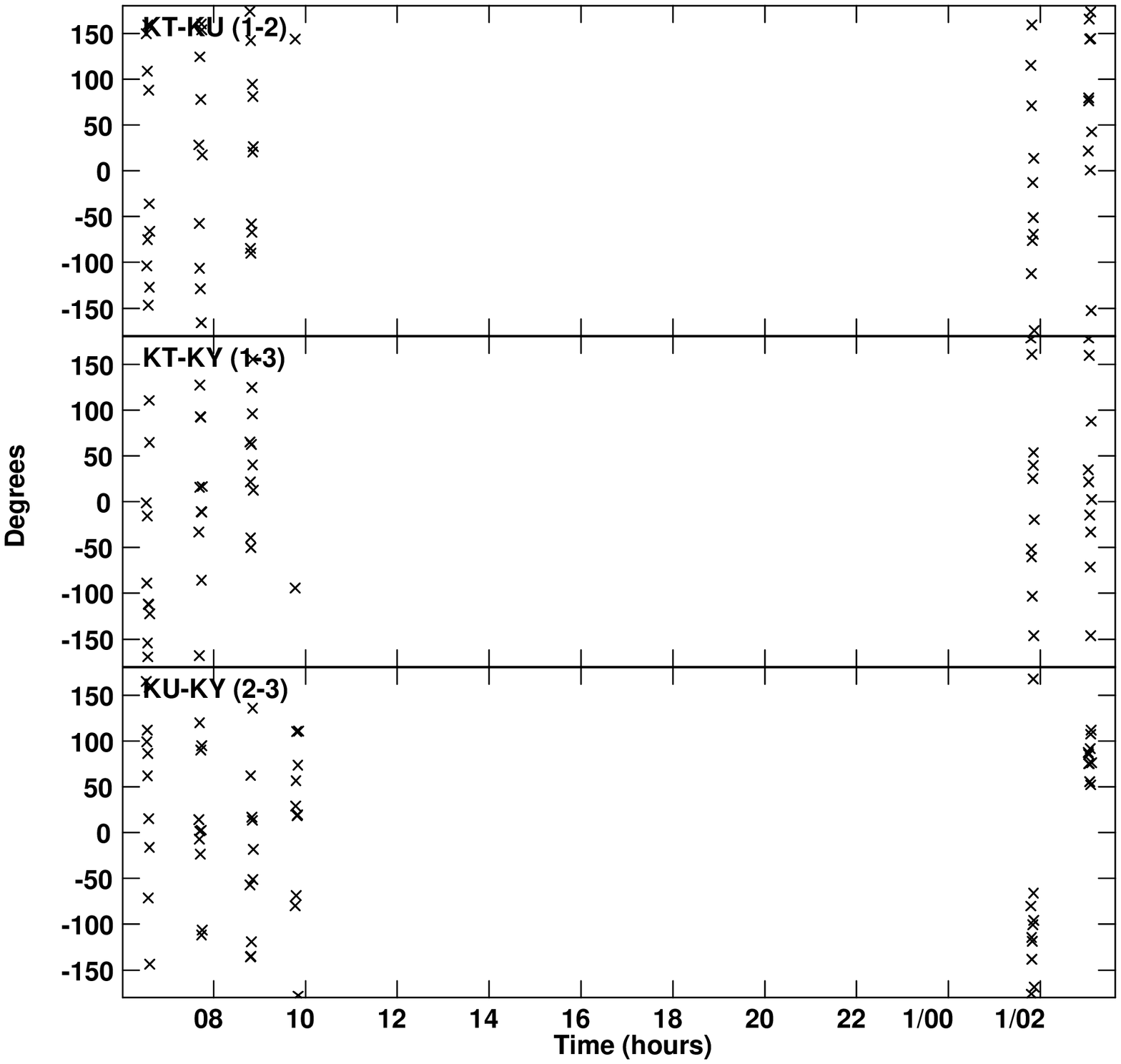}
\begin{tabular}{lll}
\hspace{0.8cm}(a) 0235+164 &\hspace{3.3cm} (b) 0420--014 &\hspace{3.3cm} (c) 0716+714\\
\end{tabular}
\includegraphics[angle=-90,width=57mm]{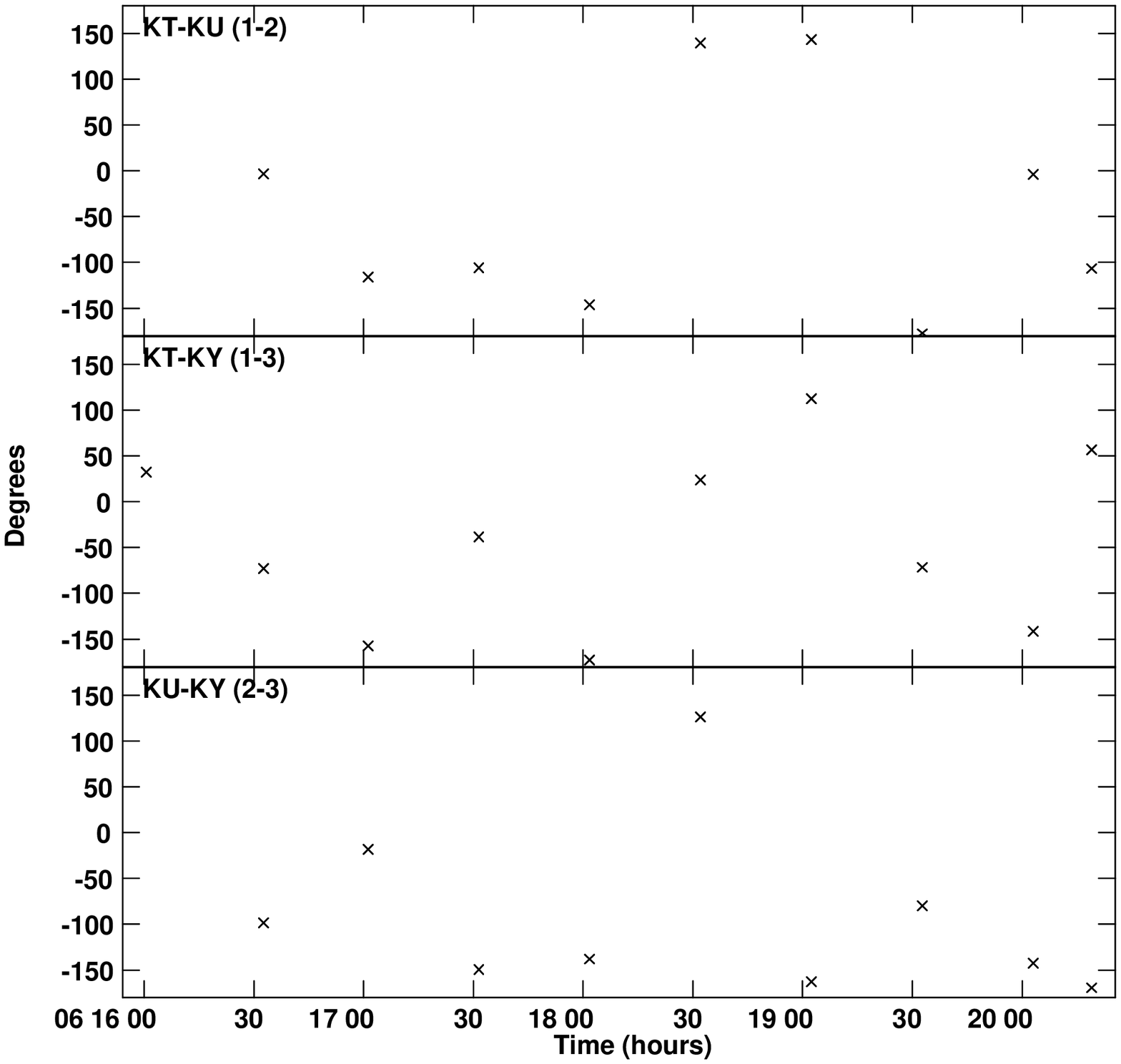}
\includegraphics[angle=-90,width=57mm]{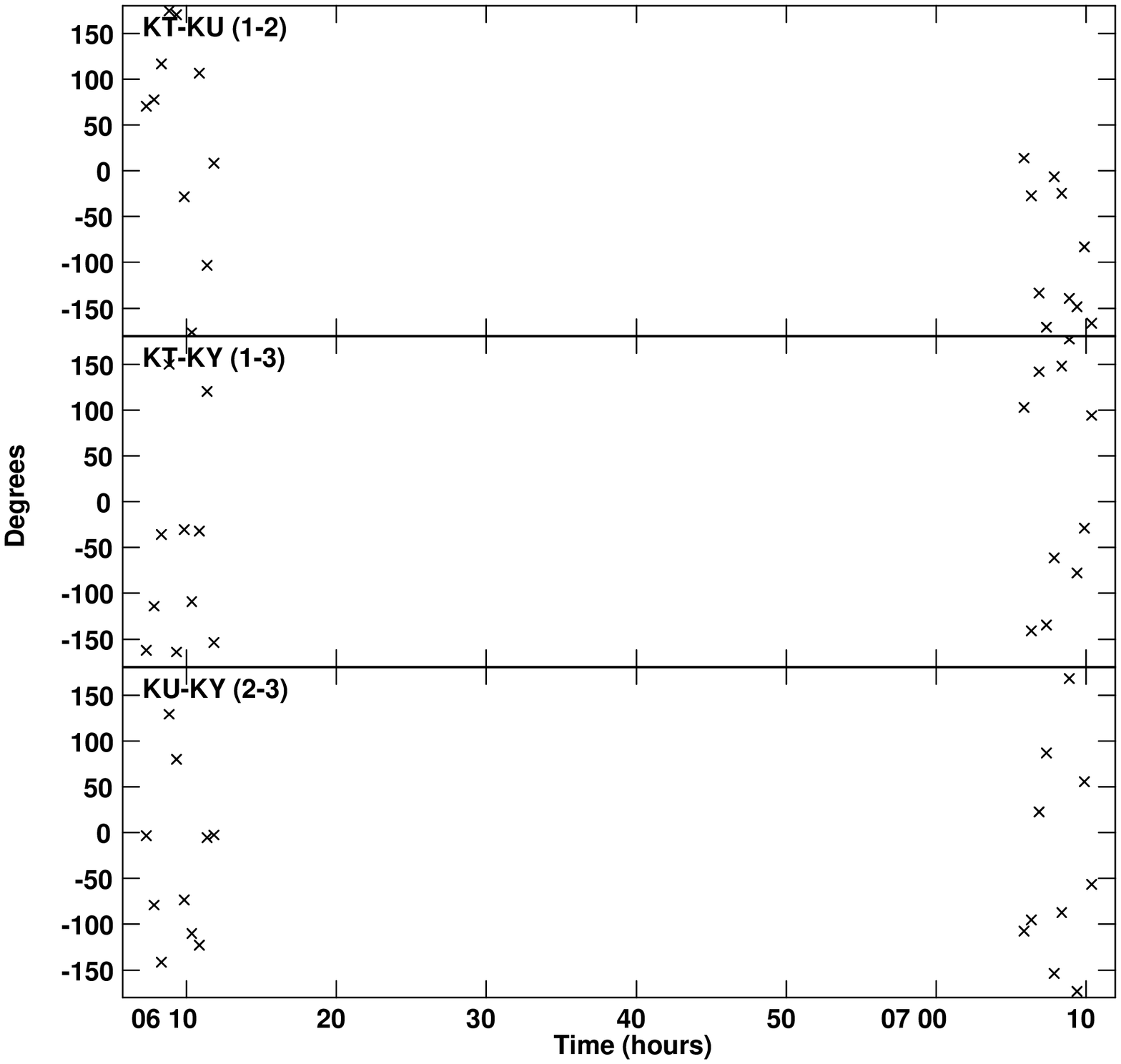}
\includegraphics[angle=-90,width=57mm]{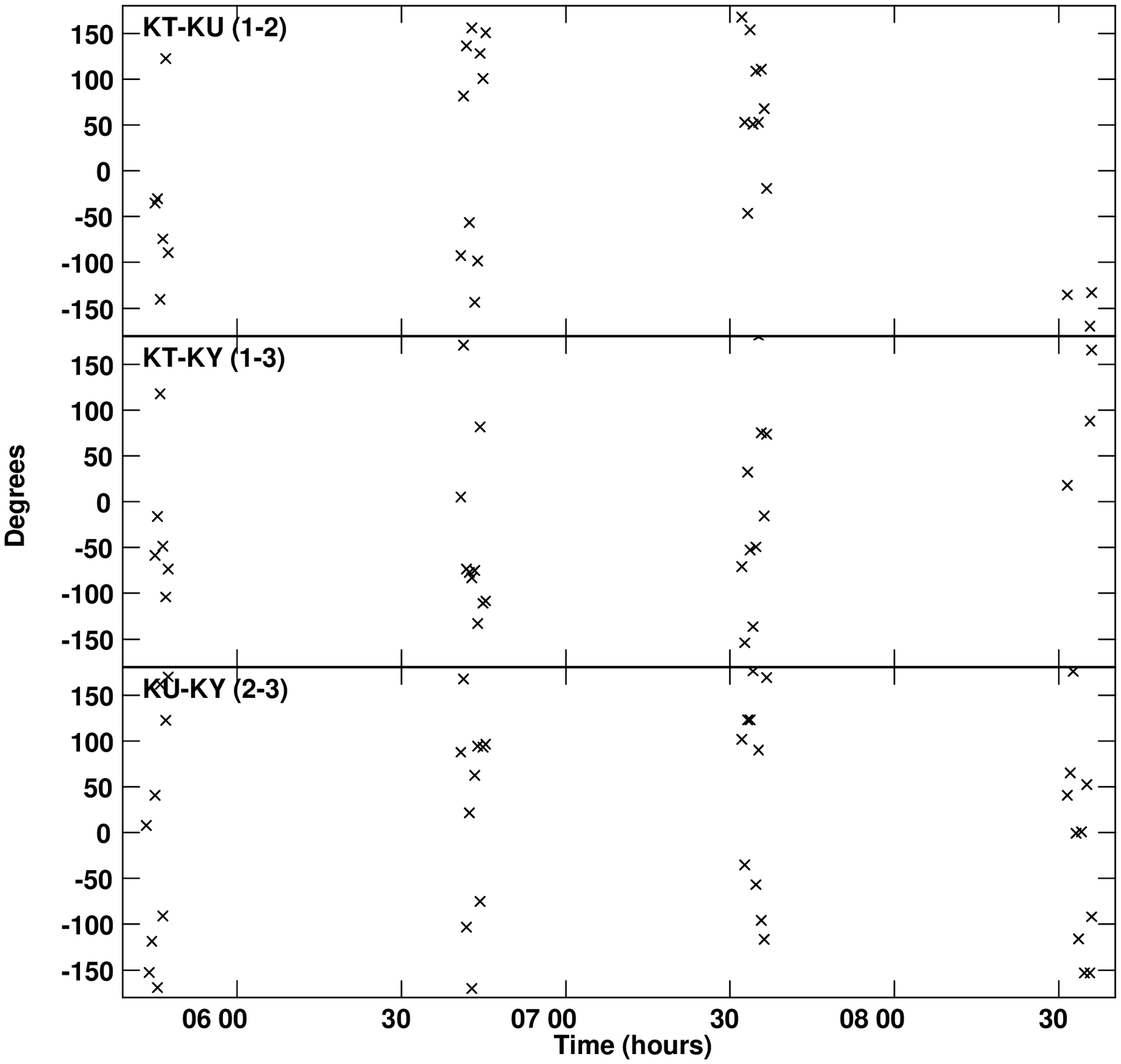}
\begin{tabular}{lll}
\hspace{0.8cm}(d) 0727--115 &\hspace{3.3cm} (e) 0735+178  &\hspace{3.3cm} (f) 0827+243\\
\end{tabular}
\includegraphics[angle=-90,width=57mm]{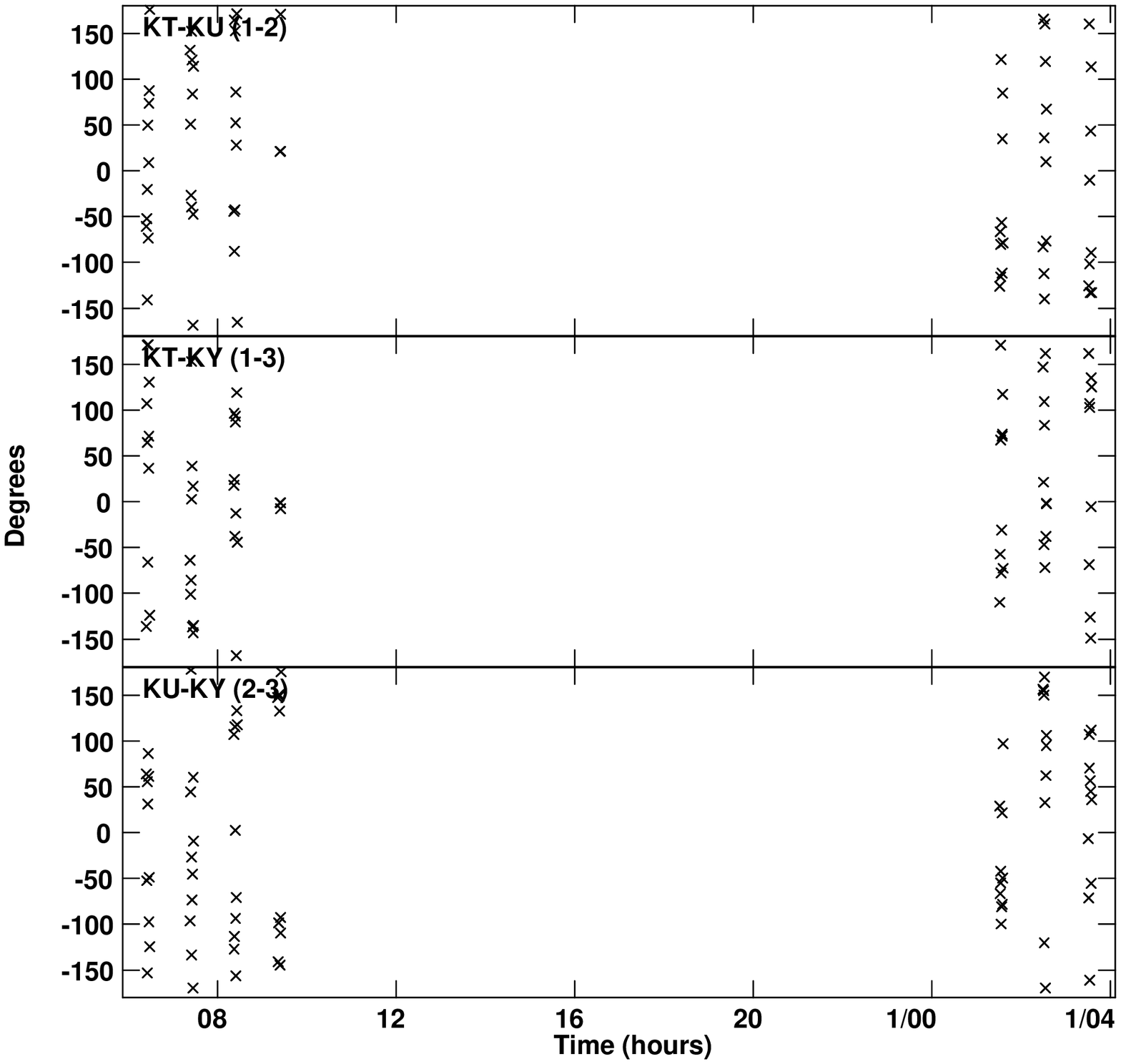}
\includegraphics[angle=-90,width=57mm]{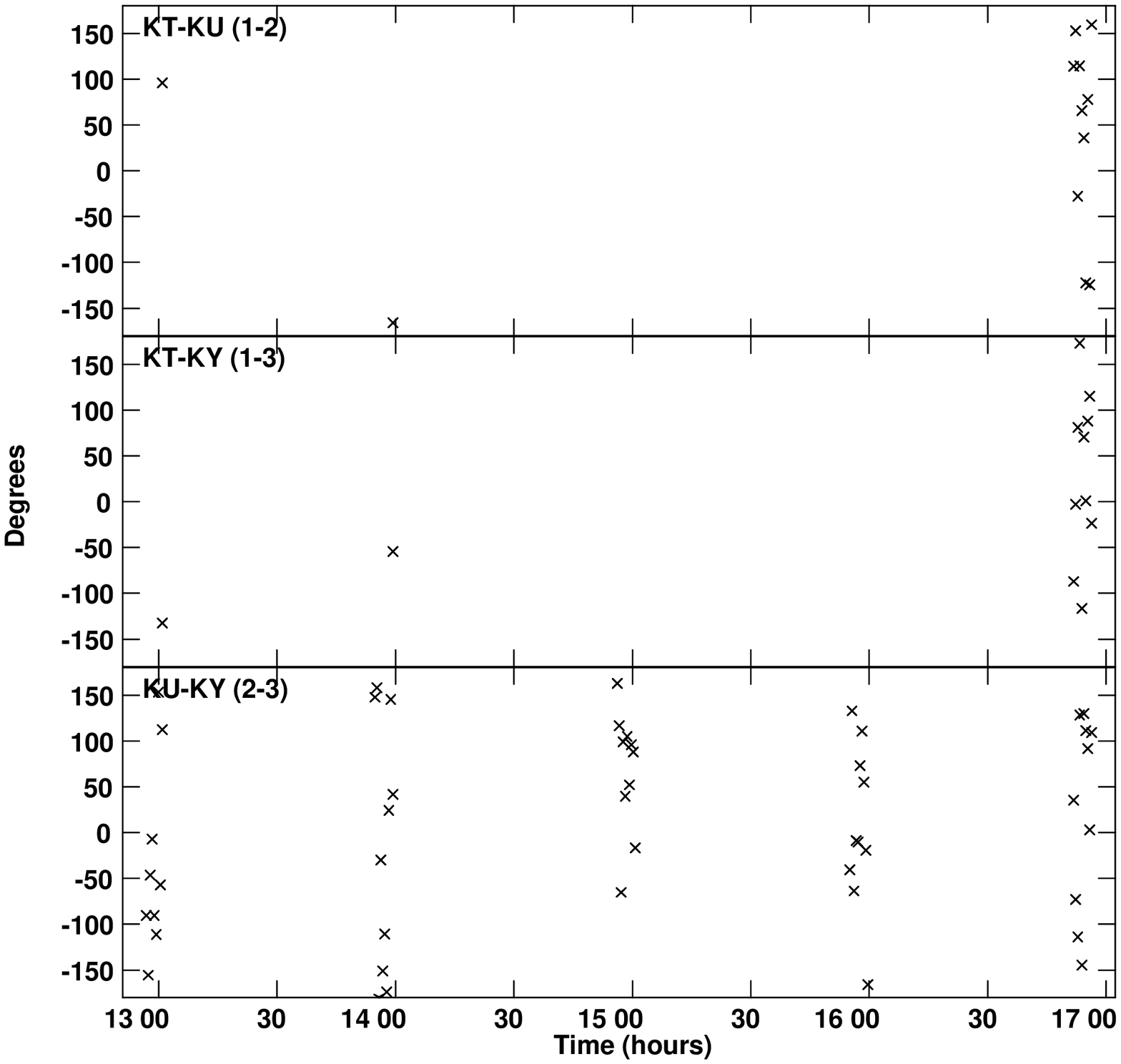}
\includegraphics[angle=-90,width=57mm]{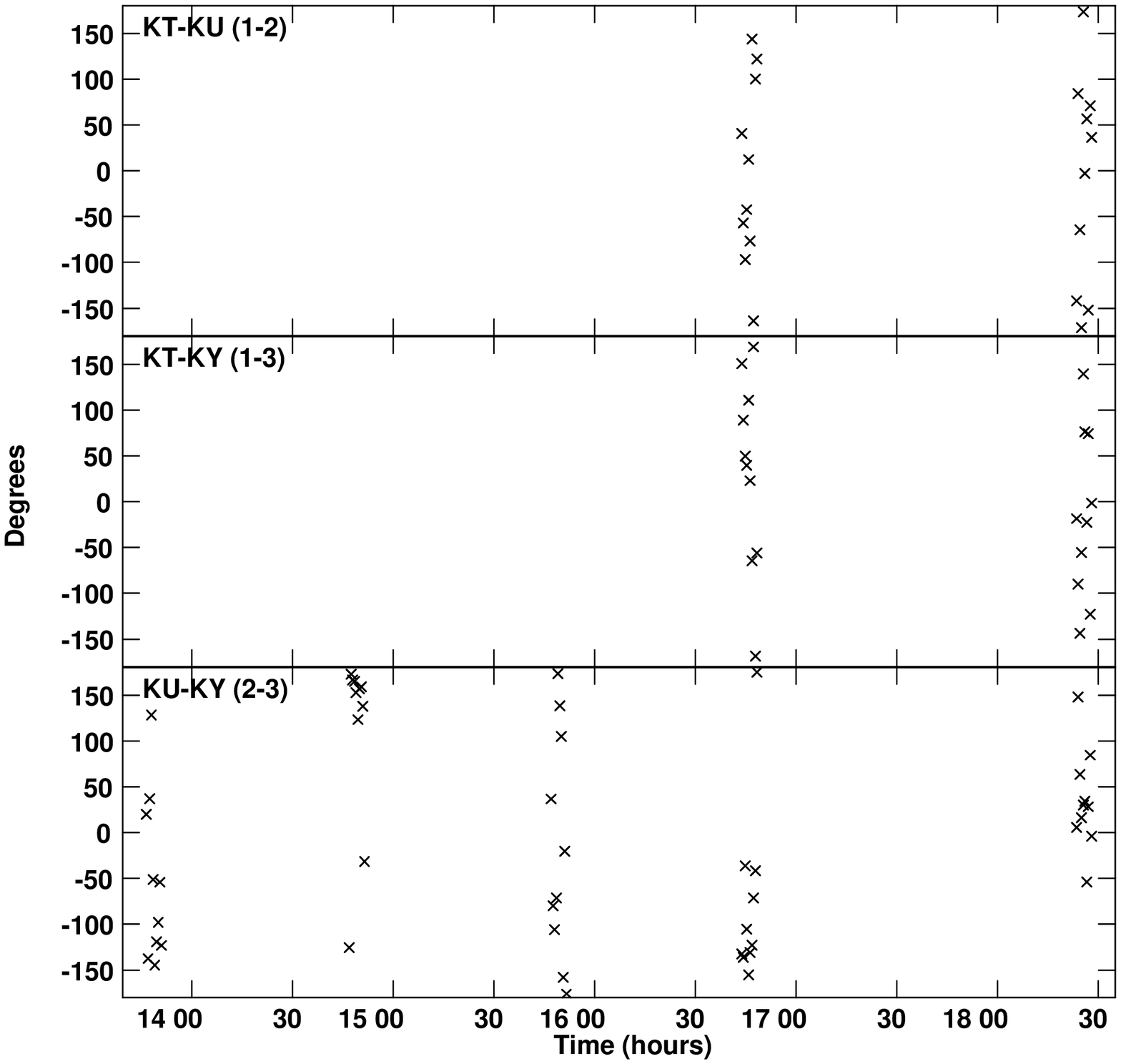}
\begin{tabular}{lll}
\hspace{0.8cm}(g) 0836+710 &\hspace{3.3cm} (h) 1510--089 &\hspace{3.3cm} (i) 1611+343\\
\end{tabular}
\includegraphics[angle=-90,width=57mm]{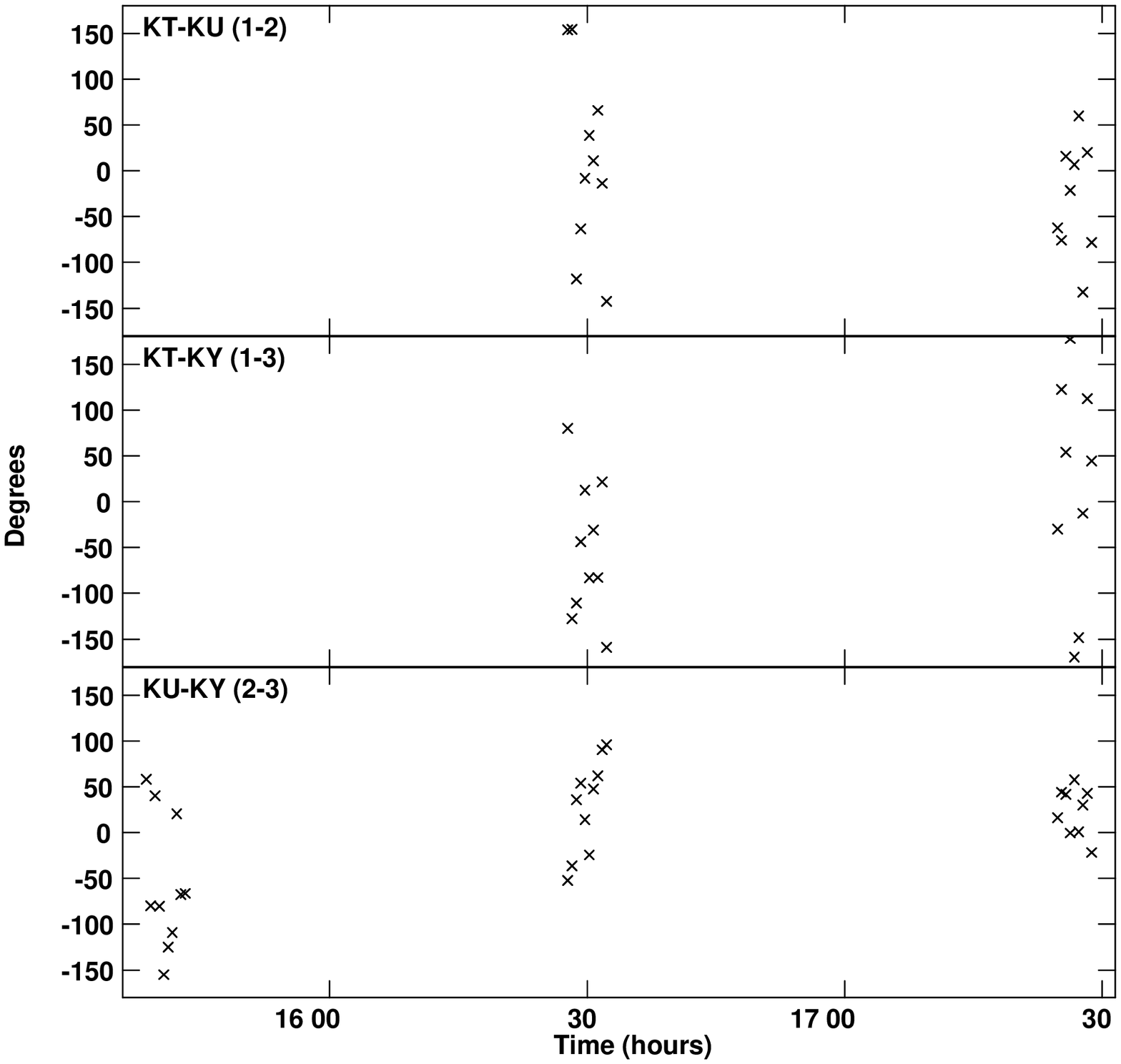}
\includegraphics[angle=-90,width=57mm]{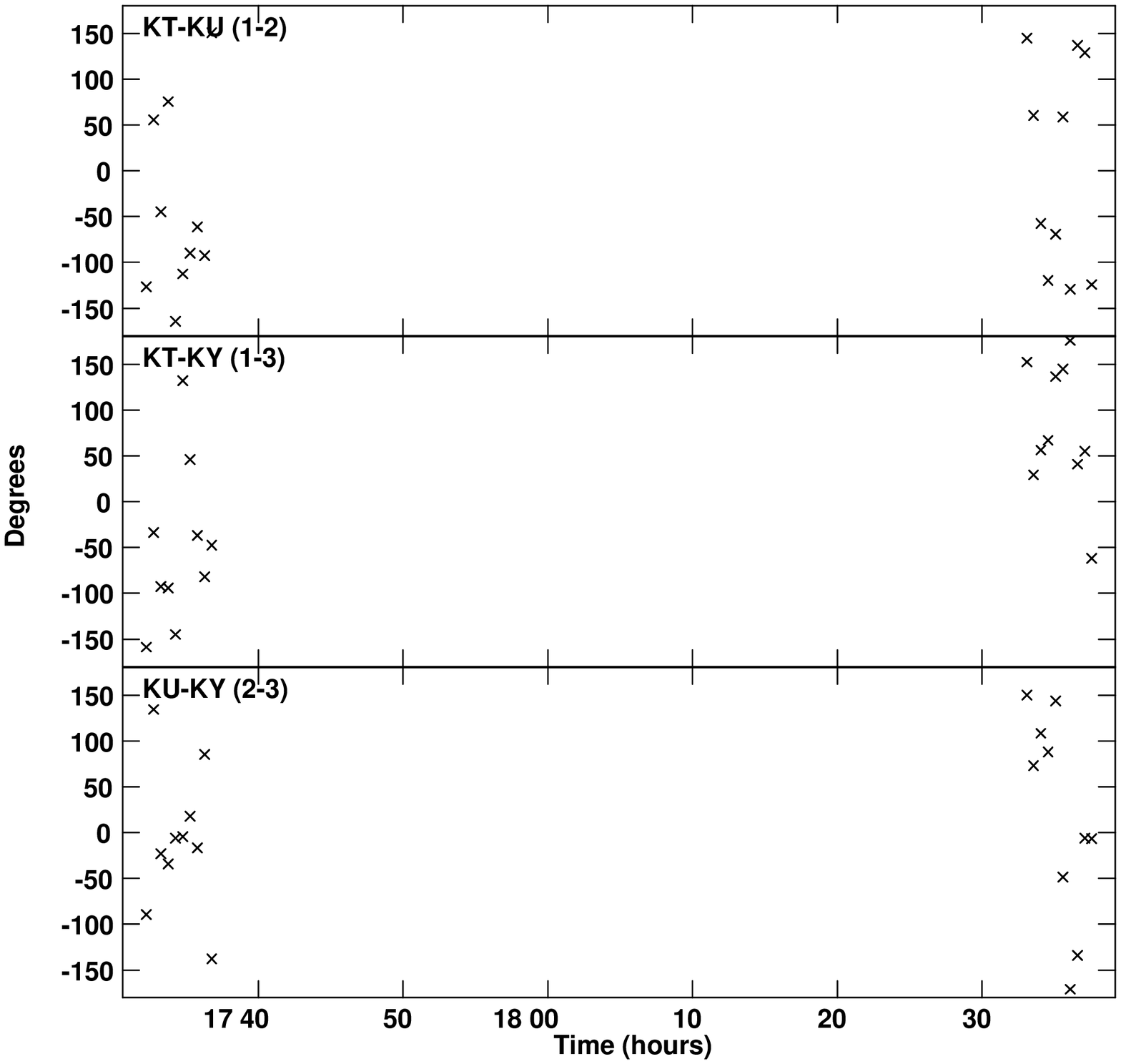}
\includegraphics[angle=-90,width=57mm]{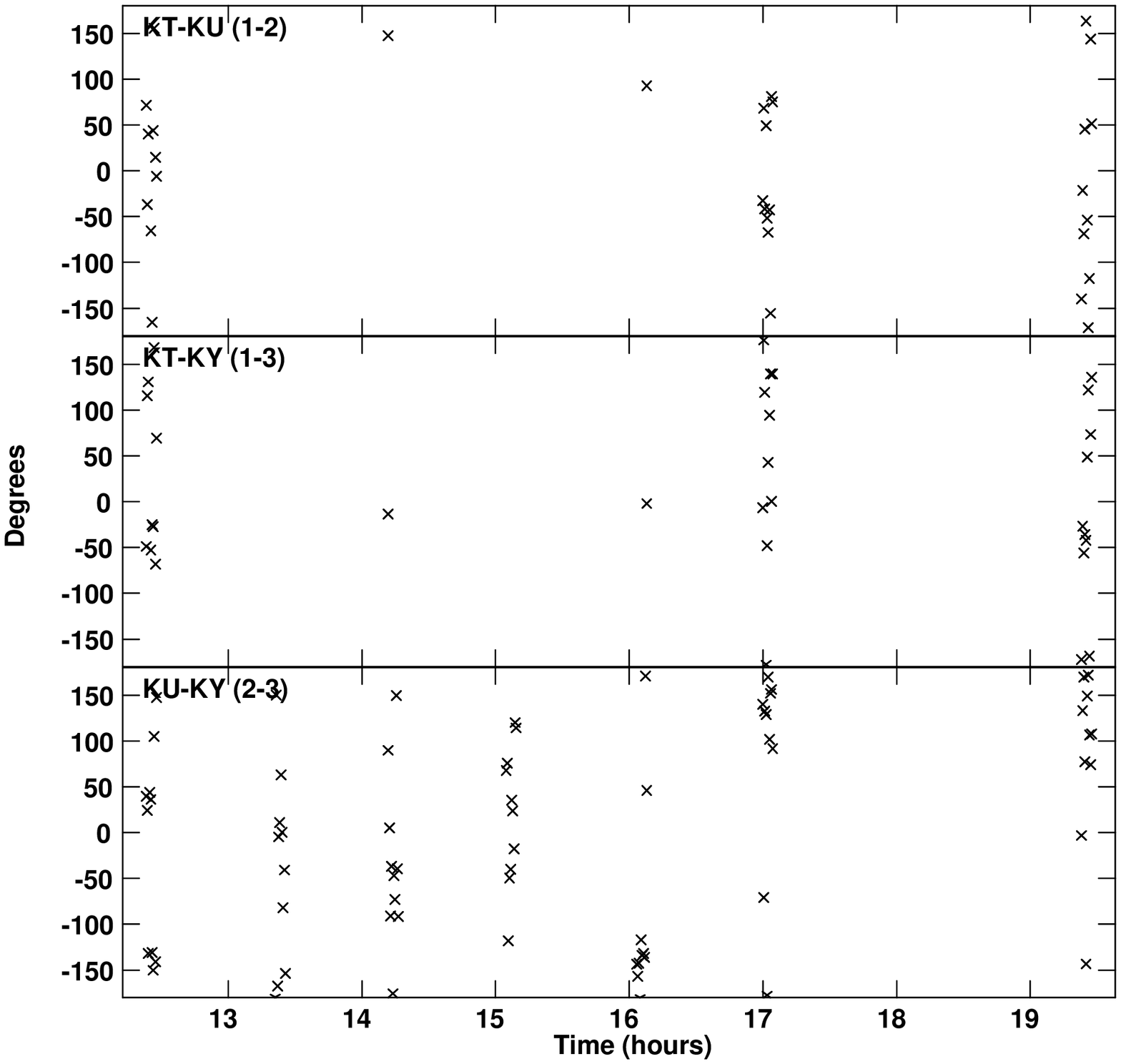}
%\begin{tabular*}{\textwidth}{c @{\extracolsep{\fill}} ccc}
\begin{tabular}{lll}
\hspace{0.8cm}(j) 1749+096 &\hspace{3.3cm} (k) 1921-293 &\hspace{3.3cm} (l) 3C345\\
\end{tabular}
\caption{FPT visibility phases for iMOGABA15 observations at 129~GHz. Each data set was calibrated using the scaled solutions from the analysis  of the same source at 22~GHz. Plots show temporal average of 30 seconds.\label{iM15-FPTphase22-129}}
%\vspace{5mm} %% add extra space ONLY when figures/tables are "colliding"!
\end{figure*}

\begin{figure*}[h]
\centering
\includegraphics[angle=-90,width=57mm]{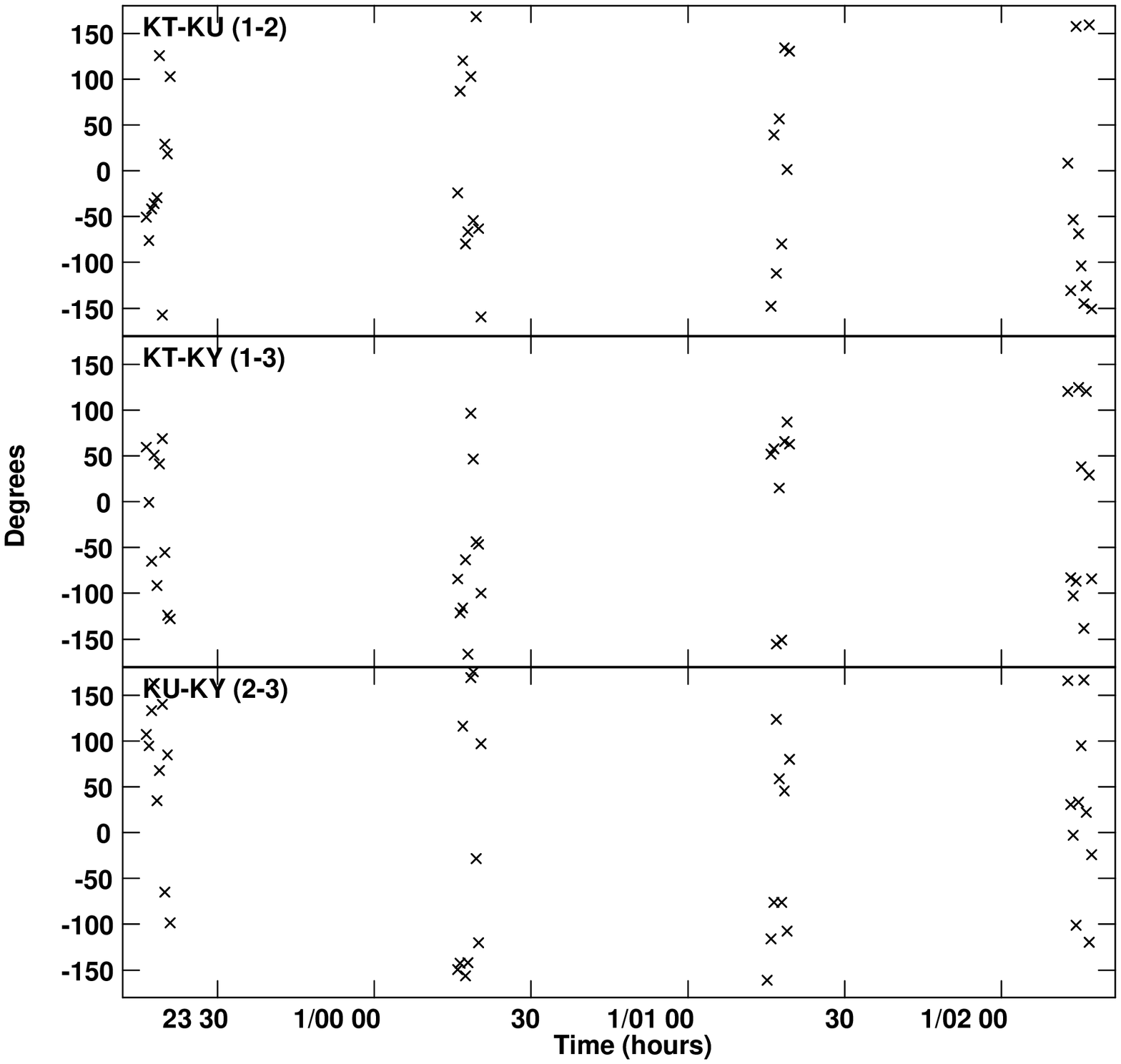}
\includegraphics[angle=-90,width=57mm]{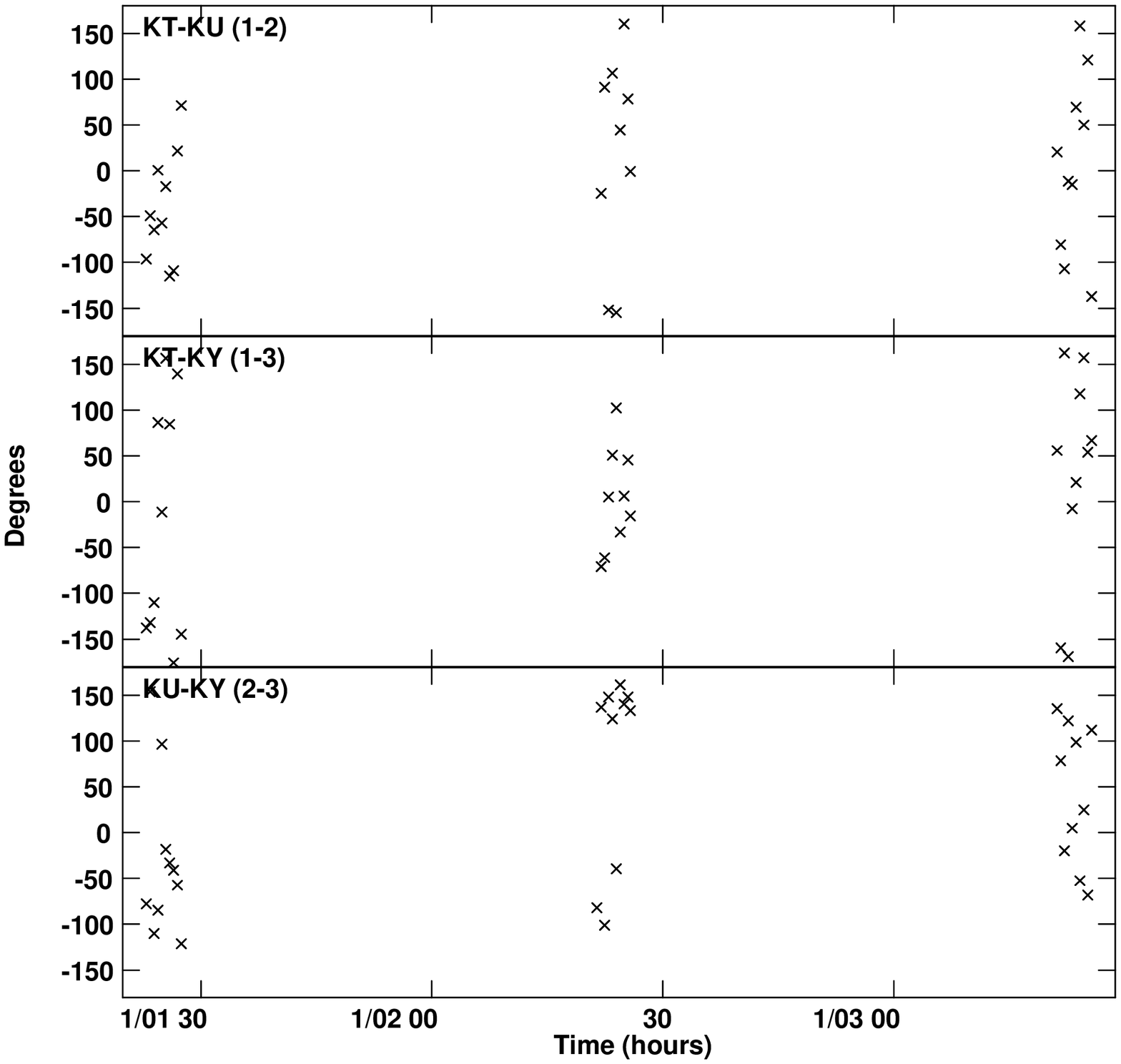}
\includegraphics[angle=-90,width=57mm]{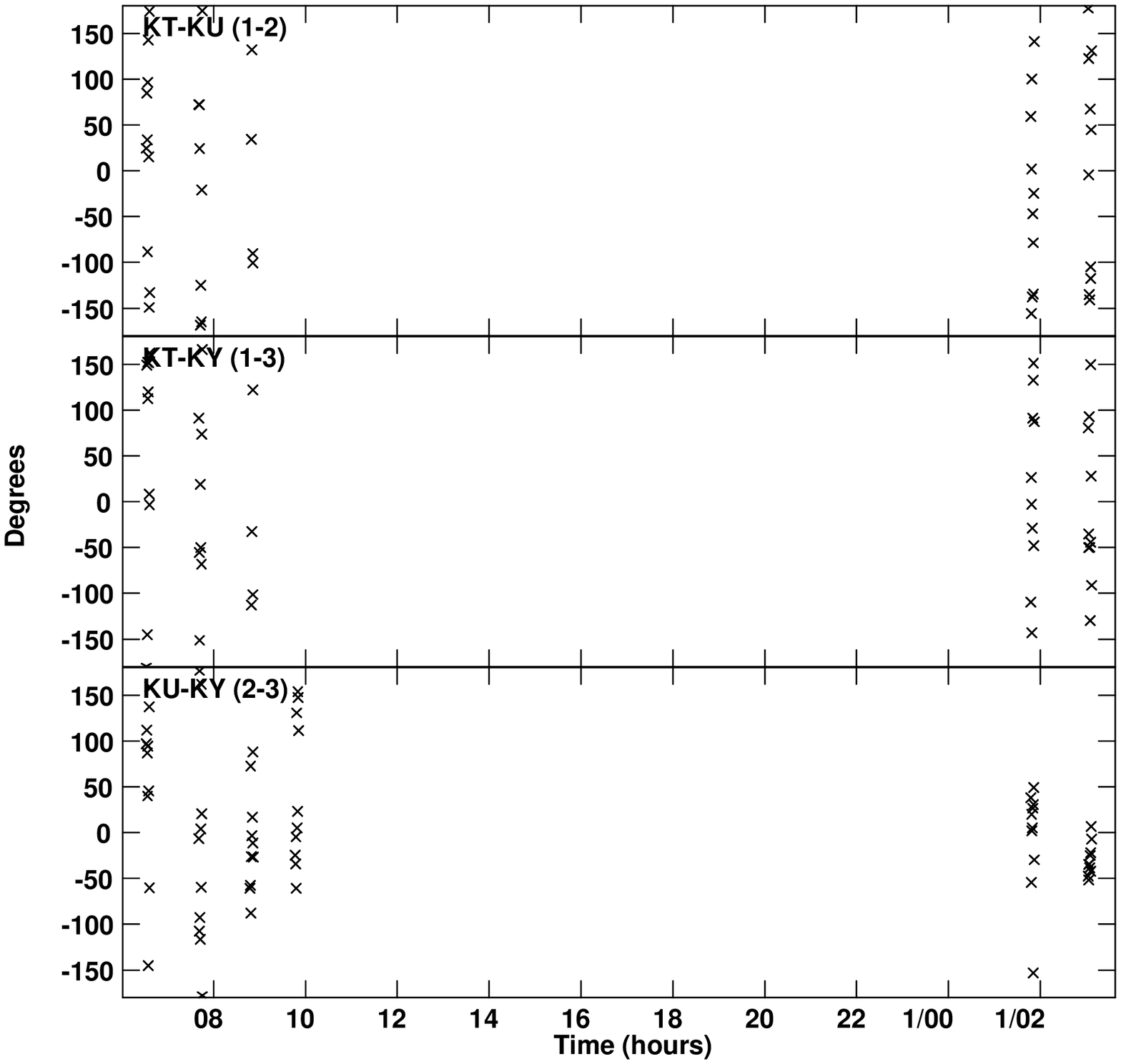}
\begin{tabular}{lll}
\hspace{0.8cm}(a) 0235+164 &\hspace{3.3cm} (b) 0420--014 &\hspace{3.3cm} (c) 0716+714\\
\end{tabular}
\includegraphics[angle=-90,width=57mm]{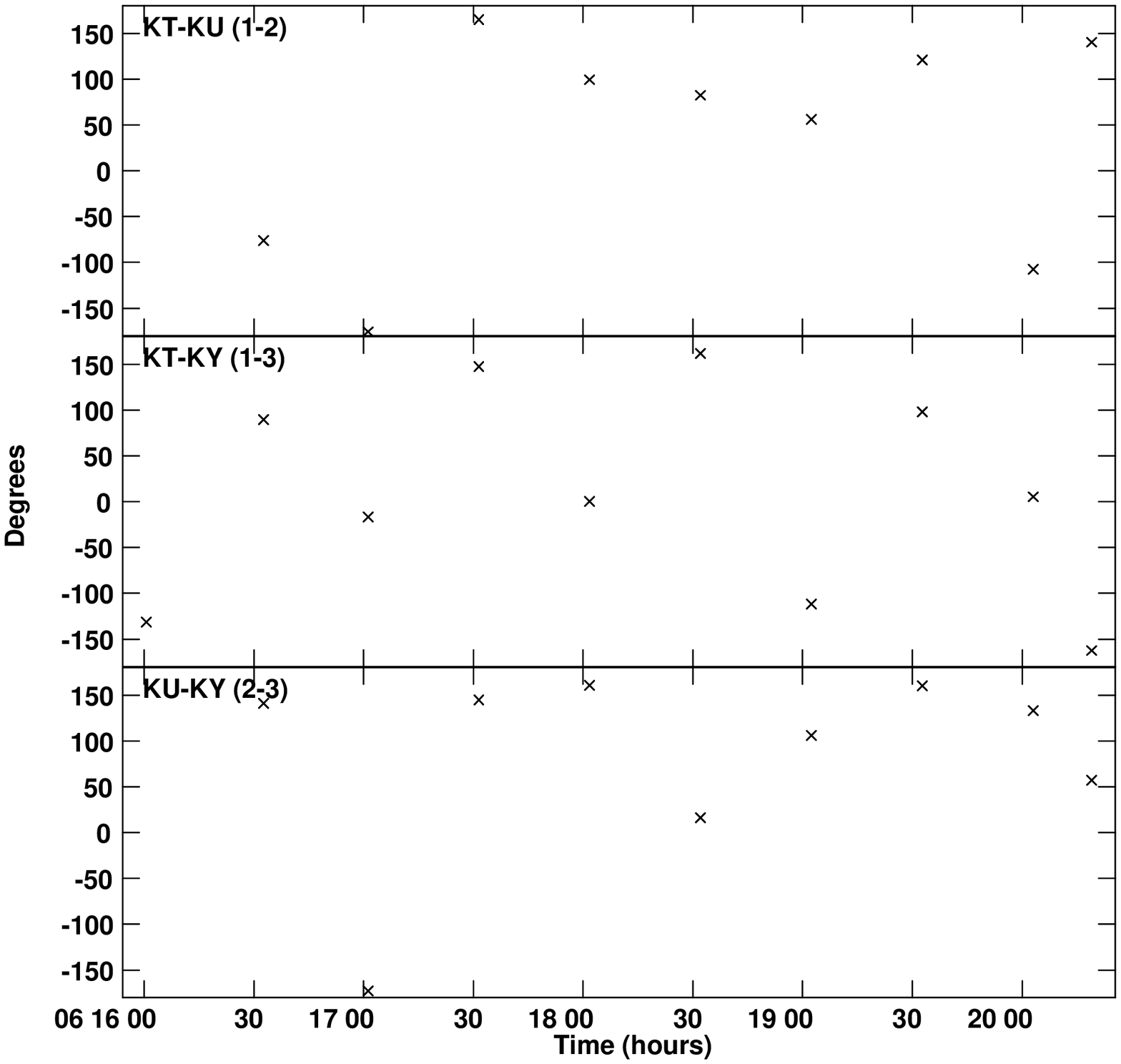}
\includegraphics[angle=-90,width=57mm]{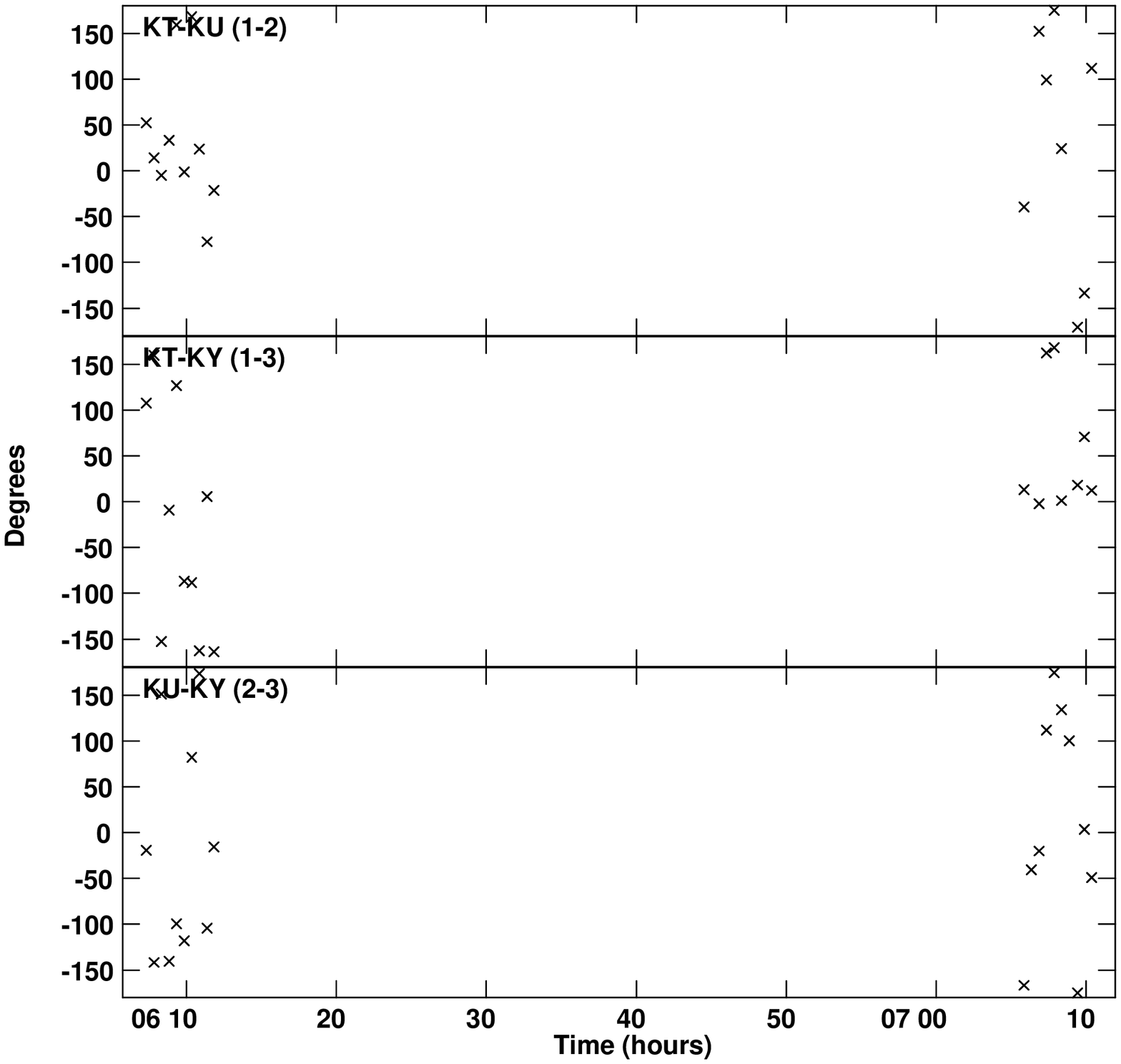}
\includegraphics[angle=-90,width=57mm]{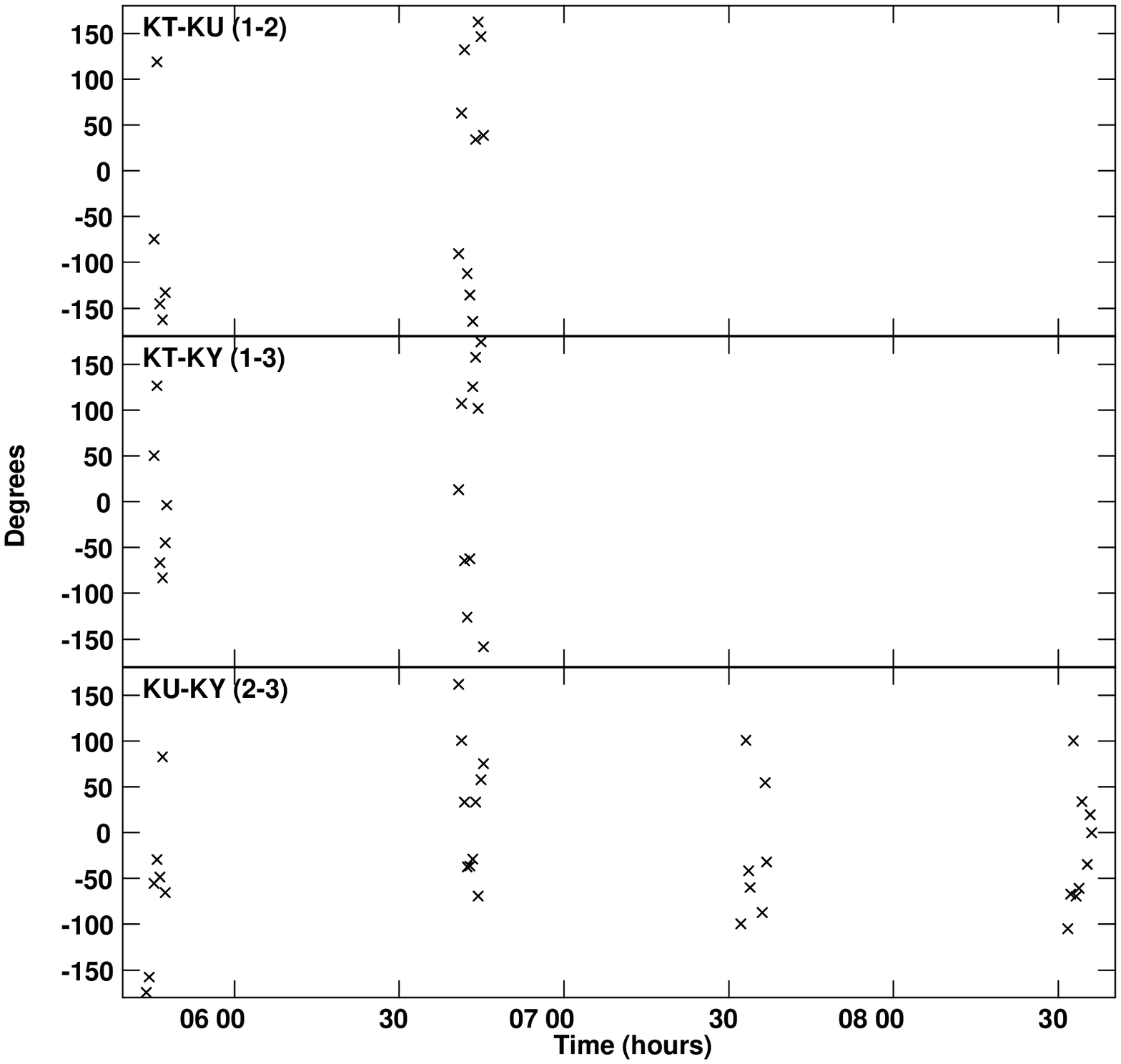}
\begin{tabular}{lll}
\hspace{0.8cm}(d) 0727--115 &\hspace{3.3cm} (e) 0735+178  &\hspace{3.3cm} (f) 0827+243\\
\end{tabular}
\includegraphics[angle=-90,width=57mm]{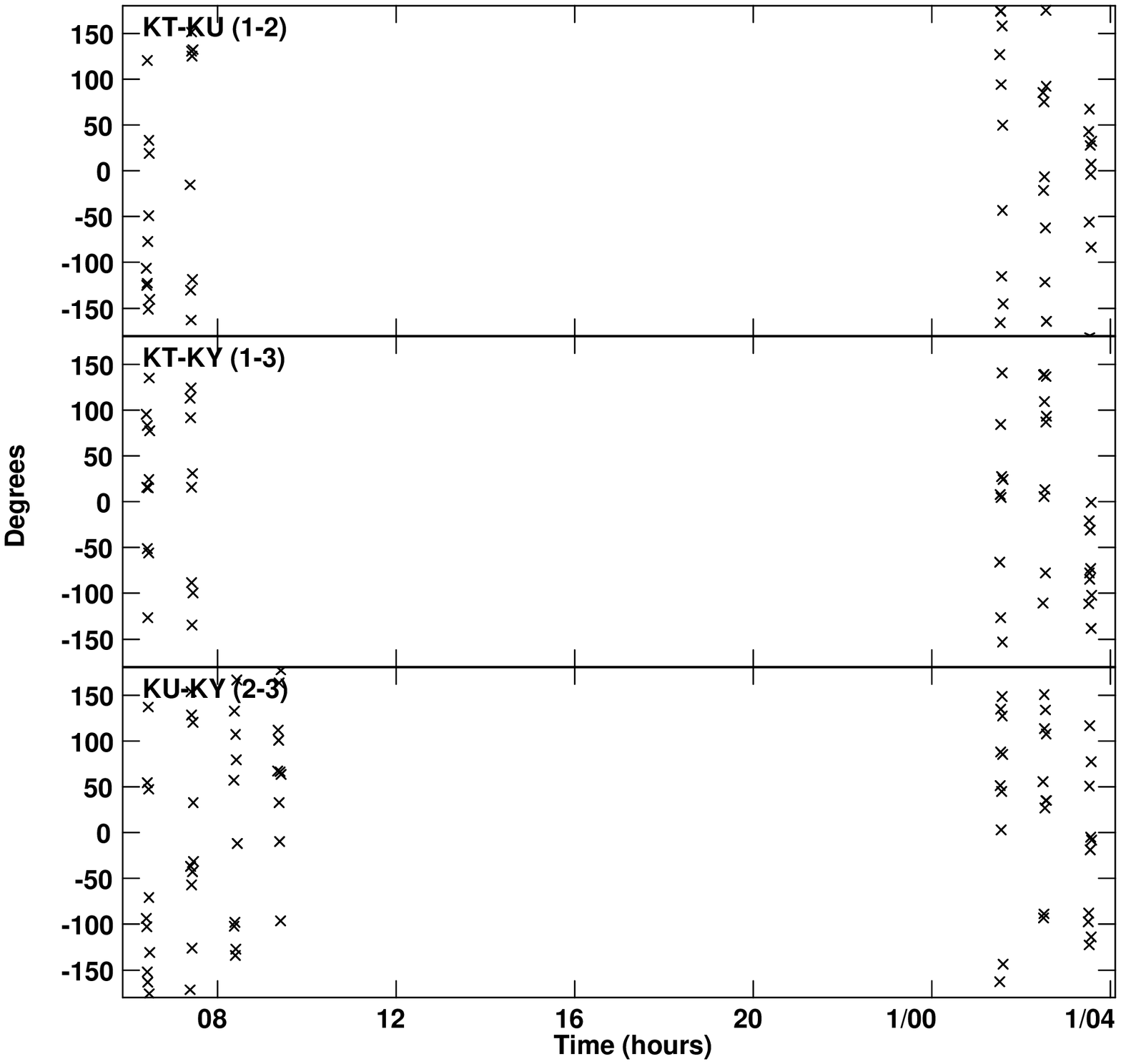}
\includegraphics[angle=-90,width=57mm]{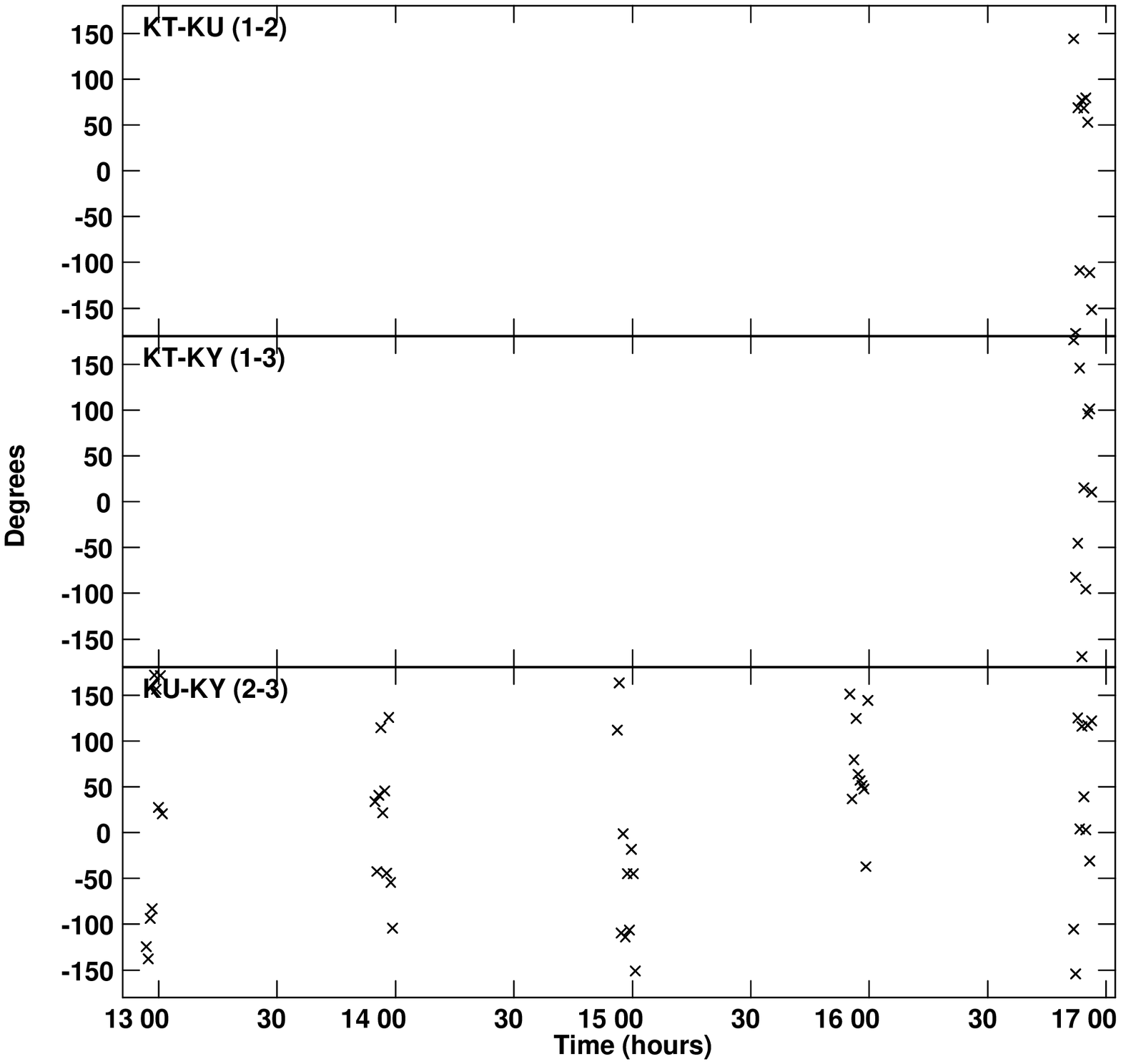}
\includegraphics[angle=-90,width=57mm]{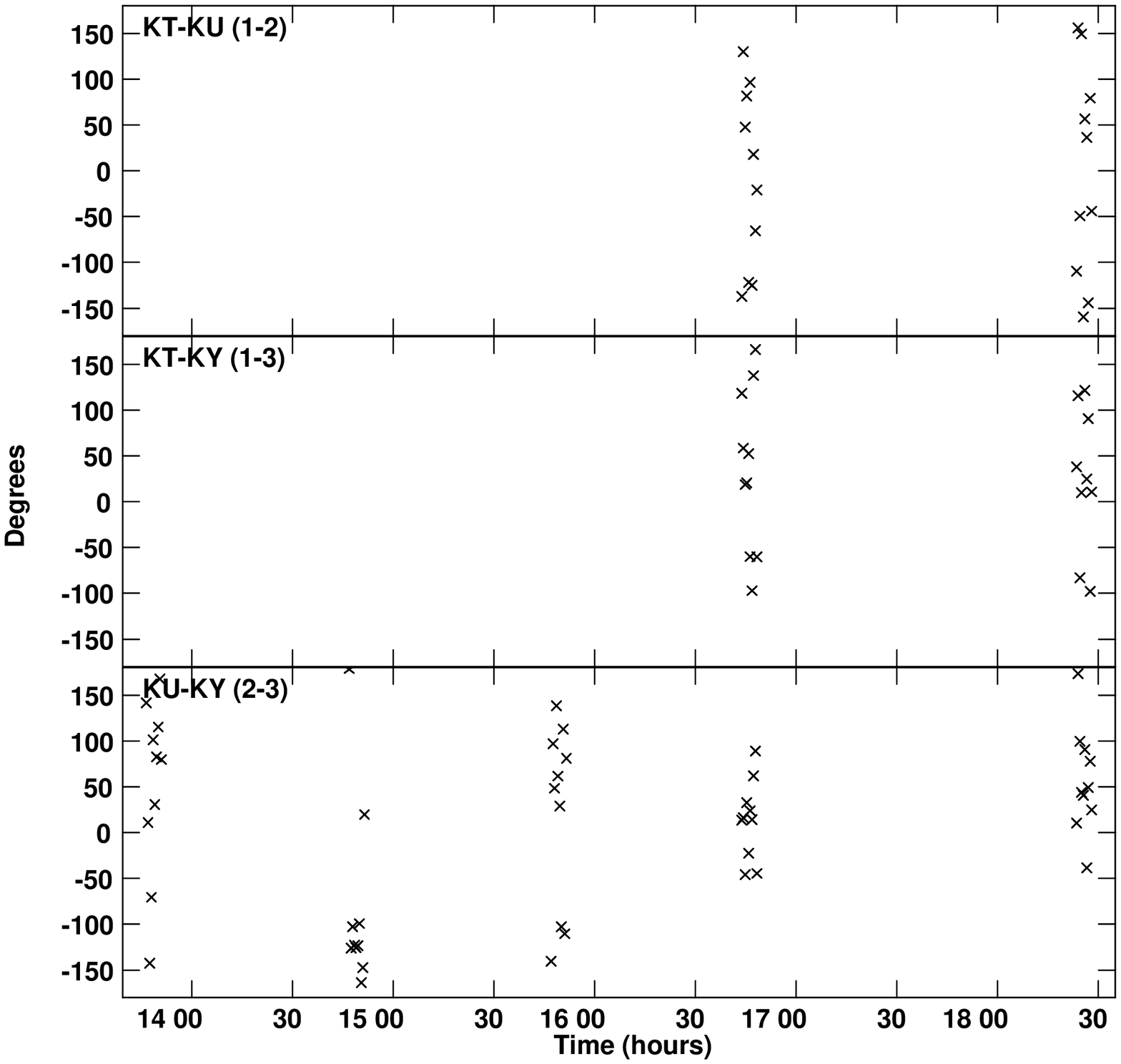}
\begin{tabular}{lll}
\hspace{0.8cm}(g) 0836+710 &\hspace{3.3cm} (h) 1510--089 &\hspace{3.3cm} (i) 1611+343\\
\end{tabular}
\includegraphics[angle=-90,width=57mm]{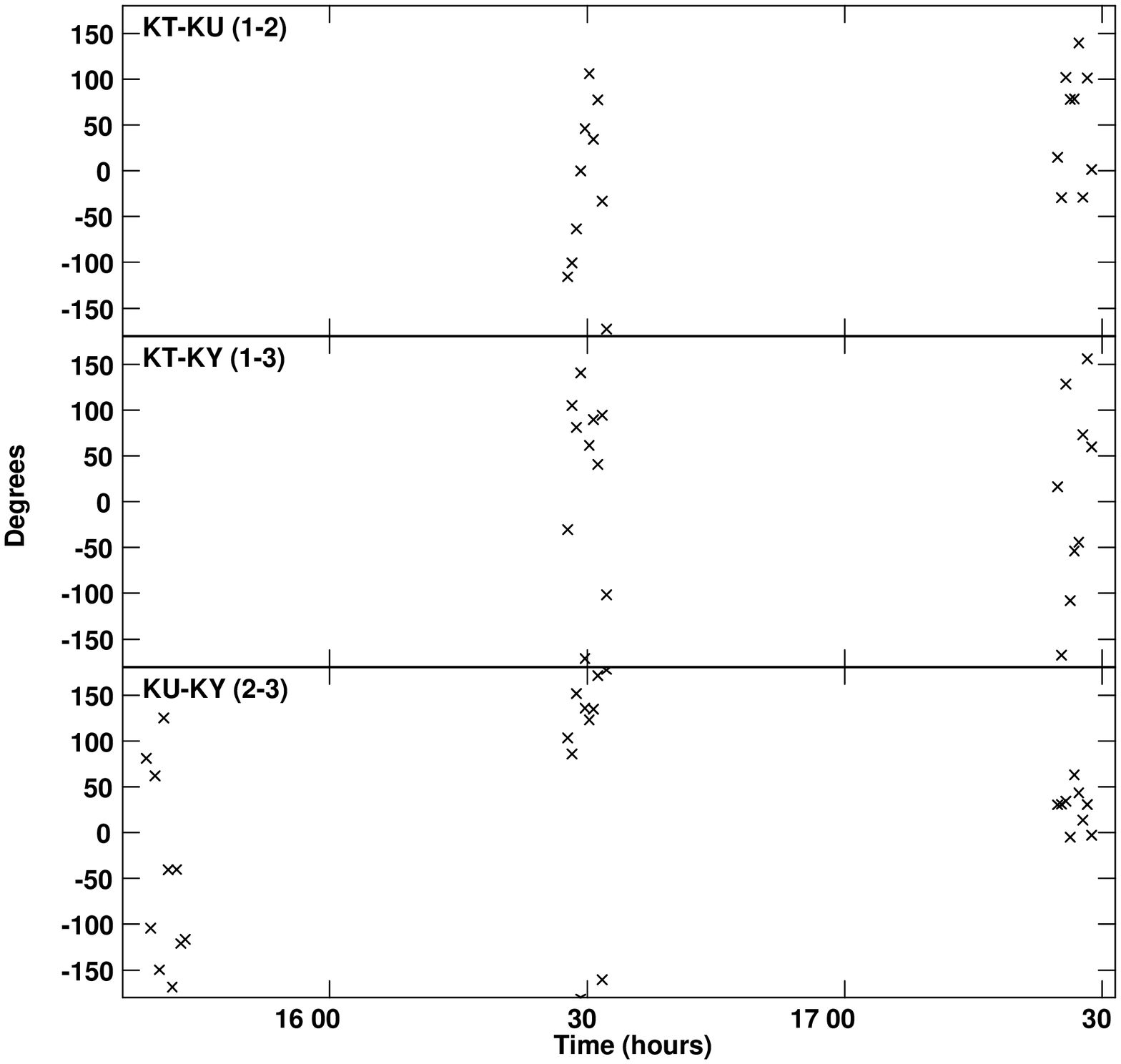}
\includegraphics[angle=-90,width=57mm]{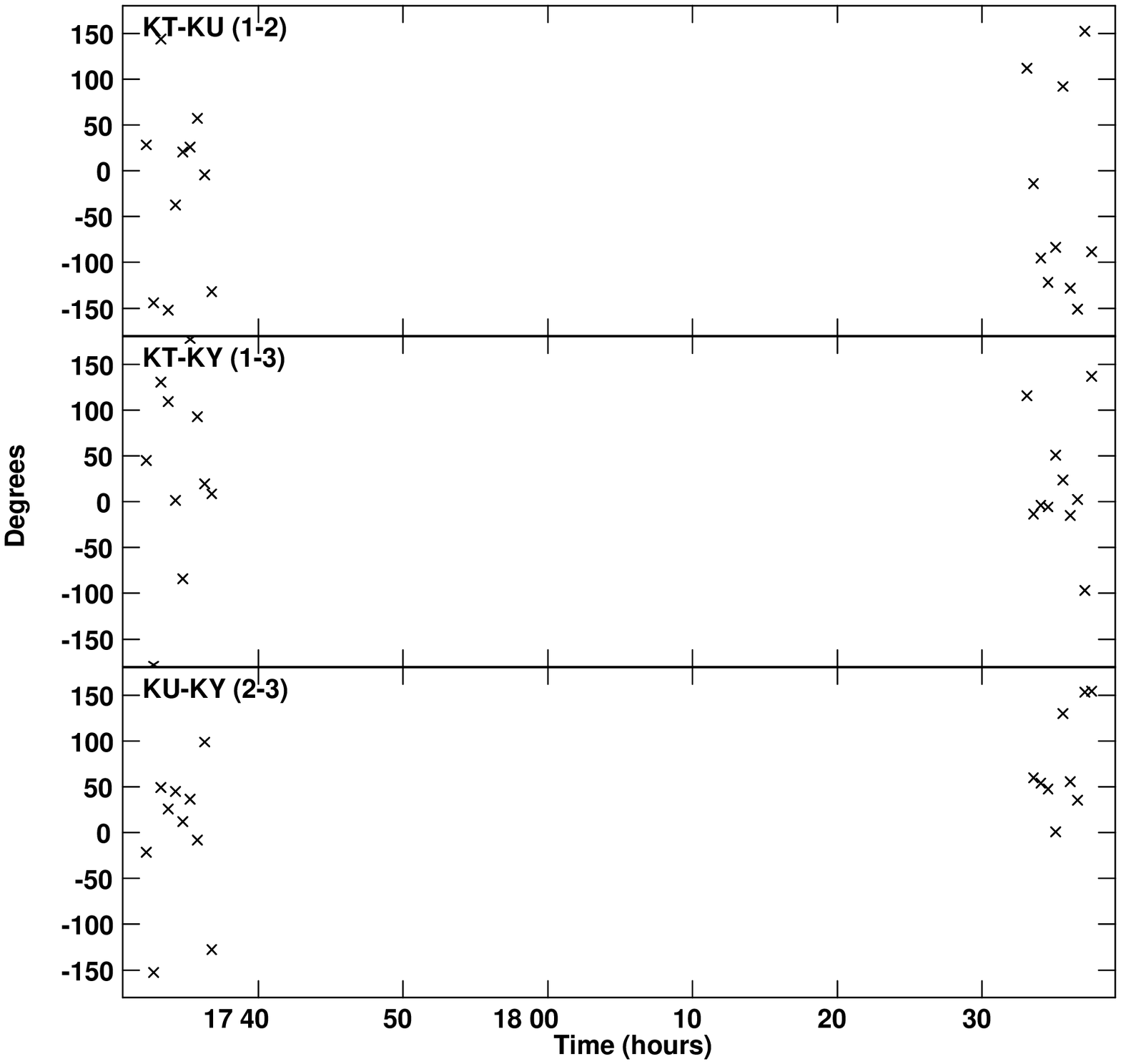}
\includegraphics[angle=-90,width=57mm]{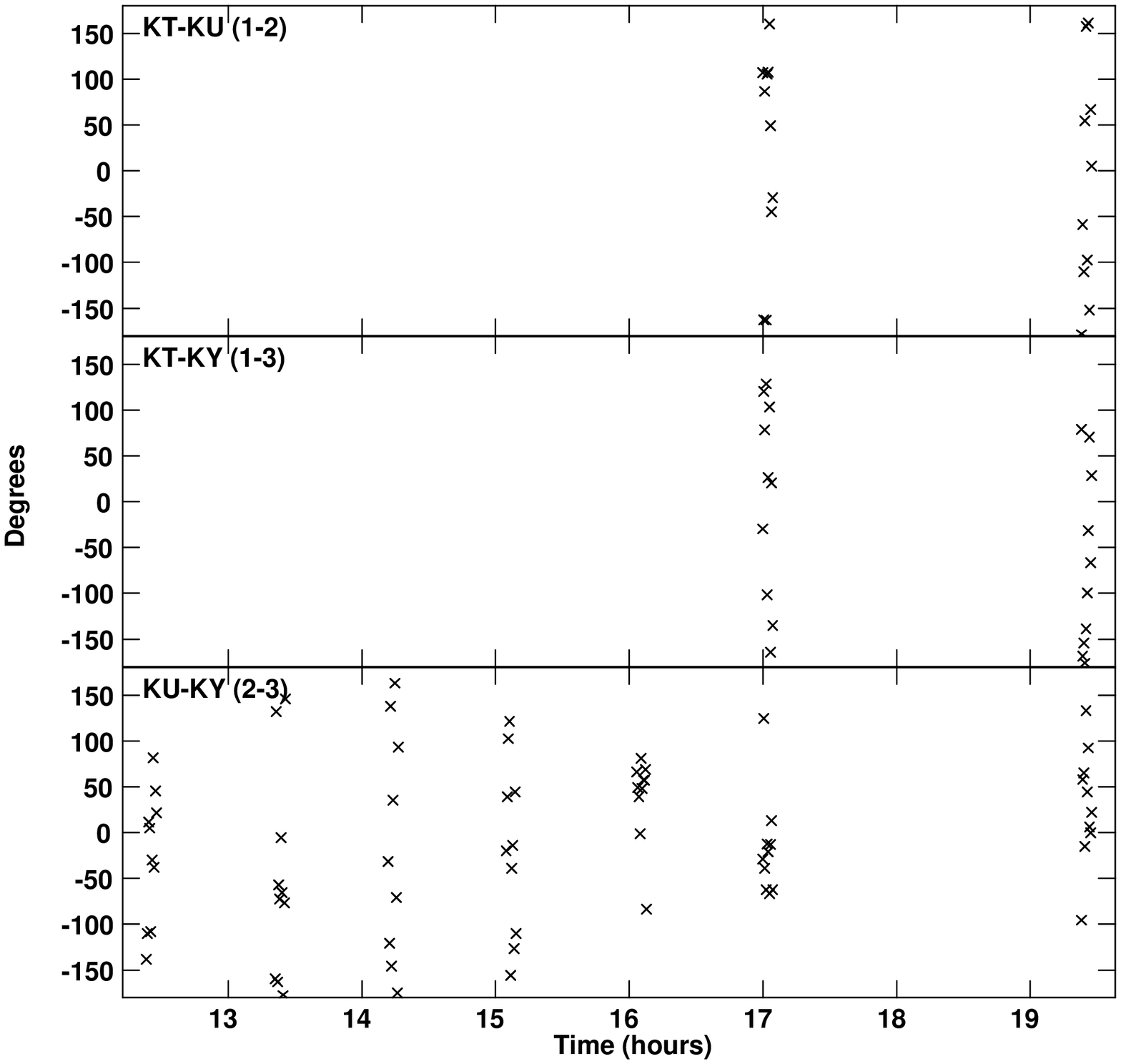}
%\begin{tabular*}{\textwidth}{c @{\extracolsep{\fill}} ccc}
\begin{tabular}{lll}
\hspace{0.8cm}(j) 1749+096 &\hspace{3.3cm} (k) 1921-293 &\hspace{3.3cm} (l) 3C345\\
\end{tabular}
\caption{Same as Figure \ref{iM15-FPTphase22-129}, but using the scaled solutions from the analysis of the same source at 43 GHz.\label{iM15-FPTphase43-129}}
%\vspace{5mm} %% add extra space ONLY when figures/tables are "colliding"!
\end{figure*}

We performed the FPT on all the sources discussed here from 22 and 43~GHz to 129~GHz. Results for the calibrated phases are shown in Figures \ref{iM15-FPTphase22-129} and \ref{iM15-FPTphase43-129}. Simple inspection reveals large phase scatter in most of the sources, with the exception of the bright source 3C454.3, possibly 3C84 and CTA102. As before, the scatter is lower for the Ulsan--Yonsei baseline. Comparison of the results transferred from 22 and 43~GHz reveal no significant differences. 

In practice, the large phase scatter results in the re-fringe fitting failing on virtually all sources. Reliable solutions for all baselines can be found only for 3C454.3 and possibly BL Lac. Based on Table \ref{theoreticalflux129b}, we consider that these detections may also be limited by the Tamna flux offset, leaving an actual SNR$^{129}_{300s}$ lower than the estimated one. We do not perform imaging as the same limitations discussed for 86~GHz also appy here.%Due to the same limitations discussed for 86~GHz, which are boosted at 129~GHz, we did not perform imaging here.

\begin{table}[t!]
\caption{iMOGABA15 Expected Detection Limits at 129~GHz\label{theoreticalflux129b}}
\centering
\begin{tabular}{ccccc}
\toprule
Source   & SEFD (Jy) & $S^{129}$ (mJy)  & SNR$^{129}_{30s}$ & SNR$^{129}_{300s}$ \\
\midrule
%Source 		& 	SEFD129 	& 	ExpectedF & 	Sigma 30 	& 	Sigma 300 	\\
0235+164 	& 	16000 	& 	490	 	& 	2 	& 	5 	\\
0420--014 	& 	18700 	& 	440 		& 	1 	& 	4 	\\
0716+714 	& 	21300 	& 	920 		& 	2 	& 	7 	\\

0727--115 	& 	26700 	& 	440 		& 	1 	& 	3 	\\
0735+178 	& 	16000 	& 	410 		& 	1 	& 	4 	\\
0827+243 	& 	21300 	& 	320 		& 	1 	& 	3 	\\

0836+710 	& 	9300 	& 	310 		& 	2 	& 	6 	\\
1510--089 	& 	26700 	& 	860 		& 	2 	& 	6 	\\
1611+343		& 	42700 	& 	650 		& 	1 	& 	3 	\\

1749+096 	& 	48000 	& 	2390 	& 	3 	& 	9 	\\
1921-293 		& 	80100 	& 	2020 	& 	1 	& 	4 	\\
3C345 		& 	45400 	& 	1200 	& 	1 	& 	5 	\\

3C454.3 		& 	40000 	& 	3500 	& 	5 	& 	15 	\\
3C84 		& 	50700 	& 	750 		& 	1 	& 	3 	\\
4C28.07 		& 	50700 	& 	630 		& 	1 	& 	2 	\\

4C38.41 		& 	50700 	& 	3030 	& 	3 	& 	10 	\\
4C39.25 		& 	42700 	& 	450 		& 	1 	& 	2 	\\
BL Lac 		& 	37400 	& 	1760 	& 	3 	& 	8 	\\

CTA102 		& 	42700 	& 	630 		& 	1 	& 	3 	\\
NRAO530 	& 	61400 	& 	880 		& 	1 	& 	2 	\\
OJ287 		& 	16000 	& 	2800 	& 	10 	& 	30 	\\
\bottomrule
\end{tabular}
\end{table}

\begin{figure*}[h]
\centering
\includegraphics[angle=-90,width=57mm]{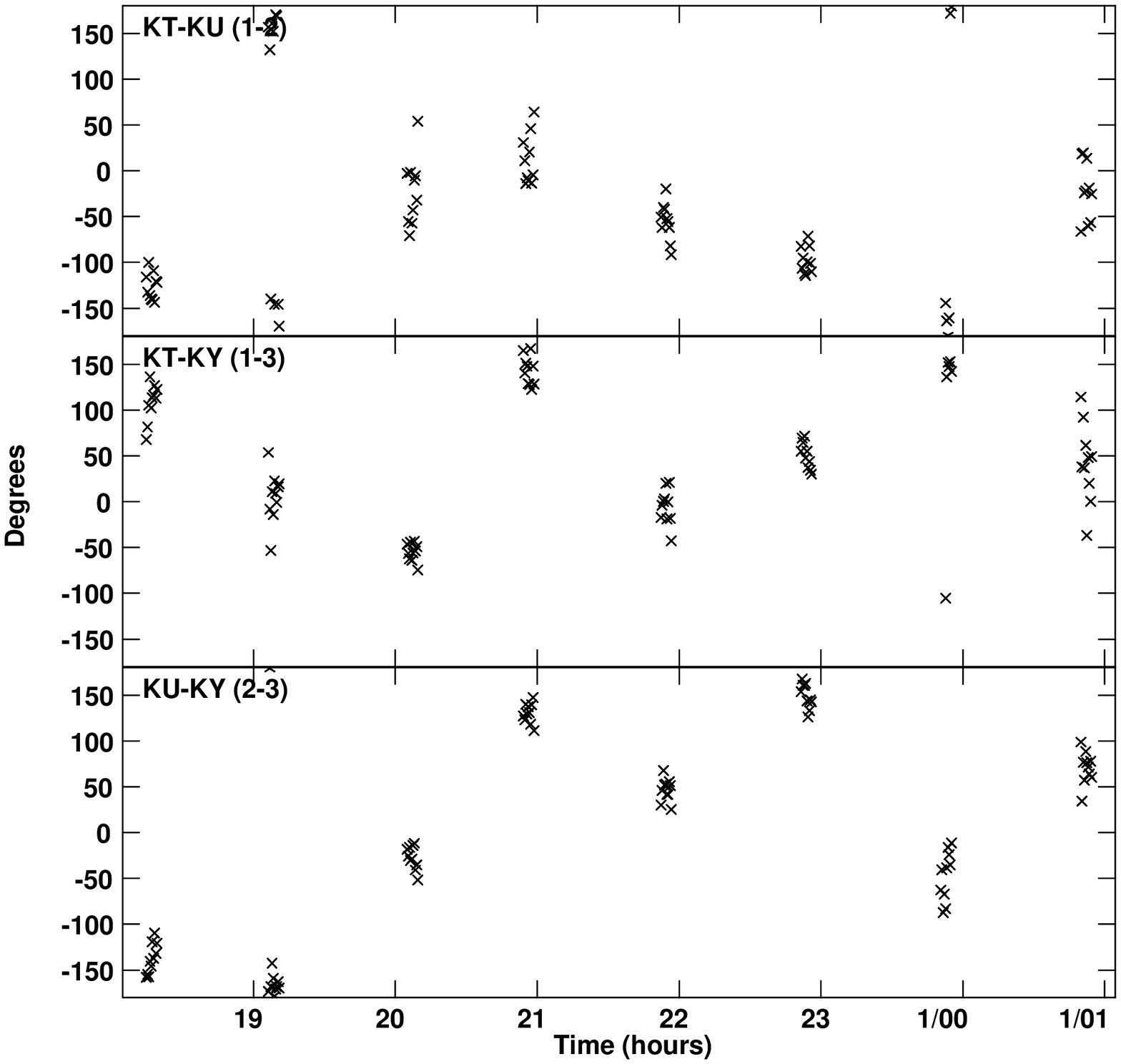}
\includegraphics[angle=-90,width=57mm]{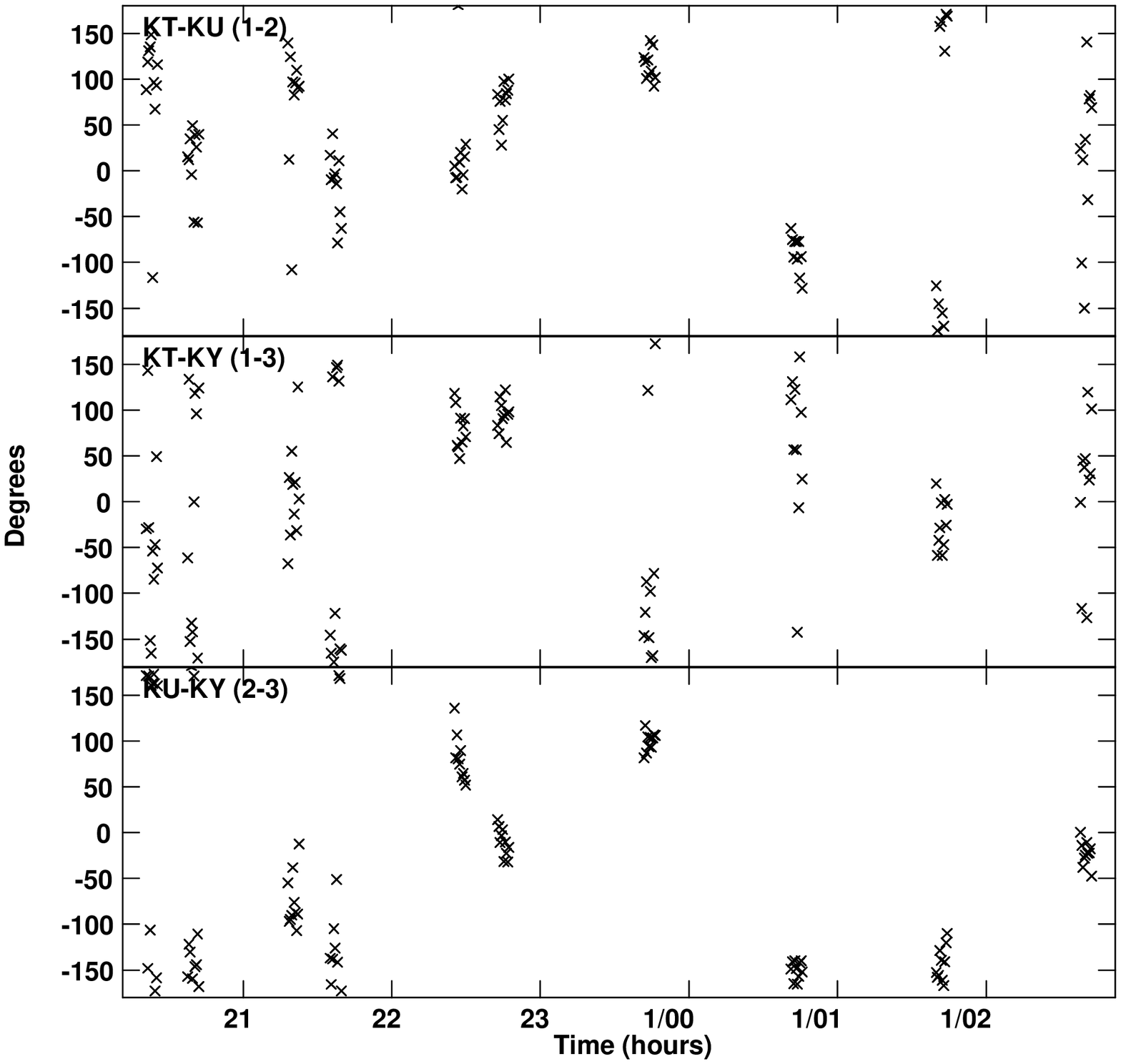}
\includegraphics[angle=-90,width=57mm]{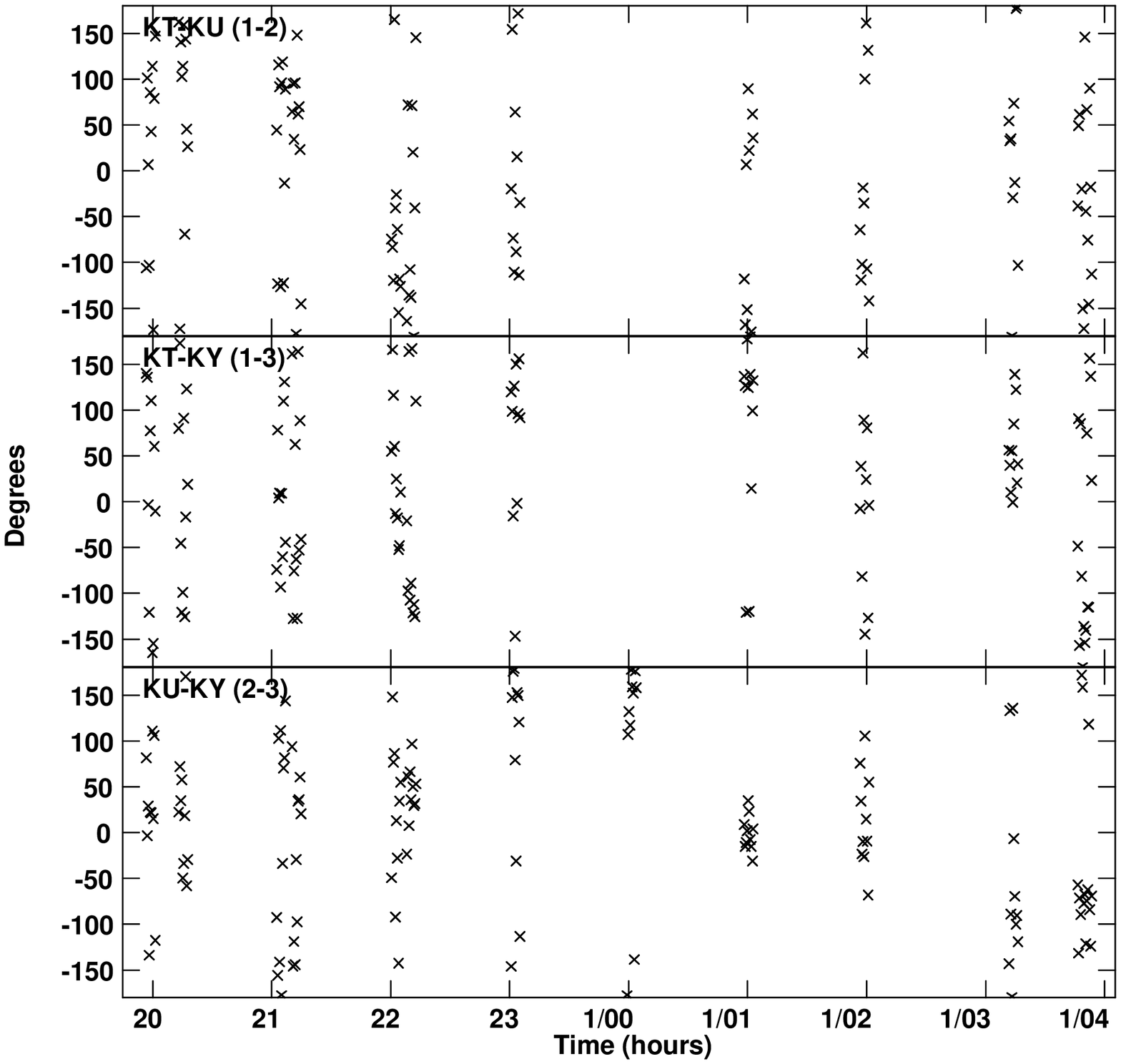}
\begin{tabular}{lll}
\hspace{0.8cm}(m) 3C454.3 &\hspace{3.3cm} (n) 3C84 &\hspace{3.3cm} (o) 4C28.07\\
\end{tabular}
\includegraphics[angle=-90,width=57mm]{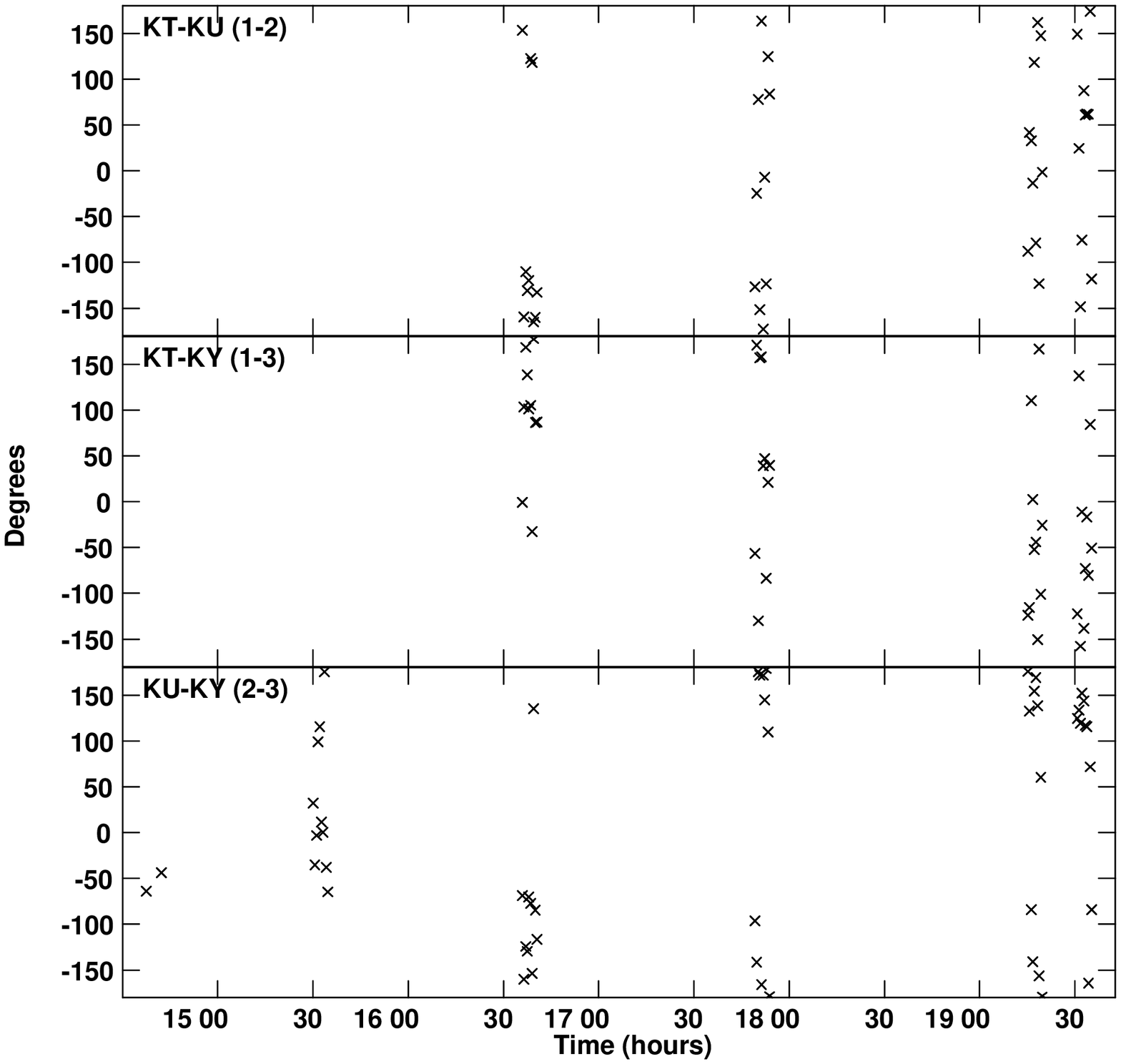}
\includegraphics[angle=-90,width=57mm]{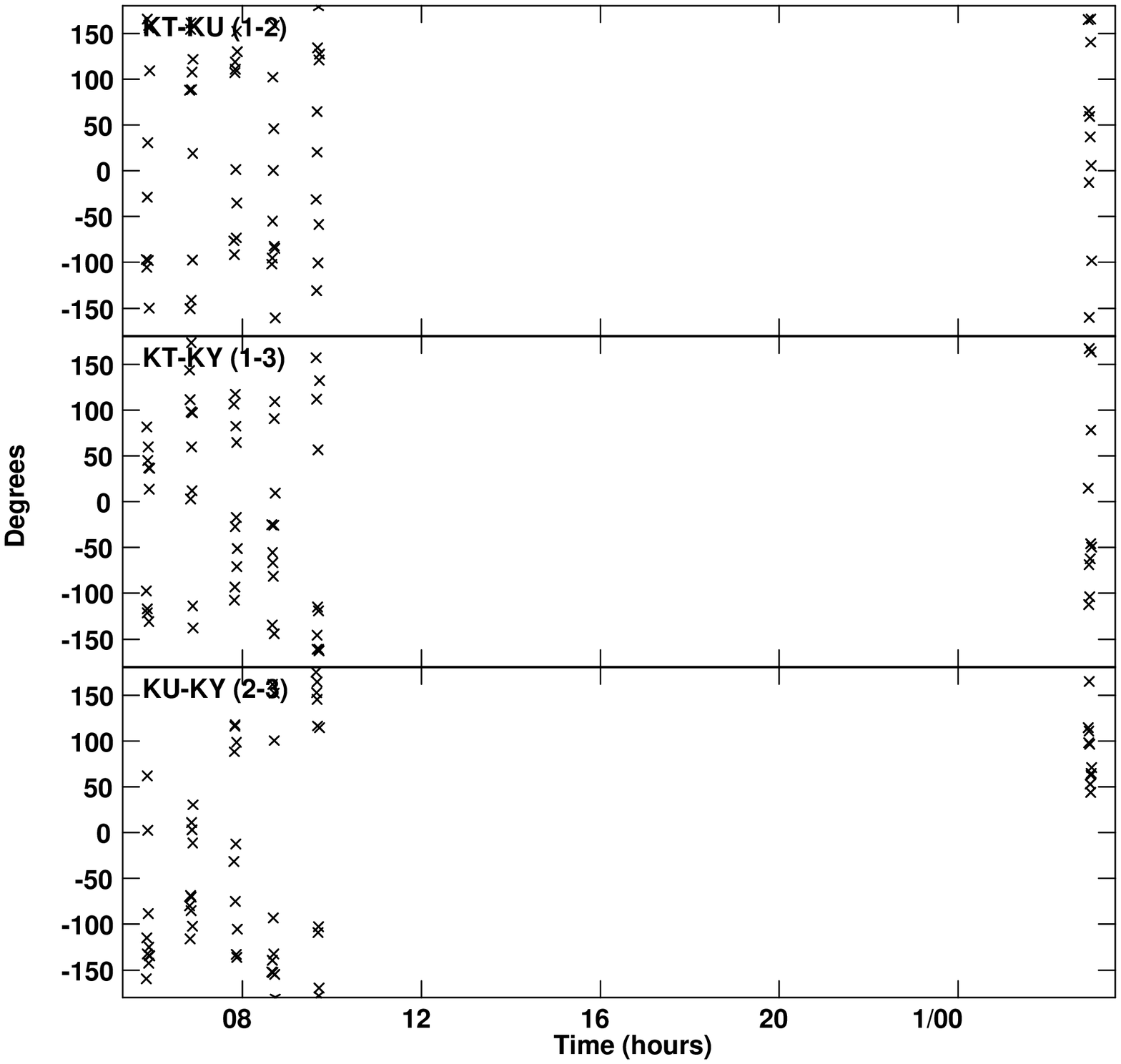}
\includegraphics[angle=-90,width=57mm]{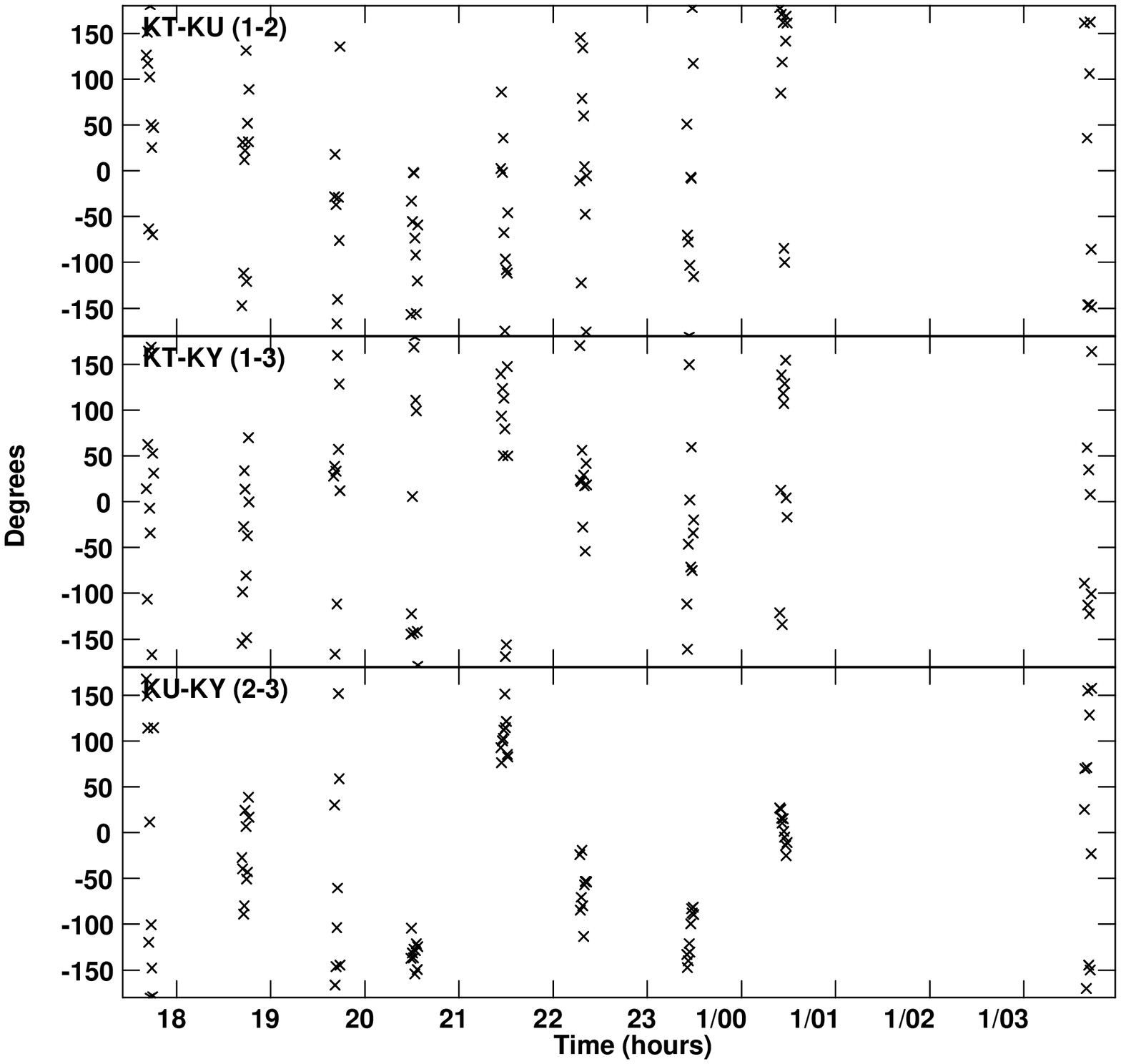}
\begin{tabular}{lll}
\hspace{0.8cm}(p) 4C38.41 &\hspace{3.3cm} (q) 4C39.25  &\hspace{3.3cm} (r) BL Lac\\
\end{tabular}
\includegraphics[angle=-90,width=57mm]{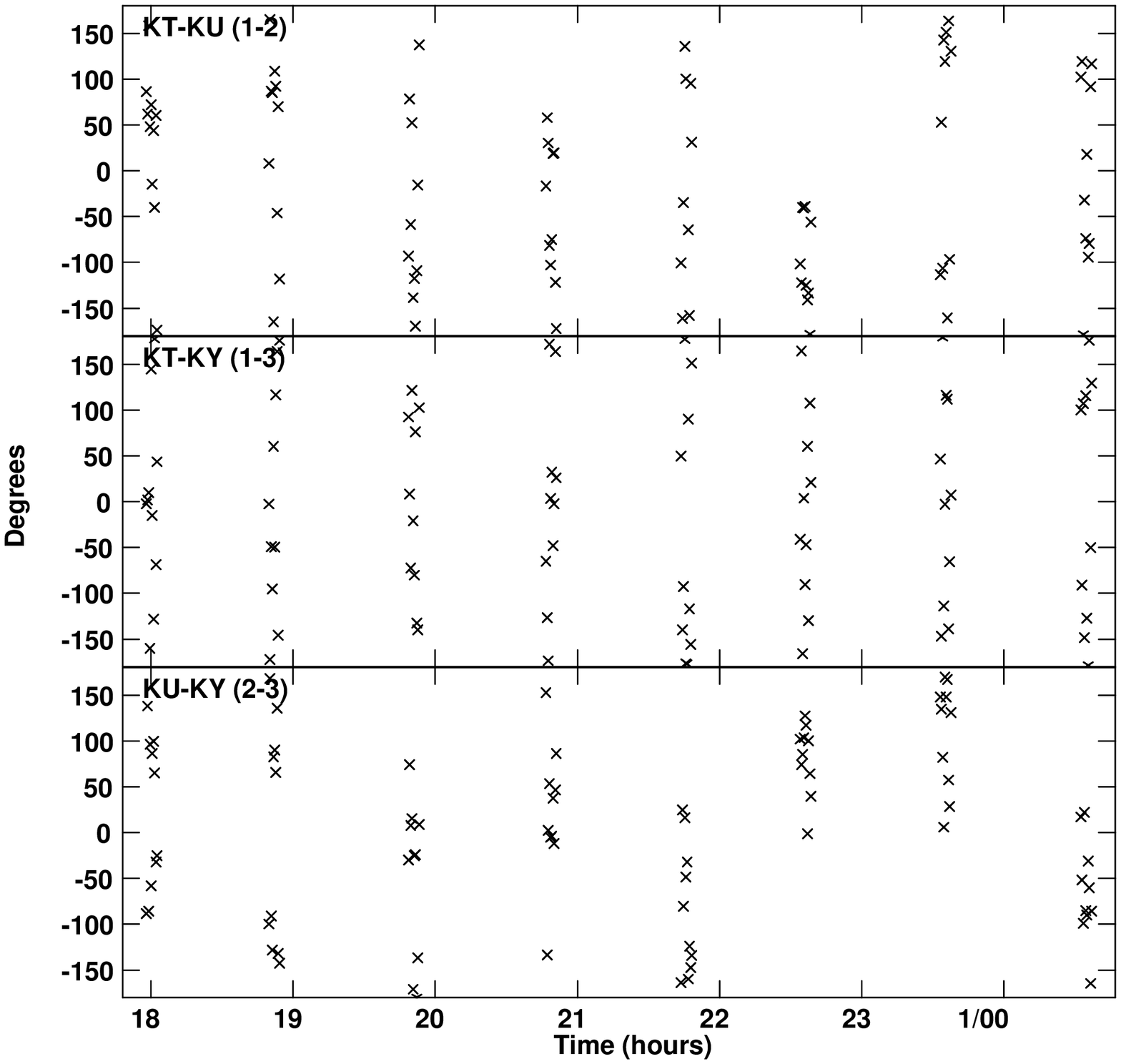}
\includegraphics[angle=-90,width=57mm]{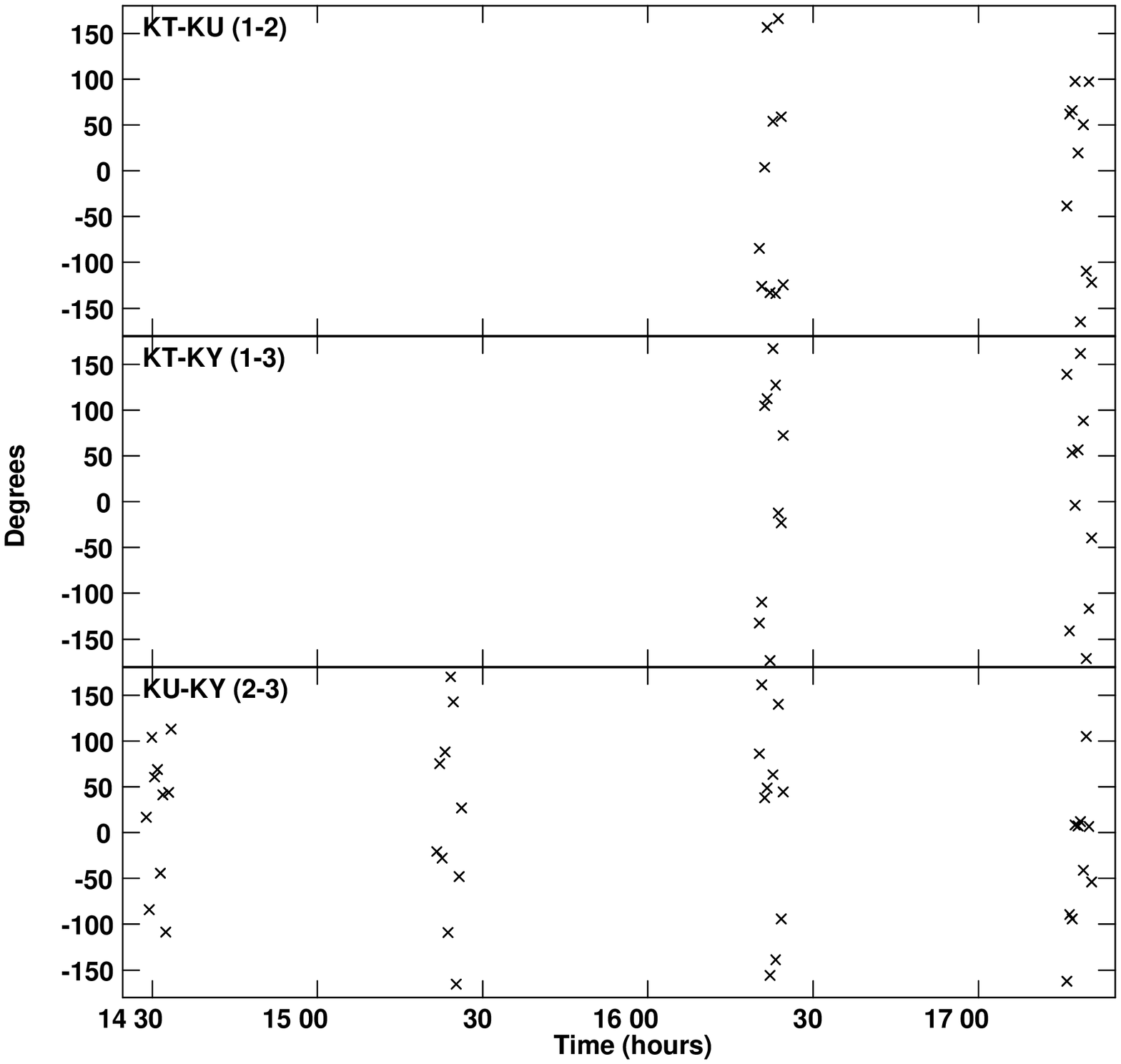}
\includegraphics[angle=-90,width=57mm]{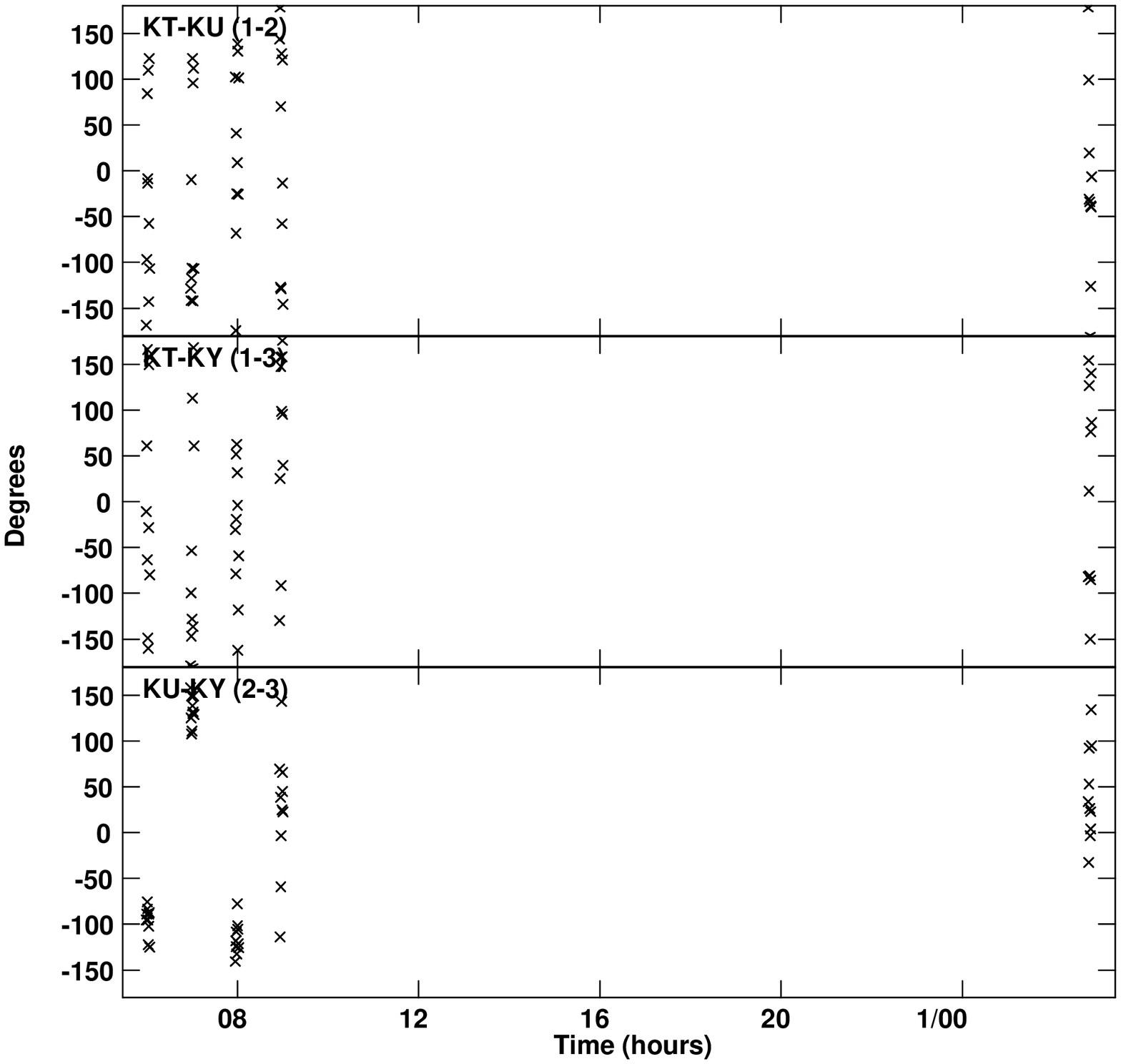}
\begin{tabular}{lll}
\hspace{0.8cm}(s) CTA102 &\hspace{3.3cm} (t) NRAO530 &\hspace{3.3cm} (u) OJ287\\
\end{tabular}
\\
\begin{flushleft}
\small{\textbf{Figure 9. } -- Cont.}
\end{flushleft}
%\vspace{5mm} %% add extra space ONLY when figures/tables are "colliding"!
\end{figure*}

\begin{figure*}[h]
\centering
\includegraphics[angle=-90,width=57mm]{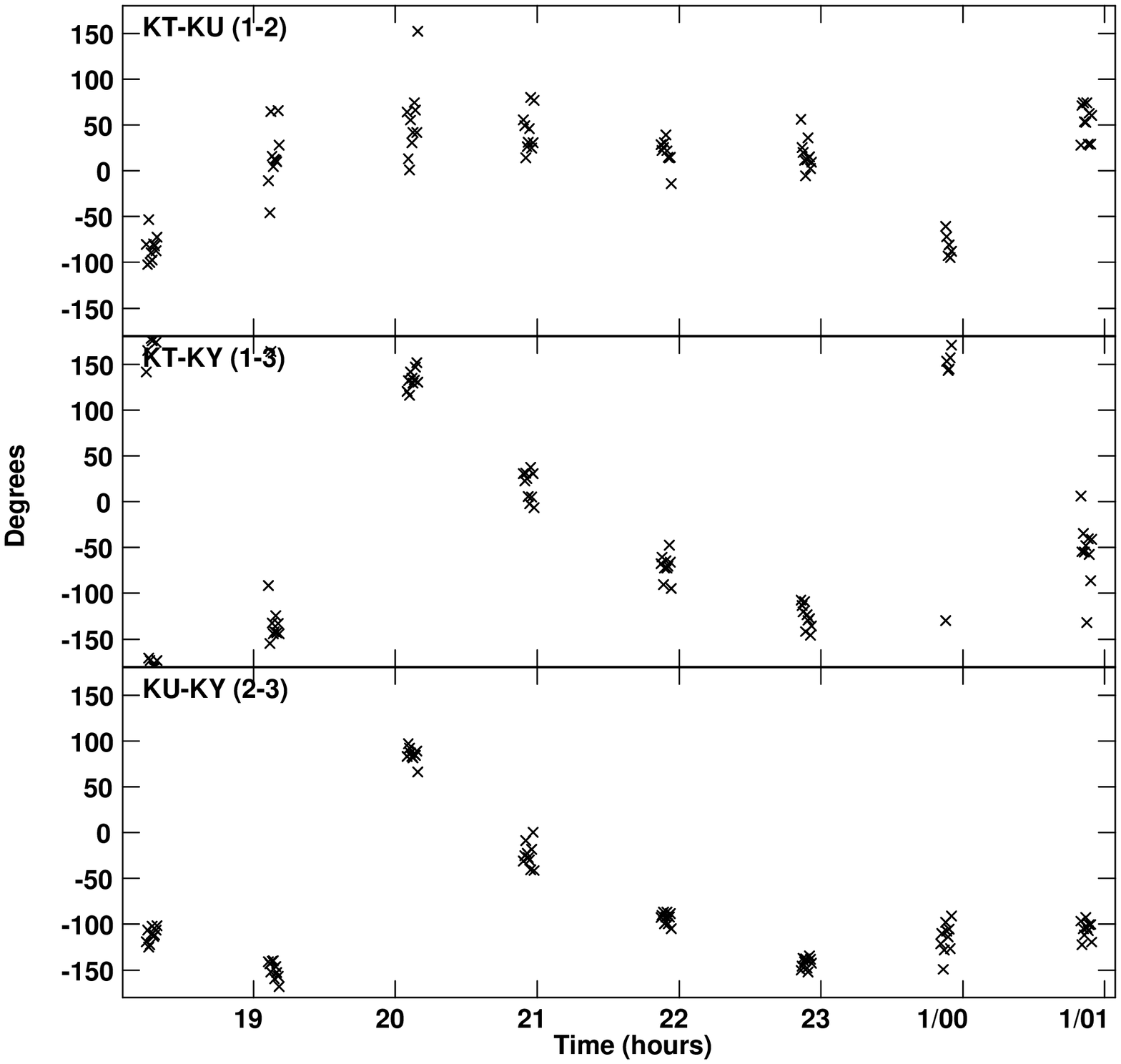}
\includegraphics[angle=-90,width=57mm]{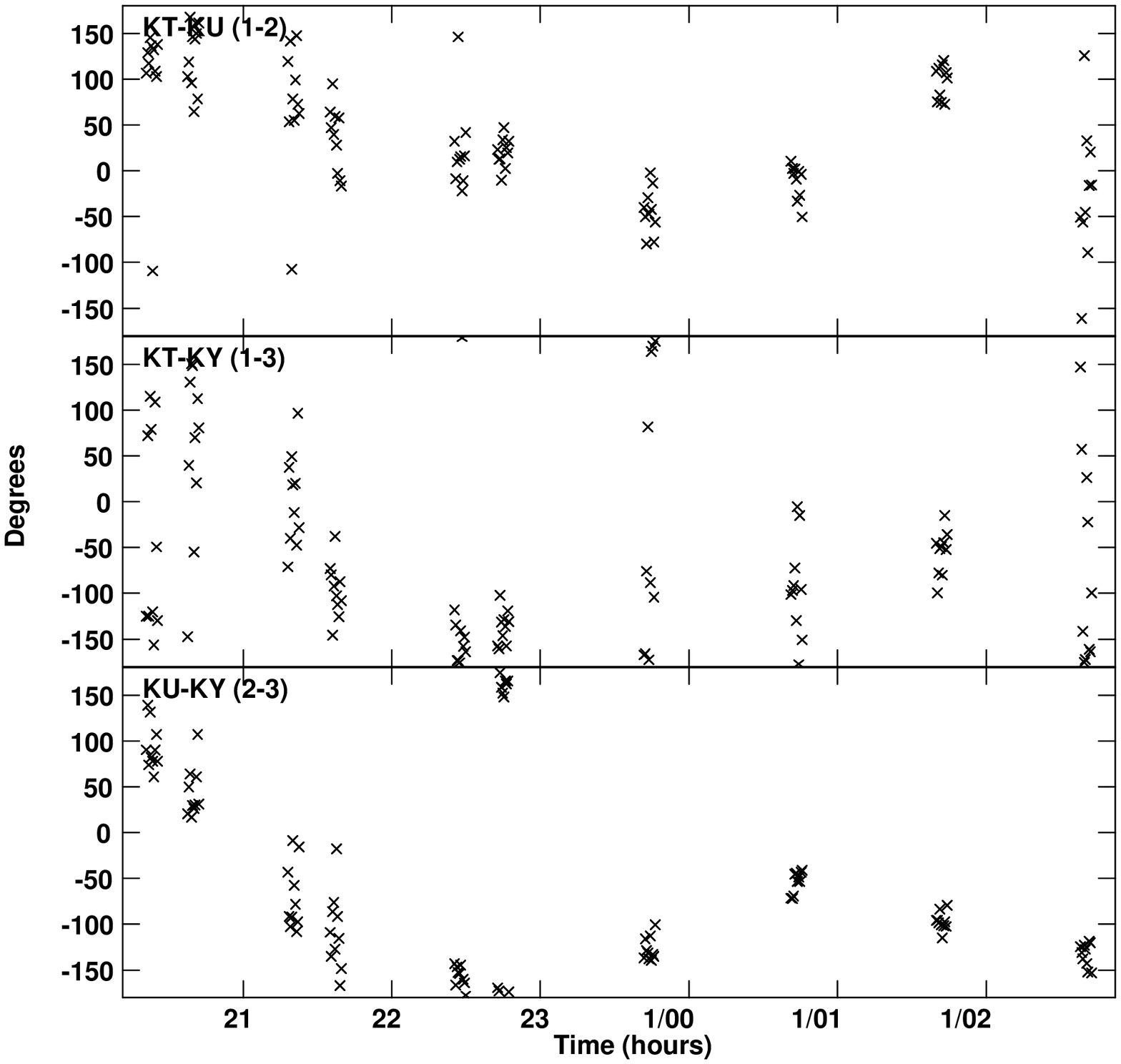}
\includegraphics[angle=-90,width=57mm]{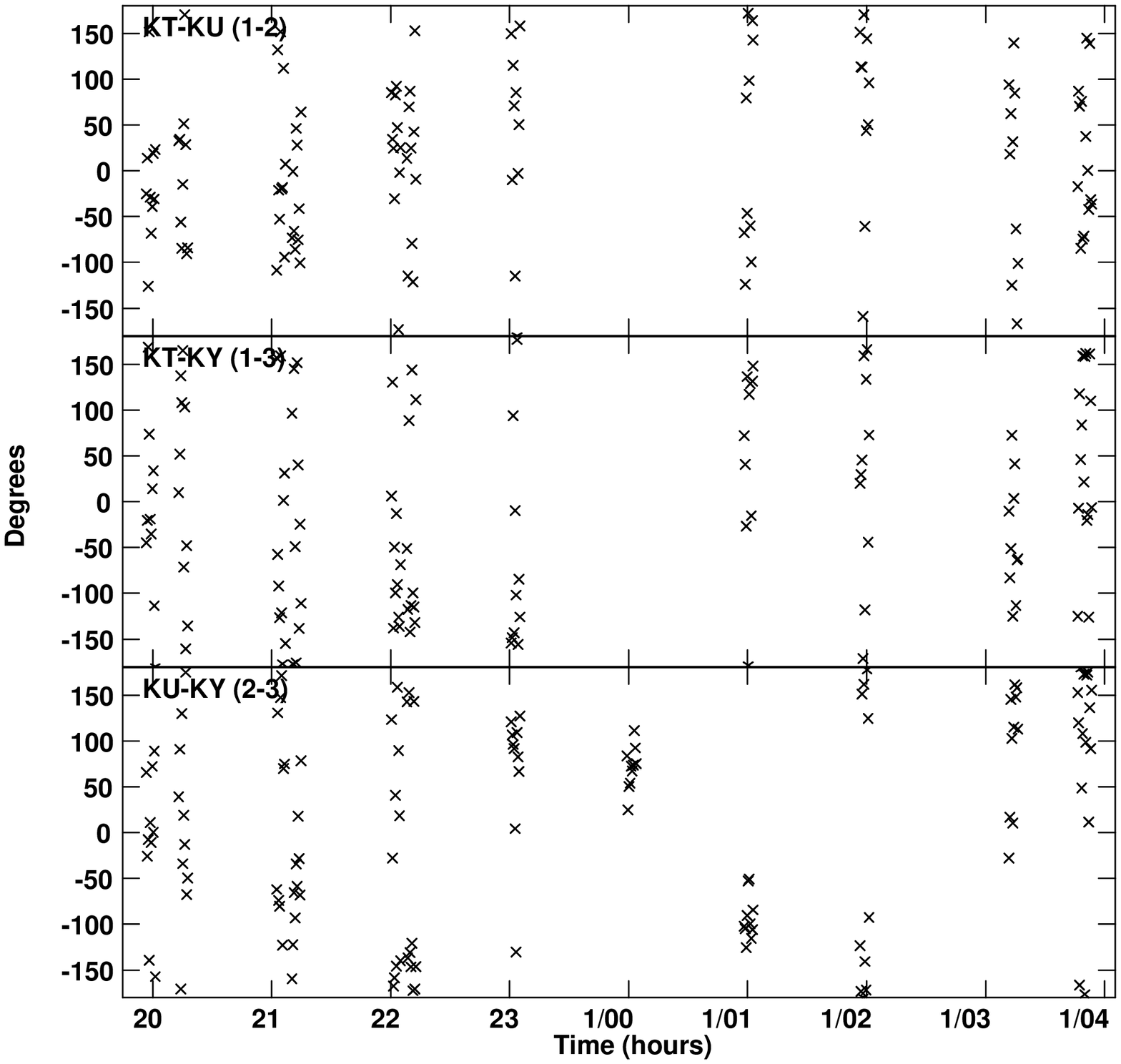}
\begin{tabular}{lll}
\hspace{0.8cm}(m) 3C454.3 &\hspace{3.3cm} (n) 3C84 &\hspace{3.3cm} (o) 4C28.07\\
\end{tabular}
\includegraphics[angle=-90,width=57mm]{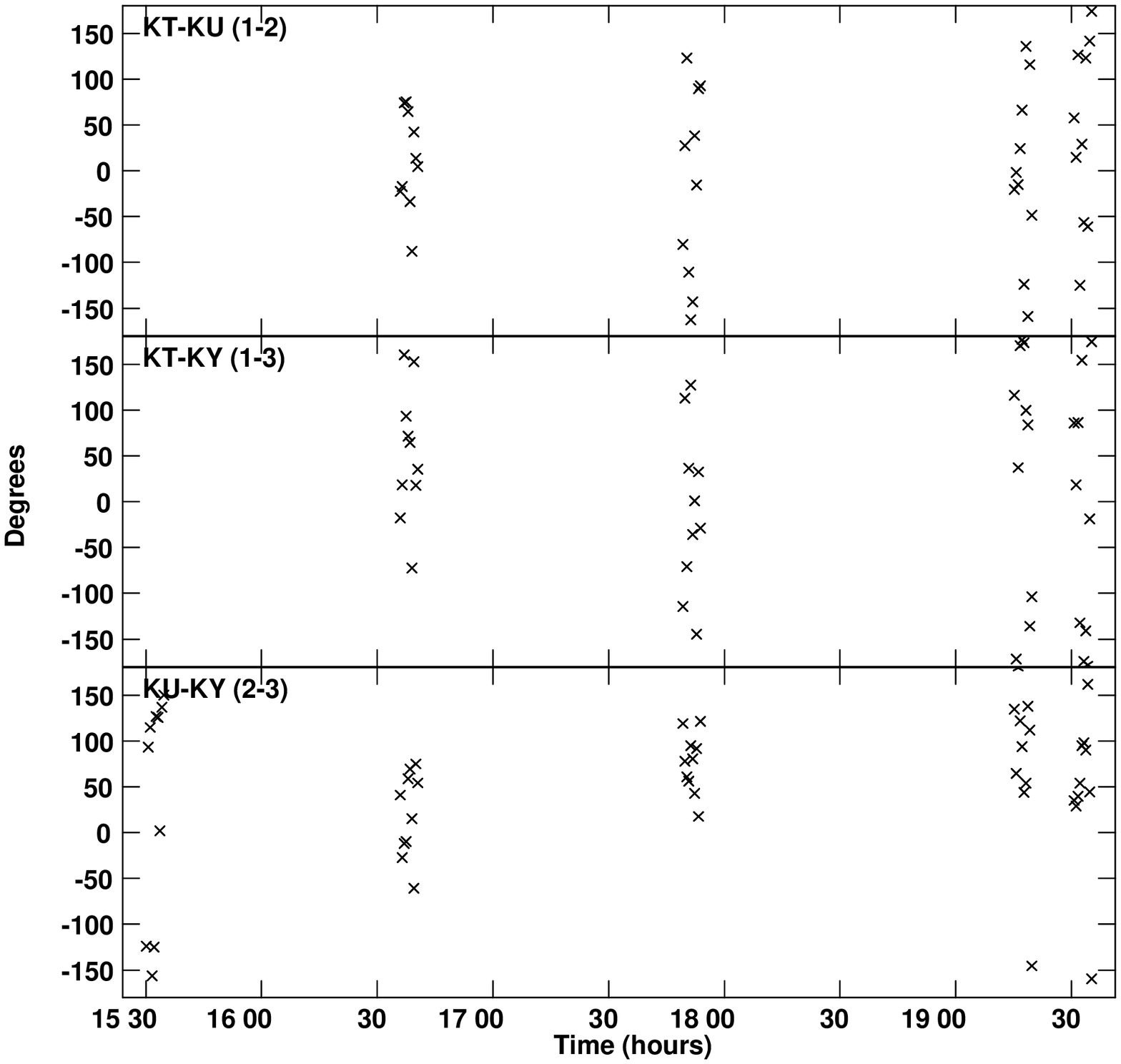}
\includegraphics[angle=-90,width=57mm]{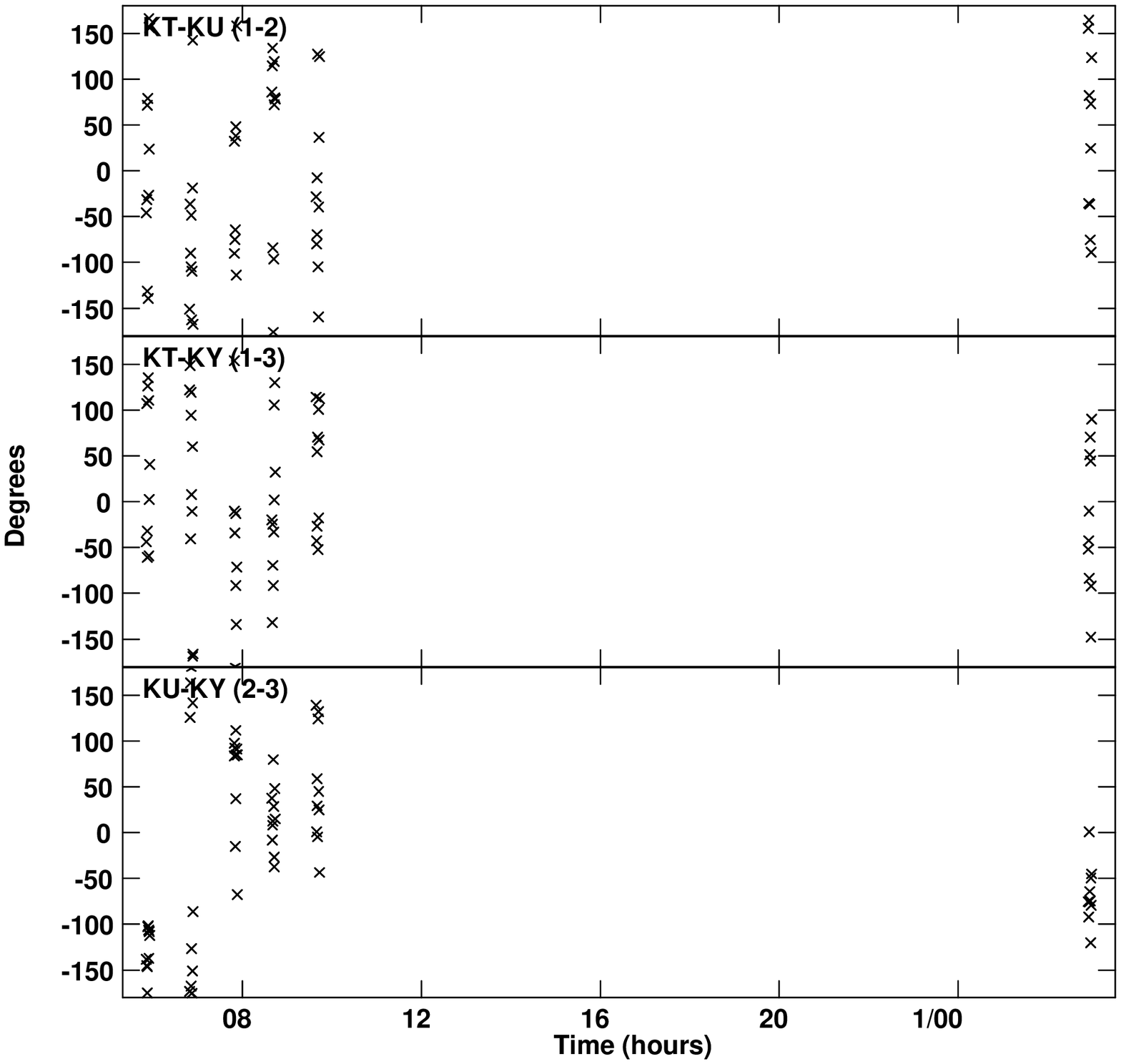}
\includegraphics[angle=-90,width=57mm]{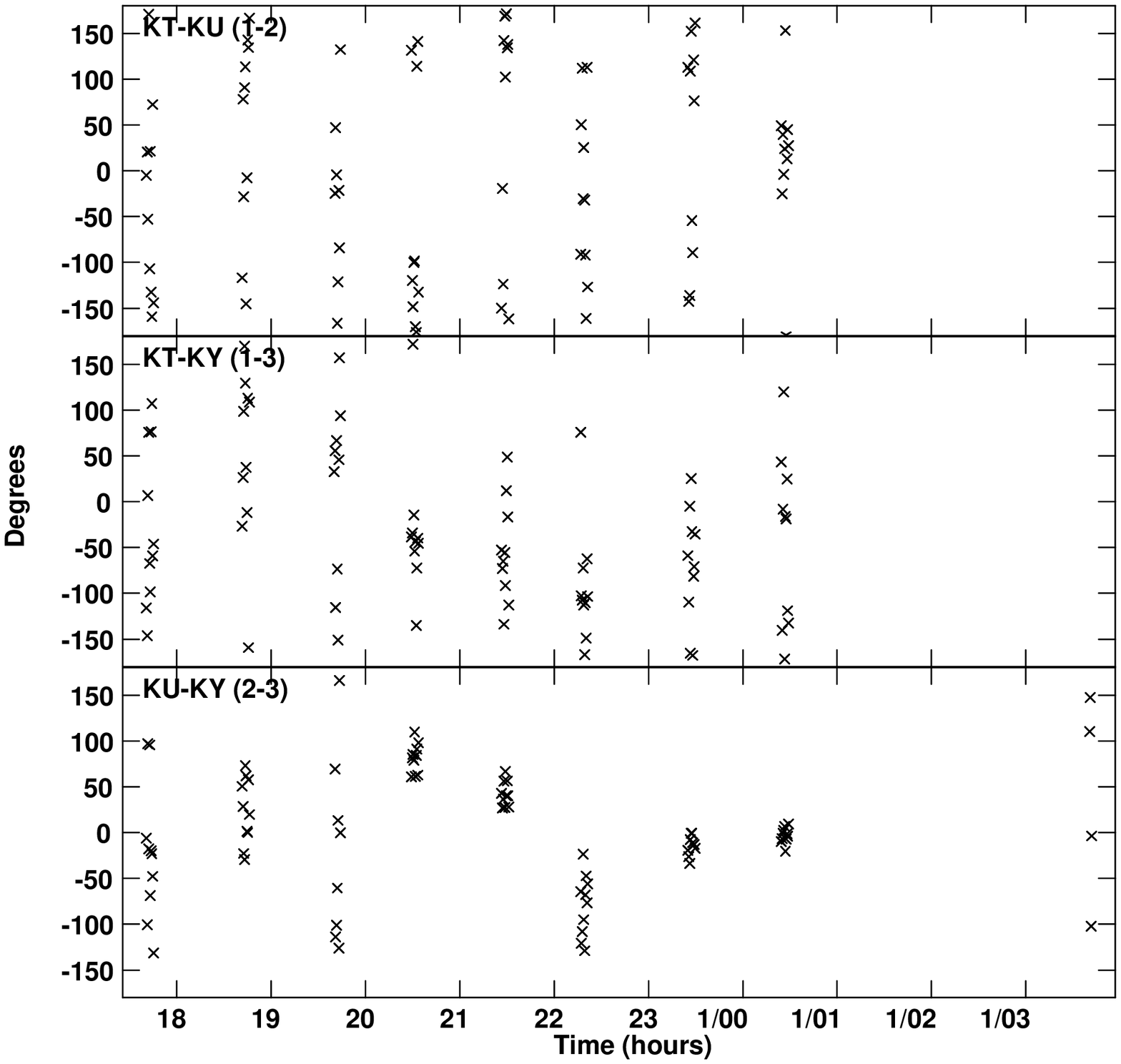}
\begin{tabular}{lll}
\hspace{0.8cm}(p) 4C38.41 &\hspace{3.3cm} (q) 4C39.25  &\hspace{3.3cm} (r) BL Lac\\
\end{tabular}
\includegraphics[angle=-90,width=57mm]{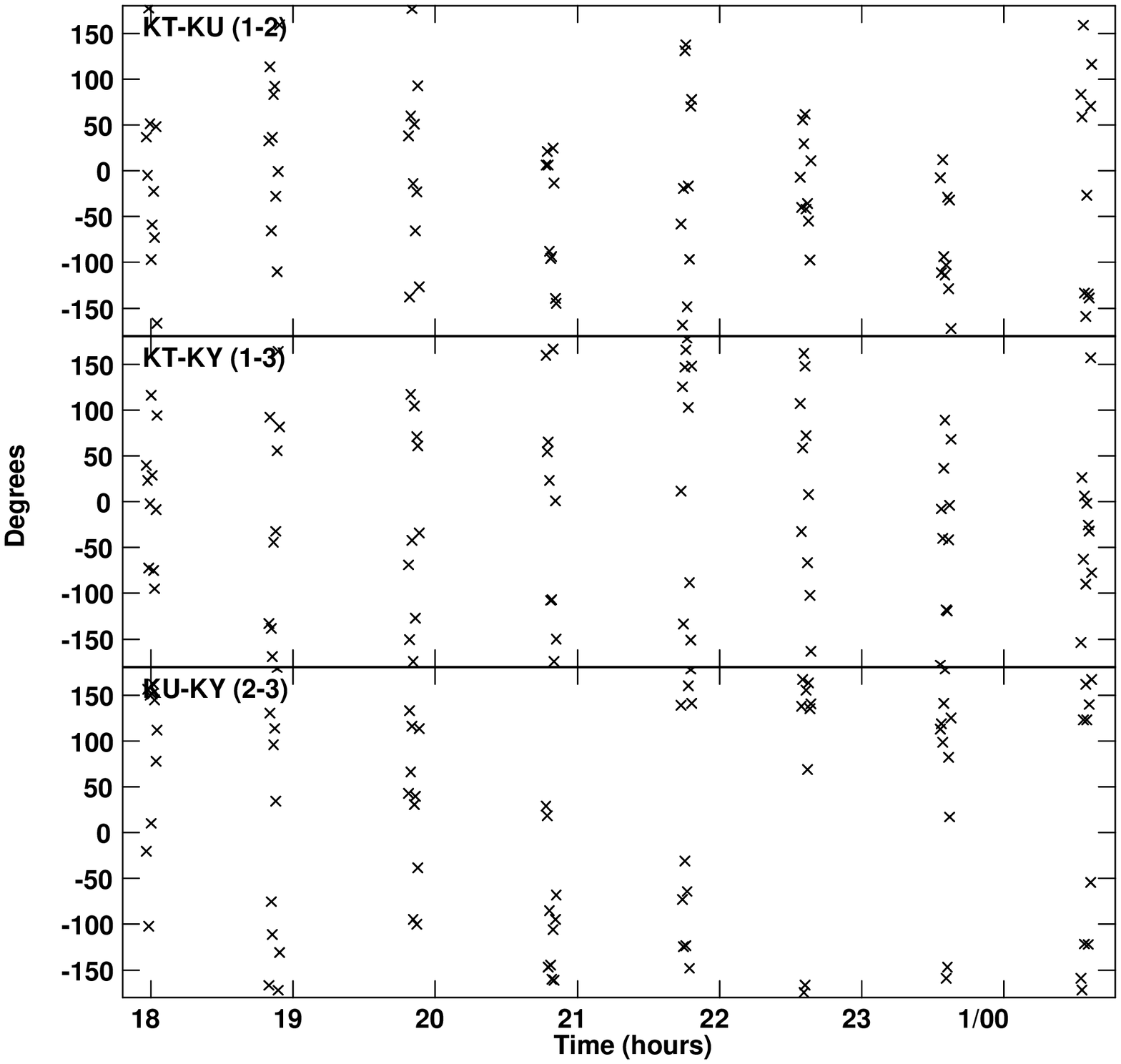}
\includegraphics[angle=-90,width=57mm]{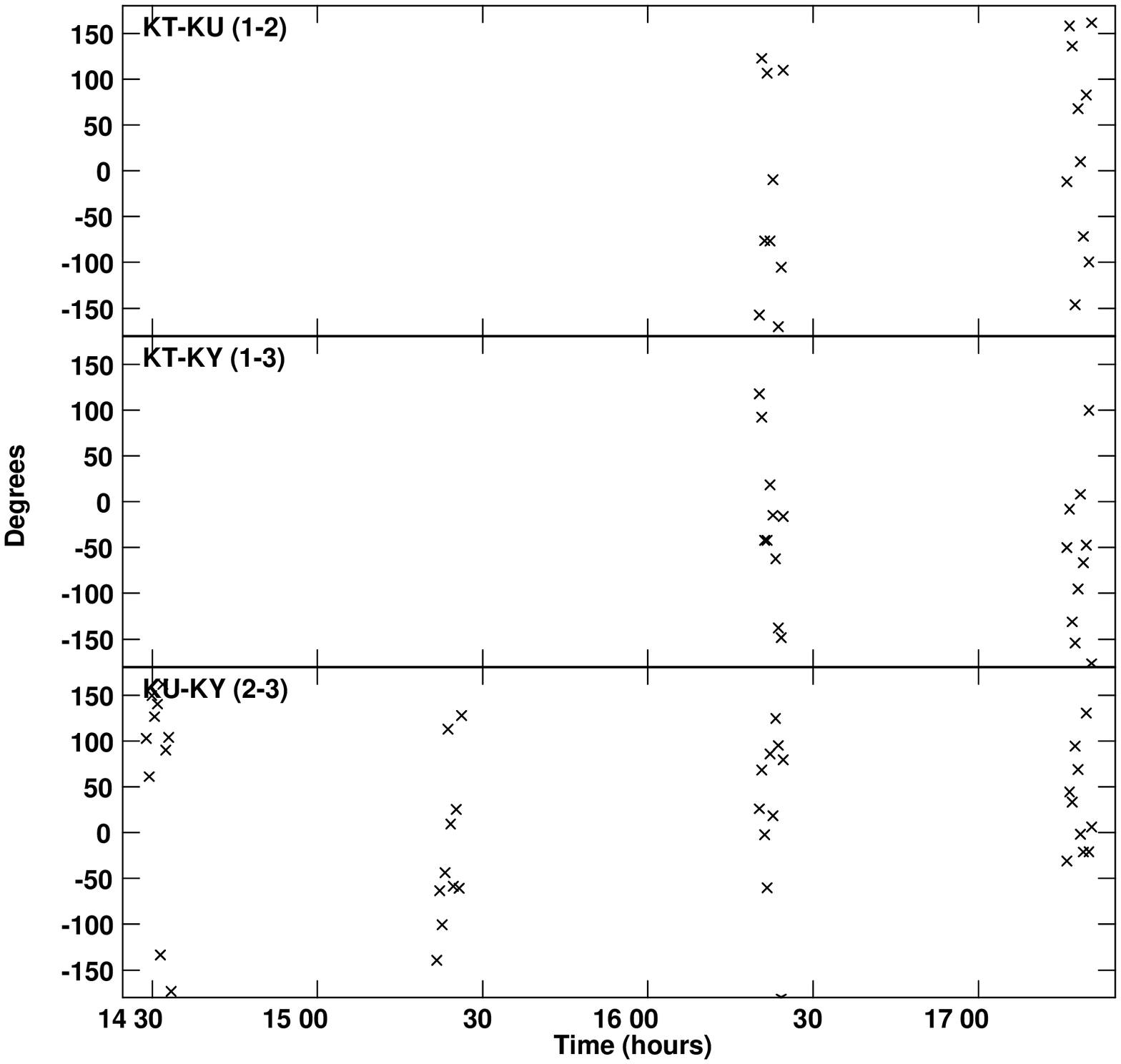}
\includegraphics[angle=-90,width=57mm]{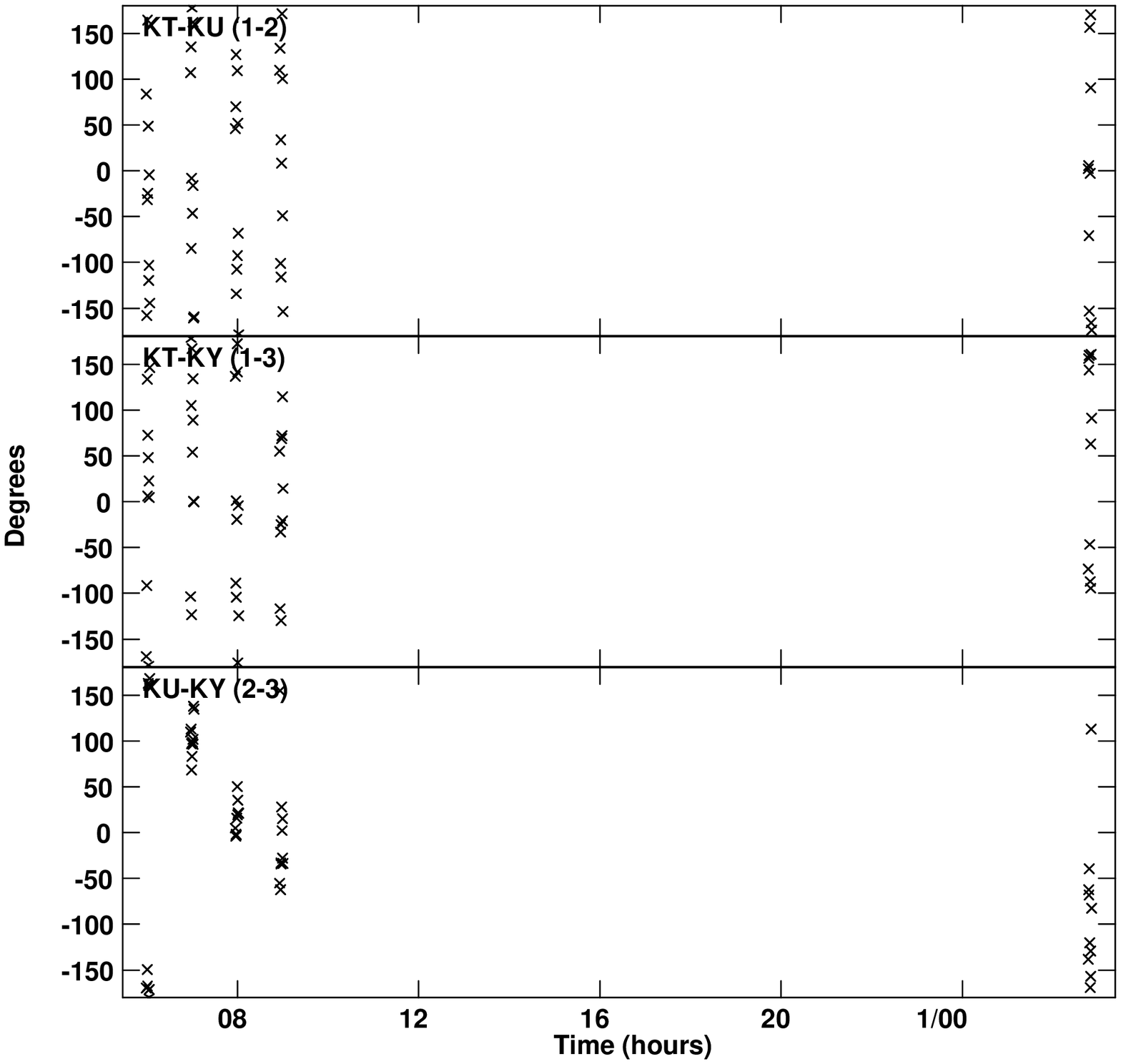}
\begin{tabular}{lll}
\hspace{0.8cm}(s) CTA102 &\hspace{3.3cm} (t) NRAO530 &\hspace{3.3cm} (u) OJ287\\
\end{tabular}
\\
\begin{flushleft}
\small{\textbf{Figure 10. } -- Cont.}
\end{flushleft}
%\vspace{5mm} %% add extra space ONLY when figures/tables are "colliding"!
\end{figure*}

\section{Discussion}
Many factors can affect the quality of the frequency phase transfer, including weather conditions (hence opacity, system temperature or system equivalent flux density), baseline length, source intensity and structure, and experimental design used (accuracy of the input model, frequency pair used). We will now  discuss these effects in more detail.

The selection of the reference antenna is important to a level similar to other VLBI analysis. The ideal reference antenna is the one with systematically lower system temperatures and instrumental errors, as the latter will still be present even after manual phase calibration corrections, and will be transferred from the lower to the higher frequencies. Taking these two factors into consideration, together with an adequate integration solution interval, we can obtain the percentage of successful phase solutions, which we initially intend to maximize. We note however that, although the number of successful solutions steadily increases from 78\% to 85\% when adopting solution intervals from 0.1 to 5 minutes, the scatter and uncertainties of the solutions increase for integration times$\>$0.5 minutes. This makes the 0.5~minutes as the the optimal solution interval \citep[see][for a more detailed description]{Wajima15}.

Weather conditions are also very important. First, poor weather increases the effective SEFD, which in turn limits the number of sufficiently high SNR reliable phase (delay/rate) solutions. For example, the good weather conditions in iMOGABA9 and a fringe fitting with solution interval of 0.5~minutes result in about 80\% good solutions, considering all frequencies. Contrary, for iMOGABA15, with poor weather, this number drops down to ~50\%. Second, the stability and clustering of these solutions at lower frequencies is crucial when extrapolating to higher frequencies. Finally, the maximum integration time that we will be able to achieve will be ultimately limited by atmospheric conditions. 

Existing observations indicate that the limiting factors in the integration time are the ionospheric and instrumental residual errors. 
In the current iMOGABA schedule however, it is not possible to adequately discuss the latter in quantitative terms. Given the ``snapshot mode'' with an on--source scan time of $\lesssim$5~minutes, and typically limited to $sim3$ scans per source over 24 hours (scan separation of about few hours), we cannot obtain coherence information for these observations. Note however that for some particular sources (e.g., 3C454.3, CTA102, BL Lac) a larger number of scans is performed, and a clear trend over various hours is seen in the phase solutions (see e.g., Figures \ref{iM15-FPTphase22-86}, \ref{iM15-FPTphase43-86}, \ref{iM15-FPTphase22-129} and \ref{iM15-FPTphase43-129}), indicating a large coherence time after the FPT. This is in agreement with existing measurements of the FPT coherence time with KVN antennas at 129~GHz of about 20~minutes \citep{Rioja14b,Rioja15}.

One of the advantages of KVN is the multi--frequency simultaneous observations.  \cite{Middelberg05} and \cite{Rioja11} considered VLBA fast frequency switching observations with 86~GHz which, unlike the simultaneous KVN observations,  introduced interpolation errors arising from the frequency switching cycle, that degrade the performance. \cite{Rioja14a} presents a comparative study between fast frequency switching VLBA and simultaneous KVN observations at 22/44~GHz. A quantitative discussion of the coherence time after FPT on multi--frequency KVN observations can be found in \cite{Rioja15}.

We calculated sensitivity limits for various combinations of frequencies and integration times and obtained the expected SNR for each source based on their expected flux.  Our results for iMOGABA9 seem to indicate that a SNR$>5$ is a minimum requirement for source detection after FPT, whereas larger values are generally preferred for as many scans as possible for later imaging. As we noted (e.g., for 0235+164 or 1343+451at 86~GHz), it is possible that some baseline--based phase solutions may have large scatter after the FPT, resulting in failed solutions after the re-fringe fitting, thus failing to obtain a proper self--calibration. Although no images were produced for iMOGABA15, results in this epoch seem consistent.

It is important to discuss the effects of the amplitude error calibration in the maps as well as in the measured fluxes. In principle, we know that the image noise level is a lower limit, as residual phase and amplitude errors affect the image quality and dynamic range. We first note that amplitude errors arising from FPT come only in the form of error propagation of the residual phase after amplitude vector averaging. An expression for the thermal noise phase error from FPT can be found in \cite{Rioja11} and takes into account the phase error of the high frequency plus a weighted error from the transferred lower frequency by a factor of $R^2/2$. However, since the KVN system consists of only three antennas, it is not possible to perform amplitude calibration. One of the ways to estimate the latter is to compare the gain performance of KVN with other arrays. Observations with KaVA, where KVN antennas are a sub--array element, can be useful in this sense and show that uncertainties in the gain calibration are less than 10\% \citep{Niinuma14}. An extensive study on amplitude calibration errors for iMOGABA at all four frequencies will be given elsewhere \citep{Wajima15}.

We also tested for stability of the solutions based on the source model accuracy. To do this, we performed the FPT again, but transferring the results from a typical fringe fitting at the lower frequency without the addition of a model image. The resulting phase solutions are very similar, if not identical. This is because, with our current setup, most of the sources are point--like or highly core--dominated. For example, 3C446, which is the most extreme case of resolved structure in our work, has a peak flux of 0.28 and an integrated flux of 0.31 Jy/beam, and more than 90\% of the flux arises as a point--source.

The comparison of the phase transferred solutions using two different reference (low) frequencies studied here can help us to estimate how to choose the ideal reference frequency for the FPT. Two elements play a key role during the FPT: the weather conditions and the spectral index of the source. For sources with a flatter spectral index, using a higher frequency as the reference is preferred. This minimizes the extrapolation effects by a small frequency ratio $R$ factor, while we do not lose a significant amount of SNR due to the decrease of the flux at higher frequencies. On the other hand, when dealing with steep spectra sources, then a lower reference frequency is be suggested, as SNR detection can be de deciding factor rather than weather conditions. This is particularly true for the weakest sources, which we aim to detect with FPT.

The analysis and discussions included in the series of iMOGABA papers, including in particular efforts towards techniques to improve the observational products, such as the FPT described here, are an important cornerstone to improve the overall performance in the next iMOGABA program observations. We can consider including also the SFPR or related strategies as a way to increase the  sensitivity of KVN for detecting and hence monitoring weak sources or under poor weather conditions. One of the possibilities is to observe a sample of pre--selected sources with the SFPR mode in another session, followed by the normal iMOGABA session. Hence we may be able to substantially increase the coherence time and, if available, to conduct relative astrometric observations on other interesting, but bright, sources.

\section{Conclusions}

We have applied the frequency phase transfer (FPT) technique to high frequency iMOGABA data in order to detect weak sources that cannot be detected and imaged with standard VLBI reduction methods. To evaluate the capabilities of this method we have tested it on two data sets, one epoch (iMOGABA9) with very good weather and high quality, and another epoch (iMOGABA15) with bad weather and several issues, including antenna offsets and amplitude problems.

Our results indicate that we are able to perform the FPT and obtain detections and imaging for weak sources in agreement with theoretical  expectations for  sensitivity limits based on integration times of 5 minutes. We have succeeded even for the iMOGABA epoch with the worst conditions to date, which demonstrates the suitability of KVN antennas for this technique.

Although a quantitative analysis of the resulting coherence time cannot be performed adequately due to the ``snapshot'' characteristics of the iMOGABA observations, we suggest that it can reach values larger than 5 minutes in our data sets. One of the reasons for this is the multi--frequency capabilities of KVN, allowing to transfer the phase solutions to higher frequencies without any time interpolation, and the resulting large errors.

We have checked the consistency of our results by performing a series of tests, including frequency phase transfer from two frequencies, 22 and 43~GHz. We obtain similar results, consistent with expectations based on source flux and frequency ratios. Comparison of these results indicate how to estimate the ideal reference frequency to be used for the FPT, depending on sources' spectral index and weather conditions. We have also checked for stability of the solutions under different reference antennas and source models. We have discussed the impact of weather, system conditions such as system temperature and system equivalent flux densities, and baseline lengths. In all cases, our results are in well agreement with expectations.

\acknowledgments
We are grateful to all staff members in KVN who helped to operate the array and to correlate the data. We are very grateful for fruitful discussions with R. Dodson and M. Rioja. The KVN is a facility operated by the Korea Astronomy and Space Science Institute. The KVN operations are supported by KREONET (Korea Research Environment Open NETwork) which is managed and operated by KISTI (Korea Institute of Science and Technology Information).

\end{document}